\documentclass[11pt]{article}

\usepackage[T1]{fontenc}
\usepackage[utf8]{inputenc}
\usepackage{lmodern}
\usepackage[margin=1in]{geometry}
\usepackage{graphicx}
\usepackage{amsmath,amssymb,amsfonts,amsthm}
\usepackage{booktabs}
\usepackage{array}
\usepackage{placeins}
\usepackage{float}
\usepackage{tikz}
\usetikzlibrary{arrows.meta,positioning}
\usepackage{authblk}
\usepackage[numbers,sort&compress]{natbib}
\usepackage[colorlinks=true,allcolors=blue]{hyperref}
\hypersetup{bookmarksdepth=subsection}

\theoremstyle{plain}
\newtheorem{lemma}{Lemma}[section]
\newtheorem{corollary}[lemma]{Corollary}

\title{Observable Spectral Framework for Noisy Amplitude Amplification: Stopping Stability and Correction}
\author[1]{Ge Zhong}
\author[1]{Charith Jayasekara\thanks{Corresponding author: \texttt{charith.jayasekara@monash.edu}}}
\author[1]{Nisal De Silva}
\author[1]{Muhammad Usman}
\author[1]{Muhammad Aamir Cheema}
\author[1]{Chetan Arora}
\affil[1]{Faculty of Information Technology, Monash University, Melbourne, Australia}
\date{}

\begin{document}
\maketitle

\begin{abstract}
Stopping rules for noisy amplitude amplification often rely on channel-specific
success-probability formulas, limiting their transfer across noise mechanisms.
We develop an observable spectral framework for stationary noisy standard
amplitude amplification with a known ideal success fraction, under explicit
leading-mode, spectral-separation, and residual-control conditions.
Within this class, the first amplification lobe
has a damped-oscillatory leading component and a bounded non-leading
contribution. We derive how frequency shift, damping, and observable phase
offset displace the leading peak, and establish sampled-gap criteria for
retaining the ideal integer stopping depth, certifying recalibration, or
selecting a robust candidate window. In single-target Grover search, frequency
and damping errors have stopping-coordinate sensitivities proportional to the
search-space size, whereas phase offsets scale with its square root; residual
modes can independently reverse the ordering of neighbouring depths.
Analytical constructions, circuit simulations, finite-shot inference, and
calibration-derived device-like trajectories examine these predictions and
their operational limits. For a 64-element search, an \emph{a posteriori}
sampled-set check selects depth five over six, with success probability
increasing from 0.595 to 0.626. The resulting theory links observable spectral
dynamics to conditional integer stopping guarantees without requiring a
separate microscopic channel solution.
\end{abstract}

\noindent\textbf{Keywords:} noisy amplitude amplification, observable spectral normal form, Grover search, quantum channel dynamics, residual bounds, sampled stopping stability, noise-aware stopping correction, finite-shot decision assessment

\medskip

\section{Introduction}
\label{sec:introduction}

Grover search and amplitude amplification are canonical quantum algorithms whose
performance depends on the success-probability trajectory and the number of
iterations executed
\cite{grover1996fast,boyer1998tight,brassard2002quantum}. Under noise, that
trajectory is determined jointly by the noisy iteration map, the initial state,
and the success measurement. Analytic expressions obtained for particular
phase-flip and depolarizing channels
\cite{azuma2005higherOrderGrover,cohn2016groverDepolarizing} generally cannot
be transferred unchanged to a different noise process. This creates a modeling problem for noise-aware algorithm execution:
one needs more structure than an arbitrary noisy channel supplies, but a useful
stopping theory should not depend on a separate closed-form solution for each
microscopic noise model. Prior work already establishes noise-induced peak
shifts and stopping corrections for specified channels; here we show how
observable spectral, residual-control, and sampled-gap conditions yield
conditional integer stopping guarantees within a defined class of noisy
dynamics, while distinguishing first-peak survival from same-step stability.

We address this problem through an intermediate, observable-level description
of \emph{stationary noisy standard amplitude amplification}. The framework
identifies a class of trajectories whose first amplification lobe is governed
by an observable-relevant, semisimple conjugate eigenvalue pair of the noisy
iterate, with non-leading contributions controlled on the same lobe. Spectral
separation, overlap, and residual bounds make the leading oscillation and its
error allowance explicit. The resulting damped-cosine normal form separates
frequency shift, damping, and observable phase offset from residual dynamics.
Section~\ref{sec:first-peak-preserving-noisy-aa} specifies the spectral
conditions defining this dynamics class, and the Supplement develops the
reduction and sufficient realizability routes. For near-term quantum devices, where a complete microscopic noise model may be unavailable but calibration data can constrain the observed success-probability trajectory, this observable-level formulation provides a route to noise-aware stopping decisions under the stated spectral and residual conditions.

The algorithmic consequence is an integer stopping decision. The continuous
first success peak and the implementable sampled maximizer are distinct, and a
useful peak may survive after the ideal integer depth ceases to be optimal.
Choosing an inappropriate depth can waste oracle calls, accumulate additional
noise, or provide a suboptimal amplification power to iterative and QPE-free
amplitude estimation
\cite{grinko2021iterativeQAE,suzuki2020amplitudeEstimationWithoutQPE}.
Within the spectral framework, continuous peak motion determines candidate
depths, while discrete probability gaps and bounded residuals determine whether
their ordering is reliable.

Two scales are therefore relevant. First-peak preservation asks whether a
useful amplification lobe survives; same-step stability asks whether the ideal
integer depth remains its best sampled choice. Frequency and damping
perturbations can shift the stopping coordinate on the search-space-size scale,
while an observable phase offset acts on its square-root scale in single-target
Grover search. A visible first peak can consequently admit substantially larger
perturbations than a fixed stopping decision. The analysis also distinguishes
horizontal displacement of the leading peak from direct value-level reordering
by non-leading modes. These mechanisms lead to retaining the ideal depth,
recalibrating to a certified nearby depth, or returning a robust local window
when a unique decision is not resolved.

The stopping theory assumes a known ideal success fraction and a stationary
noisy iterate with an observable-relevant semisimple leading spectral pair,
spectral separation, and controlled non-leading residuals over the first lobe.
Integer-depth certificates additionally require suitable sampled-probability
gap conditions. Single-target Grover search provides the main asymptotic
specialization. We assess operational relevance through circuit, finite-shot,
and backend-derived simulations, including an amplitude-estimation-localized
application.

The main contributions are as follows.
\begin{itemize}
    \item We formulate an observable spectral framework for a class of stationary
    noisy amplitude-amplification trajectories with a dominant oscillatory
    eigenvalue pair and controlled non-leading modes. We derive the resulting
    normal form and residual bounds, and give sufficient perturbative and exact
    channel realizations of the assumptions.
    \item Within this class, we derive a mechanism-resolved first-peak
    displacement law for effective-angle shift, damping, and observable phase
    offset, including its local remainder. Its single-target specialization
    quantifies the distinct stopping sensitivities and separates first-peak
    survival from fixed same-step stability.
    \item We connect the leading dynamics to integer decisions using sampled
    gaps, directional margins, and curvature-scaled residual allowances. The
    resulting rules retain the ideal stopping depth, certify a recalibrated
    depth, or return a robust local candidate window as the available
    information permits.
    \item We characterize residual-induced sampled reversal, in which
    non-leading observable contributions change the ordering of neighbouring
    depths without removing the leading first peak.
    \item We test the theoretical predictions using analytical constructions
    and simulated noise settings, assess finite-shot stopping decisions and
    matched baselines, and examine calibration and reuse costs through an
    amplitude-estimation-localized application.
\end{itemize}

Sections~\ref{sec:related-work}--\ref{sec:formulation} position
the approach, define its setting, and develop the normal form, stability,
correction, and residual-reversal results. Sections~\ref{sec:numerical-validation}
and~\ref{sec:practical-algorithmic-implications} present the validation and
operational implications before Section~\ref{sec:conclusion} concludes.

\section{Related Work}
\label{sec:related-work}

Noise-dependent success probabilities and stopping depths have been studied
under several explicit error models. Azuma developed a higher-order
perturbation expansion for Grover search with independently occurring
single-qubit phase-flip errors, examining success-probability degradation,
thresholds, and an earlier first peak under noise
\cite{azuma2005higherOrderGrover}. Cohn \emph{et al.} studied both total
(full-register) and local (per-qubit) depolarizing channels
\cite{cohn2016groverDepolarizing}. For total depolarization they obtained an analytic success trajectory and
a weak-noise correction of approximately \(-N\gamma_{\mathrm{TD}}/8\) to
the optimal iteration count, where \(\gamma_{\mathrm{TD}}\) is their
per-iterate full-register error rate; for local depolarization they developed
bounds and approximations. Our encoded-subspace realization in
Eq.~\eqref{eq:encoded-depolarizing-displacement} instead gives the continuous
first-peak displacement \(-p_N/(16\theta_N^2)\sim-Np_N/16\) for weak
\(p_N\). The corrections share an \(O(N\times\text{noise rate})\)
scale, but the full-register success probability includes a depth-dependent
maximally mixed contribution, unlike the constant background in our
encoded-subspace observable; the optimum and the continuous peak are also
distinct objects. These examples establish noise-adjusted stopping in
specified channel models. Our starting point is instead an
observable-relevant oscillatory spectral pair and controlled non-leading
contributions, defining a class on which common displacement, sampled-gap,
and correction statements can be made. The conditions needed for this reduction are specified in
Section~\ref{sec:first-peak-preserving-noisy-aa}.

Further studies treat stochastic error models
\cite{pablo1999noise,salas2008noise}, unitary gate imperfections
\cite{shapira2003effect}, and locally acting noise with time correlations
\cite{mandal2023invariance}. Shapira \emph{et al.} analyze imperfect unitary
operations, providing a related context for the coherent circuit-error test
in Subsection~\ref{subsec:coherent_circuit_validation}, although the perturbed
gates and error models differ. The time-correlated setting of Mandal
\emph{et al.} also illustrates a distinct noise regime relative to the fixed
per-iterate map adopted here. Experimental realizations include NMR,
trapped-ion, and programmable-processor searches
\cite{chuang1998experimentalFastSearch,brickman2005trappedIonGrover,
figgatt2017completeGrover,mitra2009robustNMRSearch}, with near-term and
QSVT-oriented circuit studies providing further implementation contexts
\cite{ijaz2023noise,zhang2022nisq}.

Spectral approaches provide another perspective on noisy
quantum dynamics. Reitzner and Hillery used the spectral
structure of localized-dephasing Grover evolution to
characterize its success-probability dynamics
\cite{reitzner2019groverDephasing}, while Gu et al. developed
channel-spectrum benchmarking methods based on
experimentally accessible repeated-channel signals
\cite{gu2023channelSpectrum}. Unlike spectral characterization of a
specified noisy search process or channel-spectrum benchmarking, our
framework uses an observable-relevant leading pair and controlled non-leading
residuals to derive conditional guarantees for integer stopping decisions.

Our related response-threshold study \cite{zhong2026responseThreshold} develops
a model-specific predictive theory on the standard Grover iteration grid.
For a Drude--Lorentz environment, linearized HEOM provides a priori timing
and sampled-probability responses, while finite-coupling HEOM, a same-bath
NZ2 comparator, and circuit-level validation test the threshold predictions.
Here the dynamical input is instead a leading observable spectral pair and a
controlled residual over the first lobe. The resulting statements concern
mechanism-resolved displacement, conditional sampled stability,
recalibration, robust windows, and residual-induced reversal across that
defined class. Both approaches use the same underlying integer-sampling
geometry but obtain the relevant dynamical information differently.

A second line of work changes the amplification protocol itself.
Arbitrary-phase and phase-matching methods analyze reflection-phase choices
\cite{hoyer2000arbitrary,long1999phaseMatching,long2001generalPhaseMatching,
toyama2008multiphase}; Leng, Yang, and Wang proposed a phase-modified D2p
procedure under phase noise \cite{leng2023improving}; and fixed-point
amplification avoids overshooting through modified sequences
\cite{yoder2014fixedpoint,dalzell2017fixedpoint,yan2022fixedpoint}.
Our analysis keeps the standard amplification sequence fixed and studies the
integer depth decision induced by its noisy observable trajectory.

Noise-aware depth selection provides a complementary operational perspective.
Leng, Yang, and Wang use predicted success probabilities to reduce
restart-aware noisy-search cost \cite{leng2025noiseTolerant}. Ganguly and Ahn
use depolarizing-noise information to choose hardware-aware amplification
depths \cite{ganguly2022depolarizing}. The present work targets the sampled
maximizer within a surviving first lobe and specifies when this decision is
stable or certifiably corrected under its spectral and residual assumptions.
The objectives and information required by these approaches are therefore
different, even where they lead to similar depth choices.

Robust and QPE-free amplitude estimation develops likelihoods, measurement
schedules, or noise treatments for amplitude inference
\cite{suzuki2020amplitudeEstimationWithoutQPE,grinko2021iterativeQAE,
tanaka2022noisyQAEWithoutNoiseEstimation,koh2022engineeredLikelihoods,
dalal2022noiseTailoringRAE,ramoa2025bayesianQAE}. In the application of
Subsection~\ref{subsec:diagnostic-workflow}, coarse amplitude estimation
localizes the candidate region and the stopping framework makes the
subsequent local depth decision. Table~\ref{tab:related-work-comparison}
summarizes the different dynamical inputs and algorithmic outputs.

\begin{table}[!t]
\centering
\caption{Representative noisy-Grover and amplitude-estimation approaches, organized by their dynamical inputs and principal results.}
\label{tab:related-work-comparison}
\scriptsize
\setlength{\tabcolsep}{3pt}
\renewcommand{\arraystretch}{1.08}
\begin{tabular}{>{\raggedright\arraybackslash}p{0.16\linewidth}>{\raggedright\arraybackslash}p{0.22\linewidth}>{\raggedright\arraybackslash}p{0.29\linewidth}>{\raggedright\arraybackslash}p{0.25\linewidth}}
\toprule
Work & Dynamics / information & Principal result & Analysis scope \\
\midrule
Azuma \cite{azuma2005higherOrderGrover}
& Independent single-qubit phase flips
& Perturbative success probabilities, threshold behavior, peak advance
& Specified phase-flip process \\
\addlinespace
Cohn et al. \cite{cohn2016groverDepolarizing}
& Total and local depolarizing channels
& Noise-adjusted optimal depth; local-noise bounds and estimates
& Channel-dependent probability and complexity analysis \\
\addlinespace
Zhong et al. \cite{zhong2026responseThreshold}
& Drude--Lorentz bath; linearized HEOM
& A priori response and threshold screening
& Microscopic structured-noise model \\
\addlinespace
Leng et al. \cite{leng2025noiseTolerant}
& Predicted noisy success probabilities
& Restart-aware depth selection
& Expected search cost \\
\addlinespace
Ganguly and Ahn \cite{ganguly2022depolarizing}
& Gate-level depolarizing information
& Hardware-aware depth choice
& Noise-informed compilation / execution \\
\addlinespace
Ram{\^o}a and Santos \cite{ramoa2025bayesianQAE}
& Bayesian amplitude and noise inference
& Adaptive amplitude estimation
& Parameter estimation \\
\addlinespace
Present work
& Dominant observable spectral pair; bounded residual
& Displacement, sampled certification, recalibration, robust window
& Conditional first-lobe stopping across a defined dynamics class \\
\bottomrule
\end{tabular}
\end{table}

\section{Problem Setting}
\label{sec:problem-setting}

We specify the ideal integer-sampling geometry and the stationary noisy
iterate that generates the observable success trajectory. These definitions
provide the inputs for the spectrally structured first-lobe regime developed in
Section~\ref{sec:formulation}.

\subsection{Ideal Sampled Stopping Geometry}
\label{sec:ideal-sampled-stopping}

The ideal sampling grid provides the geometric reference for subsequent
noise-dependent decisions. Let \(a_N\in(0,1)\) be the initial success probability and define
\[
    \sin^2\theta_N=a_N,
    \qquad
    \theta_N\in(0,\pi/2).
\]
The ideal success probability after \(k\) iterations is
\begin{equation}
    P_N^{(0)}(k)
    =
    \sin^2\!\bigl((2k+1)\theta_N\bigr).
    \label{eq:ideal-aa-curve}
\end{equation}
For single-target Grover search, \(a_N=1/N\) and
\(\theta_N=\arcsin(1/\sqrt N)\sim N^{-1/2}\).

Treating \(k\) as continuous, the first maximum occurs at
\begin{equation}
    s_N^0
    =
    \frac{\pi}{4\theta_N}-\frac{1}{2}.
    \label{eq:ideal-continuous-first-peak}
\end{equation}
The implementable ideal stopping step is therefore
\begin{equation}
    k_N^0
    =
    \operatorname{round}(s_N^0),
    \label{eq:ideal-sampled-stopping-step}
\end{equation}
where nearest-integer rounding uses a fixed tie-breaking convention.

Write \(d_N=s_N^0-k_N^0\), with \(|d_N|\leq 1/2\). The distance from the
continuous peak to the nearest boundary of its integer rounding cell is
\begin{equation}
    m_N
    =
    \frac{1}{2}
    -
    \left|s_N^0-\operatorname{round}(s_N^0)\right|
    =
    \frac{1}{2}-|d_N|.
    \label{eq:stroboscopic-margin}
\end{equation}
The rounding-cell geometry also appears in
Ref.~\cite{zhong2026responseThreshold}. If a perturbed continuous first
peak is \(\widetilde{s}_N=s_N^0+\Delta s_N\) and
\begin{equation}
    |\Delta s_N|<m_N,
    \label{eq:basic-margin-condition}
\end{equation}
then \(\operatorname{round}(\widetilde{s}_N)=k_N^0\). This condition
concerns the continuous peak's rounding cell; sampled stopping stability
requires separate probability comparisons in
Section~\ref{sec:fixed-sampled-stopping-stability}.

The exact right and left margins are
\[
    m_N^{+}
    =
    k_N^0+\frac{1}{2}-s_N^0,
    \qquad
    m_N^{-}
    =
    s_N^0-\left(k_N^0-\frac{1}{2}\right),
\]
with \(m_N=\min\{m_N^{+},m_N^{-}\}\). Thus a positive displacement is
allowed up to \(m_N^{+}\), whereas a negative displacement is allowed up to
\(m_N^{-}\). These directional margins give the exact continuous rounding
boundaries.

Although \(s_N^0=\Theta(\sqrt N)\) in single-target Grover search, \(m_N\)
is determined by the fractional part of \(s_N^0\) and is therefore not a smooth
asymptotic function of \(N\). Search sizes with the same asymptotic scale can
consequently have different fixed-step robustness.

\paragraph{Running example: \texorpdfstring{$N=64$}{N=64}.}
To keep the geometric, spectral, and correction quantities concrete, we use one
worked example throughout the paper. For single-target Grover search with
\(N=64\),
\[
    \theta_N=0.1253278312,\qquad
    s_N^0=5.766750,\qquad
    k_N^0=6.
\]
Hence \(d_N=-0.233250\), the early (left) directional margin is
\(m_N^-=0.266750\), the late (right) margin is \(m_N^+=0.733250\), and
the early rounding-cell boundary is \(5.5\). Section~\ref{subsec:device-like-validation}
returns to this same instance under backend-derived device-like noise, so that
the displacement, sampled-step change, correction certificate, and calibration
cost can be followed without changing examples.

\subsection{Noisy Iterate Model}

Let \(G_N\) denote the ideal standard amplitude-amplification iterate and
\(\mathcal Q_N(\rho)=G_N\rho G_N^\dagger\) its channel representation. For an
initial state \(\rho_{N,0}\) and good-subspace measurement \(M_N\), the ideal
trajectory is \(P_N^{(0)}(k)\) from Eq.~\eqref{eq:ideal-aa-curve}. We model
stationary per-iterate noise by
\begin{equation}
    \widetilde{\mathcal Q}_N
    =
    \mathcal N_N\circ\mathcal Q_N,
    \label{eq:noisy-iterate}
\end{equation}
where \(\mathcal N_N\) is CPTP in the main physical setting. After \(k\) noisy
iterations,
\begin{equation}
    \rho_{N,k}
    =
    \widetilde{\mathcal Q}_N^k(\rho_{N,0}),
    \qquad
    P_N(k)
    =
    \operatorname{Tr}(M_N\rho_{N,k}).
    \label{eq:noisy-trajectory-success}
\end{equation}

Time homogeneity permits spectral analysis of repeated noisy iterates.
The leading-mode and residual conditions in
Subsection~\ref{sec:first-peak-preserving-noisy-aa} then specify the subclass
of observable first-lobe trajectories treated by the stopping theory.

\noindent The principal symbols are introduced locally; a consolidated notation table is provided in Supplementary Table~S1.

\section{Formulation}
\label{sec:formulation}

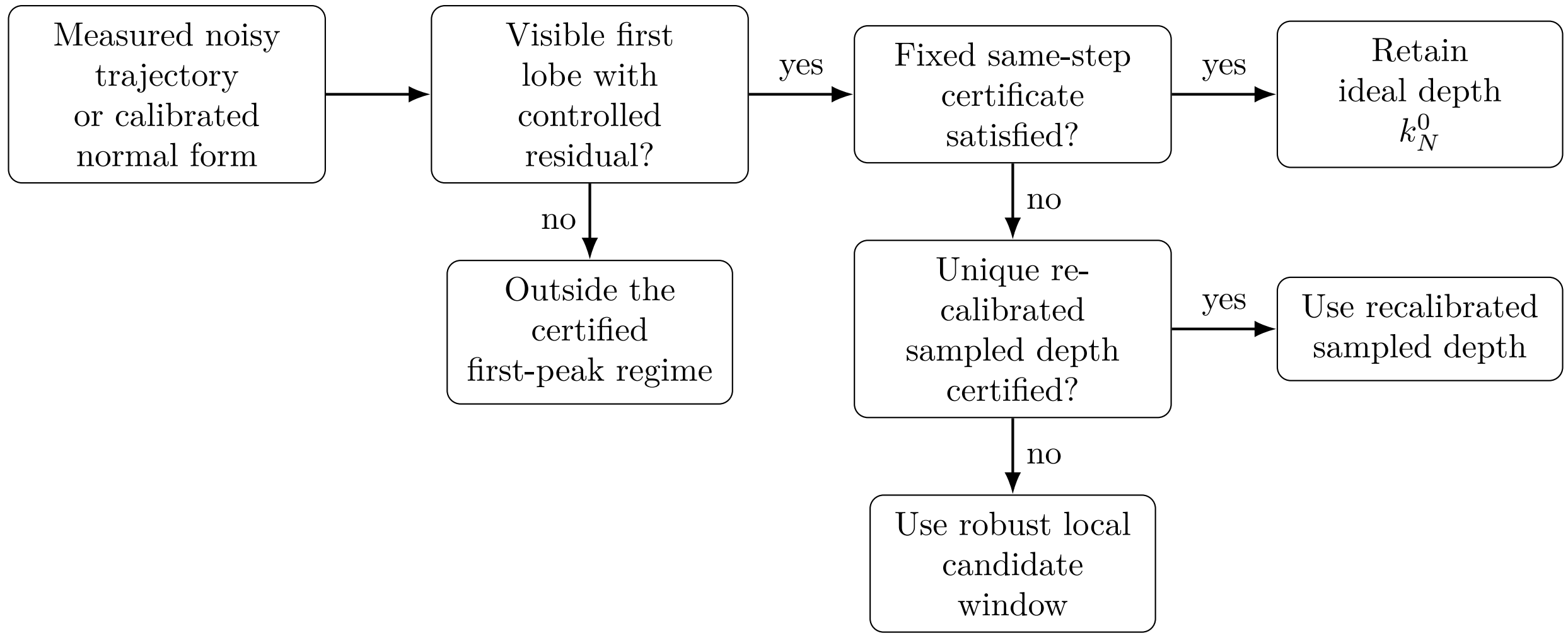
\begin{figure}[!htbp]
\centering

\begin{tikzpicture}[
    node distance=8mm and 11mm,
    >=Latex,
    every node/.style={
        font=\small,
        align=center
    },
    box/.style={
        draw,
        rounded corners,
        minimum height=10mm,
        text width=0.18\textwidth,
        inner sep=5pt
    },
    decision/.style={
        draw,
        rounded corners,
        minimum height=12mm,
        text width=0.18\textwidth,
        inner sep=5pt
    },
    action/.style={
        draw,
        rounded corners,
        minimum height=10mm,
        text width=0.16\textwidth,
        inner sep=5pt
    },
    arrow/.style={
        ->,
        thick
    }
]

\node[box] (input) {
Measured noisy trajectory\\
or calibrated normal form
};

\node[decision, right=of input] (peak) {
Visible first lobe with\\
controlled residual?
};

\node[decision, right=of peak] (fixed) {
Fixed same-step\\
certificate satisfied?
};

\node[action, right=of fixed] (retain) {
Retain ideal depth\\
\(k_N^0\)
};

\node[decision, below=of fixed] (recal) {
Unique recalibrated\\
sampled depth certified?
};

\node[action, right=of recal] (use-rec) {
Use recalibrated\\
sampled depth
};

\node[action, below=of recal] (window) {
Use robust local\\
candidate window
};

\node[action, below=of peak] (outside) {
Outside the certified\\
first-peak regime
};

\draw[arrow] (input) -- (peak);

\draw[arrow]
    (peak)
    --
    node[above]{yes}
    (fixed);

\draw[arrow]
    (peak)
    --
    node[left]{no}
    (outside);

\draw[arrow]
    (fixed)
    --
    node[above]{yes}
    (retain);

\draw[arrow]
    (fixed)
    --
    node[right]{no}
    (recal);

\draw[arrow]
    (recal)
    --
    node[above]{yes}
    (use-rec);

\draw[arrow]
    (recal)
    --
    node[right]{no}
    (window);

\end{tikzpicture}

\caption{
Diagnostic hierarchy for noisy sampled stopping.
A measured or calibrated trajectory is first checked for a reliable
first-lobe normal-form regime. If the fixed same-step certificate holds,
the ideal sampled depth \(k_N^0\) is retained. Otherwise, the fitted
leading lobe is used to test whether a unique recalibrated sampled depth
can be certified; if not, the displacement uncertainty is converted into a
robust local candidate window. The same hierarchy can be evaluated using point
estimates or finite-shot bootstrap regions.
}

\label{fig:diagnostic-hierarchy}

\end{figure}

We first identify the observable spectral conditions under which stationary
noisy iterations admit a controlled first-lobe normal form. We then derive
the peak-displacement, sampled-stability, and correction results that follow
from that structure. Figure~\ref{fig:diagnostic-hierarchy} summarizes the
resulting operational hierarchy.

\FloatBarrier

\subsection{Observable First-Peak-Preserving Regime}
\label{sec:first-peak-preserving-noisy-aa}
\subsubsection{Observable-relevant leading oscillatory pair}
\label{subsec:observable-relevant-leading-pair}

The spectral class is defined by requiring the measured trajectory on the
first-peak window to be dominated by an observable-relevant conjugate pair
\begin{equation}
    \lambda_{\pm,N}
    =
    \rho_N e^{\pm i\omega_N},
    \qquad
    \rho_N=e^{-\gamma_N},
    \qquad
    \gamma_N\geq 0.
    \label{eq:leading-oscillatory-pair}
\end{equation}
Observable relevance means that the pair has non-negligible overlap with the chosen
initial state and success measurement. Exactly stationary observable modes
(eigenvalue 1) contribute to the constant background; slowly varying modes that
are not stationary remain in the residual and are subject to its bound.

Since
\[
    P_N^{(0)}(k)
    =
    \frac{1}{2}
    -
    \frac{1}{2}\cos\bigl((4k+2)\theta_N\bigr),
\]
we parameterize the noisy frequency and observable phase as
\begin{equation}
    \omega_N
    =
    4\theta_N+\Omega_N
    =
    4(\theta_N+\delta_N),
    \qquad
    \chi_N
    =
    2\theta_N+\phi_N.
    \label{eq:frequency-phase-parameters}
\end{equation}
Thus \(\delta_N\), \(\gamma_N\), and \(\phi_N\) represent the effective-angle
shift, damping scale, and observable phase offset. They are fitted normal-form
parameters rather than independent algorithmic controls.

The relevant first-peak window is
\[
    \mathcal W_N
    =
    \left[\frac{c_1}{\theta_N},\frac{c_2}{\theta_N}\right],
    \qquad
    0<c_1<c_2,
\]
with \(c_1,c_2\) independent of \(N\). In single-target Grover search, this is an
\(O(\sqrt N)\)-iteration window. Its integer sampling set is
\(\mathcal K_N=\mathcal W_N\cap\mathbb Z_{\geq0}\). We choose the window
within a single leading first lobe and require it to contain the ideal step
when fixed-step stability is tested and both neighbouring integer candidates
of the leading peak when recalibration is used. The relevant sampled gap is
defined on a finite set with at least two candidate depths. The leading pair
is assumed semisimple. Controlled
non-leading Jordan contributions may be absorbed into the residual bound below,
whereas a Jordan block in the leading pair would produce a polynomially modulated
leading oscillation outside the present damped-cosine theory.

\subsubsection{Spectral separation and residual control}
\label{subsec:spectral-separation-residual}

Let \(\mathcal J_N\) index the observable-relevant residual modes after
removing the leading conjugate pair and the exactly stationary modes absorbed
into \(B_N\). Define
\[
    \beta_N
    =
    \max_{j\in\mathcal J_N}|\lambda_{j,N}|,
\]
with \(\beta_N=0\) if \(\mathcal J_N=\varnothing\). Nonstationary slowly
varying modes belong to \(\mathcal J_N\); they are not part of the constant
background. We assume
\begin{equation}
    g_N
    =
    \rho_N-\beta_N
    >
    0.
    \label{eq:spectral-gap}
\end{equation}
Writing the non-leading observable contribution as \(r_N(k)\), define
\[
    R_N^{\mathrm{val}}
    =
    \max_{k\in\mathcal K_N}|r_N(k)|.
\]
A spectral decomposition gives the bound
\begin{equation}
    R_N^{\mathrm{val}}
    \leq
    K_N
    \left(\frac{\beta_N}{\rho_N}\right)^{c_*/\theta_N}
    \leq
    K_N
    \exp\!\left(-\frac{c_*g_N}{\rho_N\theta_N}\right),
    \label{eq:residual-bound}
\end{equation}
where \(c_*>0\) is determined by the beginning of \(\mathcal W_N\), and \(K_N\)
collects spectral-projector conditioning and observable overlaps.

The admissible regime requires \(K_N\) not to overwhelm the relative spectral
decay; for example, uniformly conditioned blocks give \(K_N=O(1)\), while
polynomial growth is also compatible with a sufficiently persistent gap. Nearly
defective or strongly non-normal dynamics may violate this control and then require
model-specific resolvent or dynamical estimates. The residual bound and its
conditioning assumptions are derived in Supplementary Sections~S1.4--S1.6,
while the perturbative realizability route is developed separately in
Supplementary Section~S1.8.

The leading pair and residual have distinct roles: the pair controls horizontal
first-peak displacement, whereas \(R_N^{\mathrm{val}}\) controls value-level changes
in the ordering of nearby sampled steps. The sampled-gap certificate therefore
compares this residual with the local discrete probability gap, which is of order
\(m_N/N\) in the clean single-target Grover regime.

\paragraph{Perturbative realizability.}
The main results assume the observable leading-pair normal form, spectral
separation, and residual control directly. A standard sufficient route from a
small perturbation of the ideal iterate, based on Bauer--Fike and observable
overlap control, is given in Supplementary Section~S1.8.
This provides a realizability certificate for the spectral framework; the
stopping results below use the stated leading-mode and residual conditions.

\paragraph{Closed-form depolarizing realization.}
A useful exact realization of the assumptions is encoded-subspace depolarization
on the marked--unmarked Grover plane. It gives
\(\gamma_N=-\log(1-p_N)\), \(\delta_N=\phi_N=0\), and
\(R_N^{\mathrm{val}}=0\). The channel spectrum, observable-overlap calculation,
and the distinction between the continuous-cell crossing and exact sampled
equality are derived in Supplementary Section~S1.9.

\subsection{Spectral Normal Form and Peak Displacement}
\label{sec:spectral-normal-form-peak-displacement}
\subsubsection{Observable-level cosine normal form}
\label{subsec:observable-cosine-normal-form}

The spectral conditions of
Subsection~\ref{sec:first-peak-preserving-noisy-aa} translate the repeated noisy
iterate into a description of the measured values \(P_N(k)\). The leading
observable-relevant conjugate pair produces a damped cosine. Stationary
observable modes form the constant background; the other observable
non-leading modes contribute the residual:
\begin{equation}
    P_N(k)
    =
    B_N
    -
    C_N e^{-\gamma_N k}
    \cos(k\omega_N+\chi_N)
    +
    r_N(k).
    \label{eq:observable-cosine-normal-form}
\end{equation}
Here \(C_N>0\), and \(B_N\), \(C_N\), \(\gamma_N\), \(\omega_N\), and
\(\chi_N\) are observable-level normal-form parameters. We write
\[
    F_N(k)
    =
    B_N-C_Ne^{-\gamma_N k}\cos(k\omega_N+\chi_N),
    \qquad
    P_N(k)=F_N(k)+r_N(k),
\]
with
\[
    R_N^{\mathrm{val}}
    =
    \max_{k\in\mathcal K_N}|r_N(k)|.
\]
The sign convention places the ideal first peak near a cosine minimum. The
spectral-to-observable reduction is derived explicitly in Supplementary
Sections~S1.1--S1.3, where the leading pair, background contribution, and
non-leading residual modes are separated.

The ideal curve is recovered from
\[
    B_N=C_N=\frac12,
    \qquad
    \gamma_N=0,
    \qquad
    \omega_N=4\theta_N,
    \qquad
    \chi_N=2\theta_N,
    \qquad
    r_N(k)=0,
\]
which reduces Eq.~\eqref{eq:observable-cosine-normal-form} to
Eq.~\eqref{eq:ideal-aa-curve}. Using
\[
    \omega_N=4(\theta_N+\delta_N),
    \qquad
    \chi_N=2\theta_N+\phi_N,
\]
the normal form separates the effective-angle shift \(\delta_N\), damping
\(\gamma_N\), and observable phase offset \(\phi_N\).

We assume that the leading first peak remains non-degenerate:
\[
    C_N e^{-\gamma_N s_N^{\mathrm{lead}}}\geq c_0
\]
for some \(c_0>0\). This condition requires a visible oscillatory component with
non-vanishing local curvature; the background \(B_N\) may shift the curve
vertically without moving the leading peak. A sufficient perturbative route for
the persistence of the leading observable overlap is given in
Supplementary Section~S1.8.

\subsubsection{Leading first-peak equation}
\label{subsec:leading-first-peak-equation}

Ignoring \(r_N(k)\), the stationary branch of the leading damped cosine near the
ideal first peak gives
\begin{equation}
    s_N^{\mathrm{lead}}
    =
    \frac{
        \pi-\chi_N-\arctan(\gamma_N/\omega_N)
    }{
        \omega_N
    }.
    \label{eq:leading-first-peak-location}
\end{equation}
The full differentiation and stationary-branch calculation are given in
Supplementary Section~S1.7. At this peak the local
curvature is
\[
    \kappa_N
    =
    C_N e^{-\gamma_N s_N^{\mathrm{lead}}}
    \omega_N\sqrt{\omega_N^2+\gamma_N^2}.
\]
When \(\gamma_N=O(\omega_N)\),
\[
    \kappa_N
    \asymp
    C_N e^{-\gamma_N s_N^{\mathrm{lead}}}\omega_N^2.
\]
Thus, under single-target Grover scaling and non-collapsing peak height,
\(\kappa_N=\Theta(1/N)\). Setting \(\gamma_N=0\), \(\omega_N=4\theta_N\), and
\(\chi_N=2\theta_N\) in Eq.~\eqref{eq:leading-first-peak-location} recovers
\(s_N^{\mathrm{lead}}=s_N^0\).

\subsubsection{Displacement contributions from \texorpdfstring{$\delta_N$}{deltaN}, \texorpdfstring{$\gamma_N$}{gammaN}, and \texorpdfstring{$\phi_N$}{phiN}}
\label{subsec:displacement-contributions}

Substituting
\(\omega_N=4(\theta_N+\delta_N)\) and
\(\chi_N=2\theta_N+\phi_N\) into
Eq.~\eqref{eq:leading-first-peak-location}, and defining
\(\Delta s_N^{\mathrm{lead}}=s_N^{\mathrm{lead}}-s_N^0\), gives the following
expansion in the local regime
\[
    \frac{|\delta_N|}{\theta_N}
    +
    |\phi_N|
    +
    \frac{\gamma_N}{\theta_N}
    \leq \varepsilon_0,
    \qquad 0<\varepsilon_0<1:
\]
\begin{equation}
    \Delta s_N^{\mathrm{lead}}
    =
    -
    \frac{s_N^0\delta_N}{\theta_N}
    -
    \frac{\phi_N}{4\theta_N}
    -
    \frac{\gamma_N}{16\theta_N^2}
    +
    \mathcal R_N^{\mathrm{disp}} .
    \label{eq:main-displacement-formula}
\end{equation}
The remainder obeys
\[
    \mathcal R_N^{\mathrm{disp}}
    =
    O\!\left(
        \frac{1}{\theta_N}
        \left(
            \frac{|\delta_N|}{\theta_N}
            +
            |\phi_N|
            +
            \frac{\gamma_N}{\theta_N}
        \right)^2
    \right),
\]
with an \(N\)-independent local constant. Supplementary Sections~S2.1--S2.3
give the exact displacement identity, the first-order expansion, and the explicit
remainder envelope.

The three first-order terms have distinct origins:
\[
    \Delta s_N^{(\delta)}
    =
    -\frac{s_N^0\delta_N}{\theta_N},
    \qquad
    \Delta s_N^{(\phi)}
    =
    -\frac{\phi_N}{4\theta_N},
    \qquad
    \Delta s_N^{(\gamma)}
    =
    -\frac{\gamma_N}{16\theta_N^2}.
\]
The first term accumulates a frequency shift over the Grover-scale trajectory,
the second translates the observable phase, and the third arises from the tilt
of the damped envelope. The damping coefficient follows from the probability-level convention
\(\omega_N=4(\theta_N+\delta_N)\). Accordingly, \(\gamma_N\) and \(\phi_N\)
are parameters of the observable probability trajectory, not eigenangle
parameters of the state-vector Grover rotation.

For single-target Grover search,
\(\theta_N\sim N^{-1/2}\) and \(s_N^0=\Theta(\sqrt N)\), so
\[
    \Delta s_N^{(\delta)}
    =
    \Theta(N\delta_N),
    \qquad
    \Delta s_N^{(\gamma)}
    =
    \Theta(N\gamma_N),
    \qquad
    \Delta s_N^{(\phi)}
    =
    \Theta(\sqrt N\,\phi_N).
\]
These are stopping-coordinate sensitivities, not peak-height scalings. A visible
first peak may therefore cross an integer rounding boundary under perturbations
that remain small at the one-step level.

\paragraph{Running example.}
At the shifted GenericBackendV2 operating point used later in
Subsection~\ref{subsec:device-like-validation}, the following parameters are
obtained from a 13-depth sampled-trajectory fit (\(k=0,\ldots,12\))
reported in Subsection~\ref{subsec:device-like-validation}:
\[
    \gamma_N=0.023323,\qquad
    \delta_N=-0.002971,\qquad
    \phi_N=0.313048.
\]
The three first-order contributions are
\(\Delta s_N^{(\delta)}=+0.136691\),
\(\Delta s_N^{(\phi)}=-0.624457\), and
\(\Delta s_N^{(\gamma)}=-0.092803\), giving
\(\Delta s_N^{\mathrm{main}}=-0.580570\). The fitted leading peak lies at
\(s_N^{\mathrm{lead}}=5.169850\), so the exact displacement is
\(\Delta s_N^{\mathrm{lead}}=-0.596899\) and the displacement remainder is
\(\mathcal R_N^{\mathrm{disp}}=-0.016330\). Thus the observable phase offset
dominates the early shift, damping reinforces it, and the negative frequency
shift partially compensates it.
For this example the dimensionless perturbation size is approximately
\(0.523\), so the local remainder estimate applies with, for example,
\(\varepsilon_0=3/4\). The fitted leading-peak location uses the exact peak equation, while the
first-order terms decompose its displacement. Certification of the integer
stopping decision additionally requires the sampled-gap conditions of
Section~\ref{sec:fixed-sampled-stopping-stability}.

As a concrete specialization, the encoded-subspace depolarizing realization
has
\[
    \delta_N=\phi_N=0,
    \qquad
    \gamma_N=-\log(1-p_N).
\]
Substituting these parameters into Eq.~\eqref{eq:leading-first-peak-location} gives the exact continuous
first-peak displacement
\begin{equation}
    \Delta s_N^{\mathrm{lead}}
    =
    -
    \frac{1}{4\theta_N}
    \arctan\!\left(
        \frac{-\log(1-p_N)}
             {4\theta_N}
    \right).
    \label{eq:encoded-depolarizing-displacement}
\end{equation}
For weak noise,
\[
    \Delta s_N^{\mathrm{lead}}
    =
    -
    \frac{\gamma_N}{16\theta_N^2}
    +
    O\!\left(
        \frac{\gamma_N^3}{\theta_N^4}
    \right),
\]
which reproduces the damping contribution in
Eq.~\eqref{eq:main-displacement-formula}. The corresponding finite-\(N\)
sampled transition is determined by comparing neighbouring sampled
probabilities, as derived in Supplementary Section~S1.9.

\subsubsection{Combined displacement condition}
\label{subsec:combined-displacement-condition}

Define
\[
    \Delta s_N^{\mathrm{main}}
    =
    -
    \frac{s_N^0\delta_N}{\theta_N}
    -
    \frac{\phi_N}{4\theta_N}
    -
    \frac{\gamma_N}{16\theta_N^2},
    \qquad
    \Delta s_N^{\mathrm{lead}}
    =
    \Delta s_N^{\mathrm{main}}
    +
    \mathcal R_N^{\mathrm{disp}}.
\]
Before accounting for the value-level residual \(r_N(k)\), the rounding-cell
margin condition gives
\begin{equation}
    \left|
        \frac{s_N^0\delta_N}{\theta_N}
        +
        \frac{\phi_N}{4\theta_N}
        +
        \frac{\gamma_N}{16\theta_N^2}
    \right|
    +
    |\mathcal R_N^{\mathrm{disp}}|
    <
    m_N .
    \label{eq:combined-displacement-condition}
\end{equation}
Under Eq.~\eqref{eq:combined-displacement-condition},
\(\operatorname{round}(s_N^{\mathrm{lead}})=k_N^0\). The signed sum is essential:
different mechanisms may reinforce or partially cancel one another, so separate
componentwise bounds do not characterize the rounding-cell boundary.

Using the directional margins of Section~\ref{sec:ideal-sampled-stopping}, the
exact directional condition for continuous-cell membership of the leading peak is
\begin{equation}
    -m_N^{-}
    <
    \Delta s_N^{\mathrm{main}}
    +
    \mathcal R_N^{\mathrm{disp}}
    <
    m_N^{+}.
    \label{eq:directional-combined-condition}
\end{equation}
Equation~\eqref{eq:directional-combined-condition} is the exact geometric
condition for the leading peak to remain in the same nearest-integer cell, while
Eq.~\eqref{eq:combined-displacement-condition} is its uniform sufficient version.
Damping shifts the peak earlier and is therefore controlled by the left margin
\(m_N^{-}\), whereas frequency and phase shifts may act in either direction.
The sampled-value certificate is introduced once in
Section~\ref{sec:fixed-sampled-stopping-stability}.

\subsection{Fixed Sampled Stopping Stability}
\label{sec:fixed-sampled-stopping-stability}
\subsubsection{Sampled-gap certificate}

A continuous rounding-cell condition locates the leading peak, whereas an
integer stopping decision depends on the sampled probability ordering. For
the observable normal form, this ordering is determined by the leading curve
together with the non-leading residual. We express the sampled-gap condition
on a general candidate set so that it can support retaining the ideal step,
certifying a recalibrated step, and constructing a robust window.

Let \(\mathcal A_N\subseteq\mathcal K_N\) be a finite candidate set and let
\(k_N^{\dagger}\in\mathcal A_N\) be the unique sampled maximizer of the leading
curve \(F_N\) on \(\mathcal A_N\). Define
\begin{equation}
    \Delta_N(\mathcal A_N,k_N^{\dagger})
    =
    \min_{\substack{k\in\mathcal A_N\\ k\neq k_N^{\dagger}}}
    \bigl[F_N(k_N^{\dagger})-F_N(k)\bigr].
    \label{eq:generic-leading-gap}
\end{equation}

\begin{lemma}[Sampled-gap preservation]
\label{lem:sampled-gap-preservation}
Assume \(P_N(k)=F_N(k)+r_N(k)\) for \(k\in\mathcal K_N\), with
\(R_N^{\mathrm{val}}=\max_{k\in\mathcal K_N}|r_N(k)|\). If
\begin{equation}
    2R_N^{\mathrm{val}}
    <
    \Delta_N(\mathcal A_N,k_N^{\dagger}),
    \label{eq:generic-sampled-gap-condition}
\end{equation}
then \(k_N^{\dagger}\) is also the unique sampled maximizer of \(P_N\) on
\(\mathcal A_N\).
\end{lemma}
The gap-versus-residual criterion applies to any candidate set satisfying the
lemma's assumptions and supplies the common sampled-ordering condition for the
stopping decisions below.

For fixed same-step stability, take \(\mathcal A_N=\mathcal K_N\) and
\(k_N^{\dagger}=k_N^0\). Define
\begin{equation}
    \Delta_N^F
    =
    \min_{\substack{k\in\mathcal K_N\\ k\neq k_N^0}}
    \bigl[F_N(k_N^0)-F_N(k)\bigr].
    \label{eq:leading-sampled-gap}
\end{equation}

\begin{corollary}[Fixed same-step stopping]
\label{thm:fixed-same-step-stability}
If
\begin{equation}
    2R_N^{\mathrm{val}}<\Delta_N^F,
    \label{eq:value-gap-residual-condition}
\end{equation}
then the noisy sampled first-peak maximizer is unchanged:
\[
    k_N^*=k_N^0.
\]
\end{corollary}

\subsubsection{Curvature-scaled residual allowance}
\label{subsec:curvature-scaled-residual-allowance}

To make Eq.~\eqref{eq:value-gap-residual-condition} operational, write
\[
    \Delta s_N^{\mathrm{lead}}
    =
    \Delta s_N^{\mathrm{main}}
    +
    \mathcal R_N^{\mathrm{disp}},
    \qquad
    m_N^{\mathrm{rem}}
    =
    m_N-|\Delta s_N^{\mathrm{lead}}|,
\]
and let
\[
    \kappa_N=-F_N''(s_N^{\mathrm{lead}}),
    \qquad
    \mu_N(s)
    =
    \frac12-\left|s-\operatorname{round}(s)\right|.
\]
When the leading peak remains in the ideal cell,
\(\mu_N(s_N^{\mathrm{lead}})\geq m_N^{\mathrm{rem}}\).  We impose the local
sampled-gap regularity condition
\begin{equation}
    \Delta_N^F
    \geq
    c_{\mathrm{curv}}\kappa_N
    \mu_N(s_N^{\mathrm{lead}}),
    \qquad
    c_{\mathrm{curv}}\in(0,1].
    \label{eq:local-sampled-gap-regularity}
\end{equation}
For a non-degenerate damped-cosine lobe away from a sampled equality, a
Taylor expansion with controlled local skew supplies this curvature-to-gap
bound; its limitations near ties, asymmetric peaks, and flattened lobes are
given in Supplementary Section~S2.6. For the ideal
single-target curve,
\(\kappa_N=8\theta_N^2+O(\theta_N^4)\) and
\(\Delta_N^F=8\theta_N^2\mu_N(s_N^0)+O(\theta_N^4)\), so
\(c_{\mathrm{curv}}\) can be chosen close to \(1\) away from ties.

Combining Eq.~\eqref{eq:local-sampled-gap-regularity} with
Corollary~\ref{thm:fixed-same-step-stability} gives
\begin{equation}
    |\Delta s_N^{\mathrm{lead}}|
    +
    \frac{2R_N^{\mathrm{val}}}
         {c_{\mathrm{curv}}\kappa_N}
    <
    m_N .
    \label{eq:curvature-scaled-condition}
\end{equation}
Using the perturbative displacement formula yields
\begin{equation}
    \left|
        \frac{s_N^0\delta_N}{\theta_N}
        +
        \frac{\phi_N}{4\theta_N}
        +
        \frac{\gamma_N}{16\theta_N^2}
    \right|
    +
    |\mathcal R_N^{\mathrm{disp}}|
    +
    \frac{2R_N^{\mathrm{val}}}
         {c_{\mathrm{curv}}\kappa_N}
    <
    m_N .
    \label{eq:practical-same-step-condition}
\end{equation}
The first two terms consume margin through horizontal displacement, whereas the
last term is the margin reserved for residual value perturbations.  In the clean
single-target regime, \(\kappa_N=\Theta(1/N)\); asymptotic same-step stability
with non-vanishing remaining margin therefore requires
\[
    R_N^{\mathrm{val}}
    =
    o(\kappa_Nm_N)
    =
    o(m_N/N).
\]

\paragraph{Directional and discrete boundaries.}
\label{subsec:directional-margins}
The symmetric margin can be conservative when the ideal peak is not centred in
its rounding cell. After reserving
\(E_N^{\mathrm{res}}=2R_N^{\mathrm{val}}/(c_{\mathrm{curv}}\kappa_N)\)
for value-level uncertainty, a direction-aware sufficient condition is
\begin{equation}
    -m_N^-+E_N^{\mathrm{res}}
    <
    \Delta s_N^{\mathrm{lead}}
    <
    m_N^+-E_N^{\mathrm{res}}.
    \label{eq:directional-residual-stability-condition}
\end{equation}
The sign-resolved derivation, damping-only specialization, and the exact
comparison of continuous and sampled boundaries are given in Supplementary
Sections~S2.8 and
S1.9.

\subsection{Speedup--Stability Separation}
\label{sec:speedup-stability-duality}
\subsubsection{Stopping-displacement sensitivity}
\label{subsec:stopping-displacement-sensitivity}

Let
\[
    S_N=\frac{1}{\theta_N},
\]
so that \(s_N^0=\Theta(S_N)\); in single-target Grover search,
\(S_N=\Theta(\sqrt N)\). From
Eq.~\eqref{eq:main-displacement-formula}, the first-order displacement
sensitivities are
\begin{equation}
\begin{aligned}
\left|\frac{\partial\Delta s_N^{\mathrm{main}}}{\partial\delta_N}\right|
    &= \frac{s_N^0}{\theta_N}=\Theta(S_N^2),\\
\left|\frac{\partial\Delta s_N^{\mathrm{main}}}{\partial\gamma_N}\right|
    &= \frac{1}{16\theta_N^2}=\Theta(S_N^2),\\
\left|\frac{\partial\Delta s_N^{\mathrm{main}}}{\partial\phi_N}\right|
    &= \frac{1}{4\theta_N}=\Theta(S_N).
\end{aligned}
\label{eq:stopping-displacement-sensitivities}
\end{equation}
Thus frequency and damping perturbations act at the square of the
Grover stopping scale, while an observable phase offset acts at the
linear scale. For single-target Grover search these sensitivities are
\(\Theta(N)\), \(\Theta(N)\), and \(\Theta(\sqrt N)\), respectively. The
derivation of these sensitivities and the corresponding comparison between
peak-survival and no-correction scales is given in Supplementary
Subsection~S2.7.
Combining them with the directional continuous rounding margin gives the
characteristic continuous-cell no-correction scales
\[
    |\delta_N|
    =
    O\!\left(\frac{m_N}{N}\right),
    \qquad
    \gamma_N
    =
    O\!\left(\frac{m_N}{N}\right),
    \qquad
    |\phi_N|
    =
    O\!\left(\frac{m_N}{\sqrt N}\right).
\]
Together with the sampled-gap regularity condition of
Section~\ref{sec:fixed-sampled-stopping-stability}, these scales provide a
sufficient route into the no-correction regime. The \(S_N^2\) factor has a simple
origin: a frequency or damping error accumulates over \(\Theta(S_N)\) iterations
and is then converted back from phase to iteration displacement by another factor
\(\Theta(S_N)\). A direct observable phase offset requires only the second
conversion.

\subsubsection{Peak survival and the speedup--stability gap}
\label{subsec:peak-height-survival}
\label{subsec:peak-survival-weaker}

Same-step stability is stronger than preservation of a useful first peak. In the
observable normal form, the leading peak remains visible when
\[
    C_N e^{-\gamma_N s_N^{\mathrm{lead}}}\geq c_0
\]
for some \(c_0>0\), and it preserves the Grover stopping scale when
\(s_N^{\mathrm{lead}}=\Theta(S_N)\). A recalibrated rule then still uses
\(O(S_N)\) iterations, which is \(O(\sqrt N)\) in the single-target case.

For damping, peak-height survival only requires
\[
    \gamma_NS_N=O(1),
    \qquad\text{that is,}\qquad
    \gamma_N=O(S_N^{-1}),
\]
whereas the continuous-cell no-correction scale is
\[
    \gamma_N
    =
    O\!\left(\frac{m_N}{S_N^2}\right).
\]
Hence the intermediate regime
\[
    \frac{m_N}{S_N^2}
    \ll
    \gamma_N
    \lesssim
    \frac{1}{S_N}
\]
can retain a visible Grover-scale first peak after the continuous
no-correction condition has ceased to apply. Frequency and observable-phase perturbations can
produce the same separation through horizontal displacement. Under the
non-collapsing peak and Grover-scale location assumptions just stated, retaining the ideal sampled step preserves the first-peak query scale; preserving
a visible first peak at that scale does not imply same-step stability.
An \(O(1)\) peak displacement can be negligible relative to the \(O(S_N)\)
Grover stopping scale while still changing the integer decision, and residual
modes provide a second route by directly reordering neighbouring sampled values.
This separation is the speedup--stability gap. Correlations among
\(C_N,\gamma_N,\delta_N,\phi_N\) remain model-dependent.

\subsection{Stopping-Rule Correction Theory}
\label{sec:stopping-rule-correction}
\subsubsection{Recalibrated stopping}

When the ideal step \(k_N^0\) is not certified but the noisy first peak
survives, the natural correction is to follow the sampled maximizer of
the fitted leading first lobe. For a single-lobed \(F_N\), define the
local candidate set (with both adjacent integers assumed to lie in
\(\mathcal K_N\))
\[
    \mathcal C_N^{\mathrm{lead}}
    =
    \left\{
        \left\lfloor s_N^{\mathrm{lead}}\right\rfloor,
        \left\lceil s_N^{\mathrm{lead}}\right\rceil
    \right\}
    \cap
    \mathcal K_N ,
\]
and define
\begin{equation}
    k_N^{\mathrm{rec}}
    \in
    \operatorname*{arg\,max}_{k\in\mathcal C_N^{\mathrm{lead}}}
    F_N(k).
    \label{eq:recalibrated-sampled-step}
\end{equation}
A fixed tie-breaking convention is used if the two neighbouring sampled
values are equal. The displacement formula locates the candidate pair,
while the sampled comparison selects the corrected integer. For a
locally symmetric peak this reduces to nearest-integer rounding, but
that reduction is not assumed for a general damped trajectory.

Define the leading recalibrated gap by
\[
    \Delta_N^{\mathrm{rec}}
    =
    \min_{\substack{k\in\mathcal K_N\\ k\neq k_N^{\mathrm{rec}}}}
    \left[
        F_N(k_N^{\mathrm{rec}})-F_N(k)
    \right].
\]

\begin{corollary}[Recalibrated first-peak stopping]
\label{thm:recalibrated-stopping}
Assume \(s_N^{\mathrm{lead}}\in\mathcal W_N\), and that
\(k_N^{\mathrm{rec}}\) is the unique sampled maximizer of \(F_N\) on
\(\mathcal K_N\). If
\begin{equation}
    2R_N^{\mathrm{val}}<\Delta_N^{\mathrm{rec}},
    \label{eq:recalibrated-residual-condition}
\end{equation}
then
\[
    k_N^*=k_N^{\mathrm{rec}}.
\]
\end{corollary}

This is Lemma~\ref{lem:sampled-gap-preservation} with
\(\mathcal A_N=\mathcal K_N\) and
\(k_N^{\dagger}=k_N^{\mathrm{rec}}\).

Recalibrated stopping does not require the noisy peak to remain in the
ideal rounding cell.  It only requires the peak to remain identifiable
and the residual to be smaller than its local sampled gap.  Since
\(s_N^{\mathrm{lead}}=\Theta(S_N)\) in the first-peak-preserving
regime, \(k_N^{\mathrm{rec}}=\Theta(S_N)\); thus recalibration preserves
the Grover query scale even when \(k_N^{\mathrm{rec}}\neq k_N^0\).
Finite-shot inference is specified in
Subsection~\ref{subsec:finite-shot-stopping-certification} and Supplementary
Subsection~S4.12; calibration design and amortized cost
are quantified in Subsection~\ref{subsec:calibration-cost} and Supplementary
Subsection~S4.15.

\paragraph{Running example.}
For \(N=64\), the displaced leading peak gives the candidate pair \(\{5,6\}\)
and selects \(k_N^{\mathrm{rec}}=5\). The numerical certificate and finite-shot
fallback for this same instance are reported in
Subsection~\ref{subsec:device-like-validation}.

\subsubsection{Robust-window stopping}
\label{subsec:robust-window-stopping}

When the exact noisy peak is uncertain, suppose instead that
\[
    |\Delta s_N^{\mathrm{lead}}|
    \leq
    \widehat{\Delta}_N,
\]
where one admissible bound is
\[
    \widehat{\Delta}_N
    =
    \left|
        \frac{s_N^0\delta_N}{\theta_N}
    \right|
    +
    \left|
        \frac{\phi_N}{4\theta_N}
    \right|
    +
    \frac{\gamma_N}{16\theta_N^2}
    +
    |\mathcal R_N^{\mathrm{disp}}|.
\]
Define the extremal sampled candidates compatible with this displacement
bound by
\[
    \ell_N
    =
    \left\lfloor
        s_N^0-\widehat{\Delta}_N
    \right\rfloor,
    \qquad
    u_N
    =
    \left\lceil
        s_N^0+\widehat{\Delta}_N
    \right\rceil .
\]
A symmetric robust window is
\[
    \mathcal I_N(W_N)
    =
    \{k\in\mathbb Z:\ |k-k_N^0|\leq W_N\},
\]
with
\begin{equation}
    W_N
    \geq
    \max\{
        k_N^0-\ell_N,\,
        u_N-k_N^0
    \}.
    \label{eq:robust-window-radius}
\end{equation}
This construction contains both integer neighbours of every admissible
continuous peak location and therefore contains
\(k_N^{\mathrm{rec}}\). The corresponding containment proof is given in
Supplementary Section~S3.1. Let
\[
    k_N^{\mathrm{win}}
    \in
    \operatorname*{arg\,max}_{k\in\mathcal I_N(W_N)\cap\mathcal K_N}
    P_N(k).
\]

\begin{corollary}[Robust-window first-peak stopping]
\label{thm:robust-window-stopping}
Assume \(k_N^{\mathrm{rec}}\in\mathcal I_N(W_N)\cap\mathcal K_N\), and that
\(k_N^{\mathrm{rec}}\) is the unique sampled maximizer of \(F_N\) in the
window. Define
\[
    \Delta_N^{\mathrm{win}}
    =
    \min_{\substack{
        k\in\mathcal I_N(W_N)\cap\mathcal K_N\\
        k\neq k_N^{\mathrm{rec}}
    }}
    \left[F_N(k_N^{\mathrm{rec}})-F_N(k)\right].
\]
If
\begin{equation}
    2R_N^{\mathrm{val}}<\Delta_N^{\mathrm{win}},
    \label{eq:window-residual-condition}
\end{equation}
then
\[
    k_N^{\mathrm{win}}=k_N^{\mathrm{rec}}.
\]
\end{corollary}

This is Lemma~\ref{lem:sampled-gap-preservation} restricted to the robust
candidate set \(\mathcal I_N(W_N)\cap\mathcal K_N\).

The window construction guarantees containment of the leading-curve
floor--ceiling candidates, including \(k_N^{\mathrm{rec}}\), not automatically the
true noisy maximizer outside the window. If \(W_N=O(1)\), only a constant
number of candidate depths is added; more generally, any \(W_N=o(S_N)\) is subleading relative to the
\(O(S_N)\) amplification cost.

If sign information or asymmetric uncertainty bounds are available, the robust
window can be tightened separately on its early and late sides, provided it
contains every admissible floor--ceiling candidate pair and satisfies the same
sampled-gap certificate. The complete direction-aware construction is given in
Supplementary Section~S3.2.

\subsection{Residual-Induced Sampled Reversal}
\label{sec:residual-induced-reversal}
\subsubsection{Value-to-step conversion}
\label{subsec:value-to-step-conversion}

Let \(k_{\mathrm{lead}}\) be the sampled maximizer of the leading curve
\(F_N\), and let \(k\) be a competitor.  Define
\[
    \Delta_N^F(k)
    =
    F_N(k_{\mathrm{lead}})-F_N(k),
    \qquad
    D_N^r(k;k_{\mathrm{lead}})
    =
    r_N(k)-r_N(k_{\mathrm{lead}}).
\]
Then
\[
    P_N(k_{\mathrm{lead}})-P_N(k)
    =
    \Delta_N^F(k)-D_N^r(k;k_{\mathrm{lead}}).
\]
Hence only residual differentials matter for sampled stopping: a
common vertical shift is irrelevant, whereas a competitor overtakes the
leading step once its residual advantage exceeds the leading sampled
gap.

\subsubsection{Sampled reversal condition}
\label{subsec:sampled-reversal-condition}

For a competing step \(k_{\mathrm{comp}}\neq k_{\mathrm{lead}}\), sampled
reversal occurs exactly when
\begin{equation}
    r_N(k_{\mathrm{comp}})
    -
    r_N(k_{\mathrm{lead}})
    >
    \Delta_N^F(k_{\mathrm{comp}}).
    \label{eq:sampled-reversal-condition}
\end{equation}
In the fixed same-step setting,
\(k_{\mathrm{lead}}=k_N^0\).  The nearest competitor is typically one
of \(k_N^0\pm1\), and local quadraticity gives
\[
    \Delta_N^{\mathrm{near}}
    \asymp
    \kappa_N\mu_N(s_N^{\mathrm{lead}}).
\]
Using
\(m_N^{\mathrm{rem}}=m_N-|\Delta s_N^{\mathrm{lead}}|\) recovers the
curvature-scaled allowance in
Eq.~\eqref{eq:curvature-scaled-condition}.

The uniform condition
\(2R_N^{\mathrm{val}}<\Delta_N^F\) rules out every reversal, but its
failure only removes the certificate; an actual reversal additionally
requires the destabilizing sign in
Eq.~\eqref{eq:sampled-reversal-condition}.  For the ideal
single-target curve,
\[
    P_N^{(0)}(k)
    =
    1-4\theta_N^2(k-s_N^0)^2
    +O\!\left(\theta_N^4|k-s_N^0|^4\right),
\]
so the nearest-competing gap satisfies
\[
    \Delta_N^F(k_{\mathrm{comp}})
    \approx
    8\theta_N^2m_N
    \approx
    \frac{8m_N}{N}.
\]
Therefore the critical direct residual differential is
\[
    D_{N,\mathrm{crit}}^r
    \approx
    \frac{8m_N}{N}.
\]
If a symmetric adversarial perturbation lowers \(k_N^0\) and raises the
competitor by the same amplitude \(A_N^r\), the direct differential is
\(2A_N^r\), giving
\[
    A_{N,\mathrm{crit}}^{r,\mathrm{sym}}
    \approx
    \frac{4m_N}{N}.
\]
The direct residual construction, its spectral parametrization, and the
large-\(N\) validation protocol are given in Supplementary
Sections~S4.2--S4.4.

\subsubsection{Spectral interpretation}
\label{subsec:spectral-residual-parametrization}

The value-level reversal condition can be connected to spectral separation because
non-leading observable modes decay relative to the leading pair at a rate governed
by \(q_N=\beta_N/\rho_N\). A positive spectral gap alone is not sufficient: over
the \(O(1/\theta_N)\) first-peak window, the surviving residual differential must
be smaller than the local sampled gap, which is
\(\Theta(\kappa_Nm_N)=\Theta(m_N/N)\) in clean single-target Grover scaling.
Consequently, a slowly decaying non-leading mode can change the sampled maximizer
even while the leading first peak remains visible.

The explicit residual-mode parametrization, critical spectral coefficient, and the
resulting sufficient spectral-gap condition are given in Supplementary
Section~S4.3. The main-text point is
therefore operational: displacement control locates the leading lobe, while
residual control is separately required to preserve its sampled-value ordering.

\section{Numerical Validation}
\label{sec:numerical-validation}

The numerical study examines the predictions of the observable normal form
and the sampled decisions supported by its residual conditions. Five tests
address mechanism-resolved displacement, the separation of peak survival and
same-step stability, residual-induced reordering, parameter inference from
circuit and finite-shot data, and calibration-derived device-like trajectories
with matched baselines. Normal-form and residual constructions directly probe
the analytical conditions; circuit and backend-derived studies assess their
use through fitted observables and sampled residual checks. Detailed protocols
and secondary scans are given in Supplementary Sections~S4 and~S5.

\subsection{Mechanism and margin checks}
\label{subsec:numerical-single-mechanism-scaling}
\label{subsec:numerical-combined-displacement}
\label{subsec:numerical-margin-sensitive-stopping}

We test the mechanism decomposition within the observable normal form:
whether the three perturbation parameters produce the displacement scales
predicted by Eq.~\eqref{eq:main-displacement-formula}, and whether their joint
effect follows the signed displacement coordinate.
For single-target Grover search, the single-mechanism scans use
\(\delta_N=c/N\), \(\gamma_N=c/N\), and \(\phi_N=c/\sqrt N\), one at a time.
A separate \(N=256\) scan varies \((\delta_N,\phi_N)\) jointly, and a dense
search-size scan with \(\delta_N=\pm0.60/N\) tests the effect of the directional
rounding margin.  The full parameter choices are given in Supplementary
Subsections~S4.5--S4.7.

Figure~\ref{fig:mechanism-scaling} shows that the normalized exact displacements
approach the predicted constants for all three mechanisms.  The joint scan
collapses onto a diagonal band of nearly constant signed displacement, while the
dense \(N\)-scan shows that the same asymptotic perturbation law can either
preserve or change the sampled step depending on the arithmetic position of the
ideal peak inside its rounding cell.  The joint and margin-sensitive scans are shown in Supplementary
Figures~S3 and~S4.

\begin{figure}[t]
    \centering
    \includegraphics[width=\linewidth]{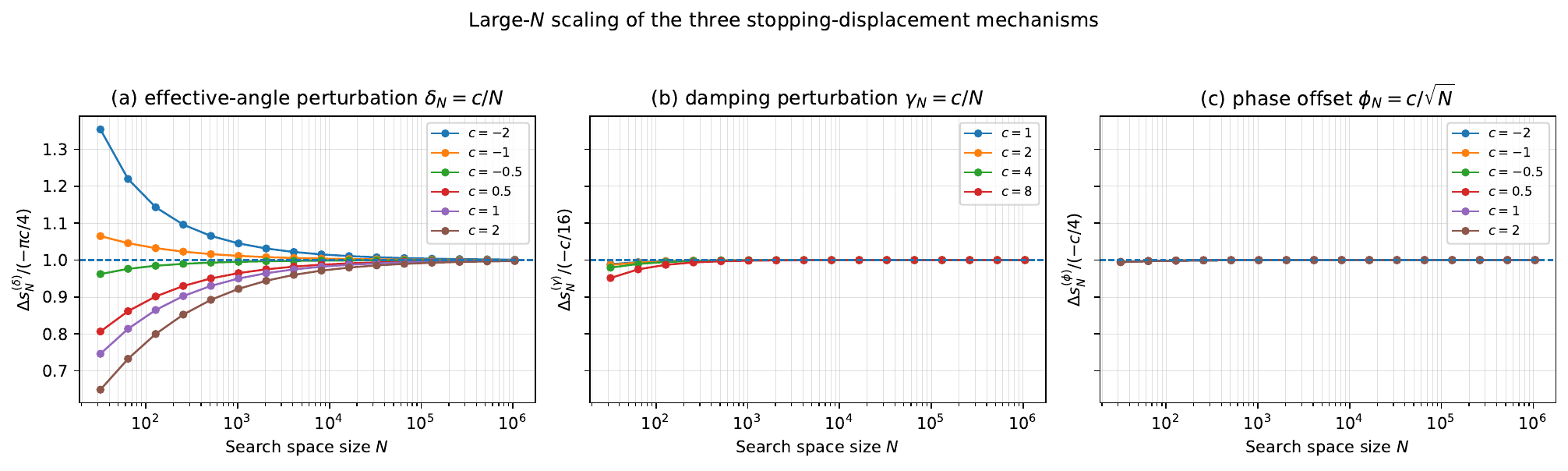}
    \caption{
    Single-mechanism large-\(N\) scaling.  The normalized displacements for
    \(\delta_N=c/N\), \(\gamma_N=c/N\), and \(\phi_N=c/\sqrt N\) approach the
    predicted limits \(-\pi c/4\), \(-c/16\), and \(-c/4\), respectively.
    }

\label{fig:mechanism-scaling}
\end{figure}

\subsection{Speedup--stability separation}
\label{subsec:numerical-peak-vs-stopping}

We test the distinct noise scales predicted for a surviving leading first
peak and for preservation of the ideal integer sampling decision.
We use damping-only trajectories over increasing search size and compare the
first-peak survival scale with the no-correction scale predicted by the
Section~\ref{sec:speedup-stability-duality} analysis. The sampled stopping step is
evaluated directly from the noisy trajectory.

\begin{figure}[t]
    \centering
    \includegraphics[width=\linewidth]{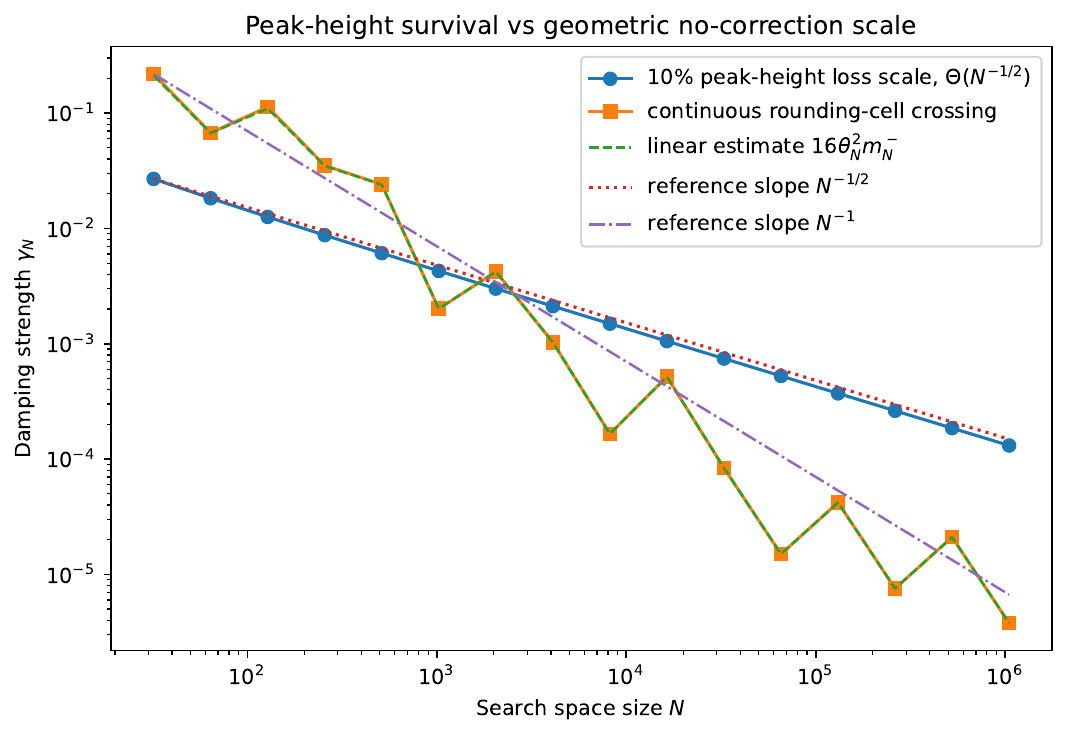}
    \caption{
    Peak-height survival versus the geometric no-correction scale under damping.
    A constant fractional peak-height loss occurs at the \(N^{-1/2}\) scale,
    whereas the rounding-margin threshold follows the sharper
    margin-modulated \(N^{-1}\)-type scale. Sampled decisions are evaluated
    directly as specified in Section~\ref{sec:fixed-sampled-stopping-stability}.
    }

\label{fig:peak-vs-stopping}
\end{figure}

Figure~\ref{fig:peak-vs-stopping} shows the two scales separating with \(N\): a
non-collapsed first peak persists on an \(N^{-1/2}\)-type damping scale, whereas
the no-correction scale follows the sharper margin-modulated \(N^{-1}\)-type
behavior. The two asymptotic envelopes differ, but their finite-\(N\)
separation is modulated by the arithmetic margin \(m_N\), so the plotted
no-correction threshold is non-monotone and does not define a single
crossover search size.

\subsection{Residual-induced sampled reversal}
\label{subsec:numerical-residual-reversal}

This experiment isolates the value-level failure mode that is not captured by
horizontal peak displacement: a non-leading residual can reverse the ordering
of neighbouring sampled depths while the leading peak remains geometrically
well located.
At \(N=256\), the leading curve is held fixed and the residual amplitude is
varied through the local sampled-gap threshold.  A second scan varies the
residual spectral ratio \(q=\beta/\rho\) to control how strongly the residual
survives to the first-peak window.

\begin{figure}[t]
    \centering
    \includegraphics[width=\linewidth]{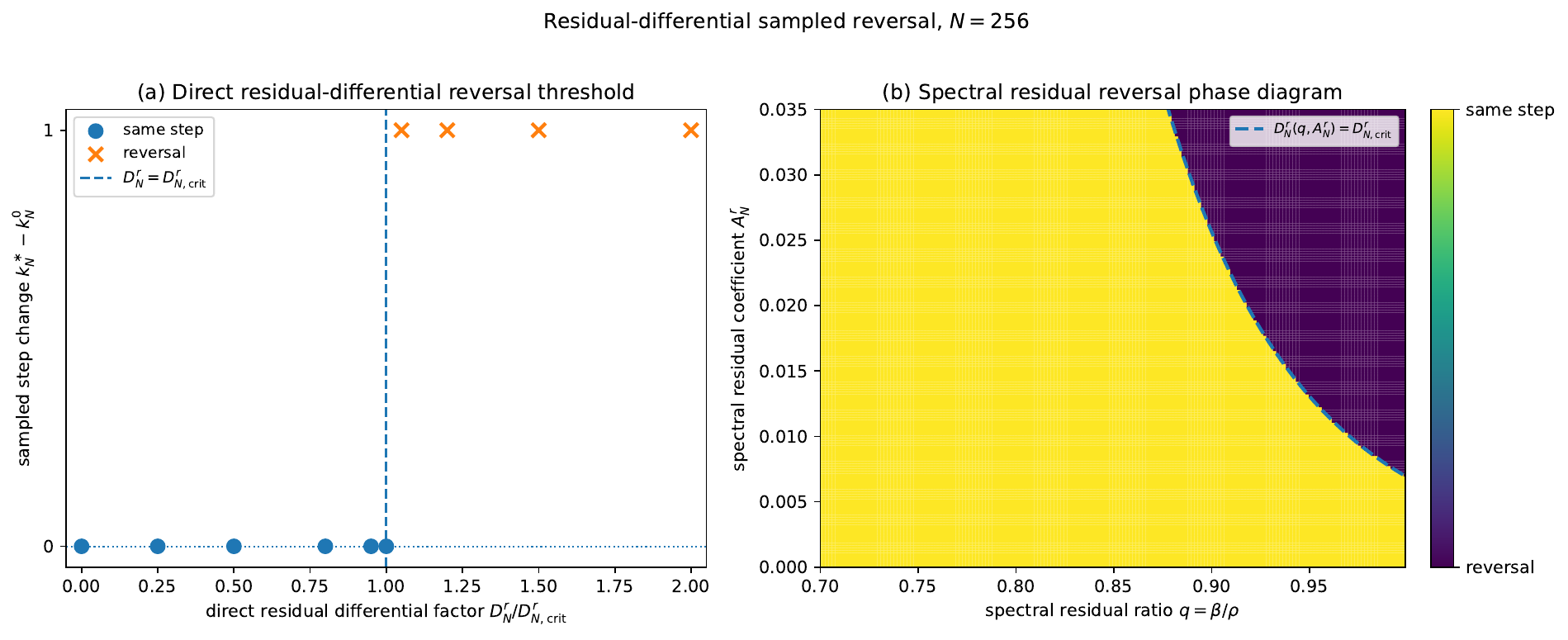}
    \caption{
    Residual-induced reversal at \(N=256\).  The sampled step changes when the
    residual differential exceeds the leading local gap; the spectral ratio
    \(q=\beta/\rho\) controls how strongly the residual survives to the first peak.
    }

\label{fig:residual-reversal}
\end{figure}

Figure~\ref{fig:residual-reversal} resolves the predicted transition: the
sampled step changes when the residual differential exceeds the leading local
gap, and modes with \(q\) closer to one require a smaller initial coefficient to
produce the reversal.  A larger-\(N\) supporting scan is reported in
Supplementary Figure~S2.

\subsection{Circuit and finite-shot robustness}
\label{subsec:coherent_circuit_validation}
\label{subsec:finite-shot-stopping-certification}

We test whether the same stopping quantities remain observable in an explicit
Grover circuit and whether the retain--recalibrate--window hierarchy survives
finite measurement noise.
For \(n=6\), \(N=64\), the marked-state reflection is given a coherent phase
over-rotation \(\varepsilon\), while the diffuser remains ideal.  The resulting
statevector trajectories are fitted to the observable normal form.  We then add
binomial sampling noise at
\(M\in\{512,2048,8192,32768\}\) shots per sampled depth and use
\(B_{\mathrm{boot}}=1000\) parametric-bootstrap refits.  Complete fitting and
bootstrap details are given in Supplementary
Subsections~S4.10 and~S4.12.

The exact circuit trajectories fit with root-mean-square residual below
\(10^{-9}\).  The observed sampled transition occurs near
\(\varepsilon\simeq0.154\), close to the first-order prediction near 0.150.
In the finite-shot bootstrap decisions, \(\varepsilon=0.10\) supports
retaining the ideal depth 6, \(\varepsilon=0.15\) returns the robust window
\(\{5,6,7\}\), and \(\varepsilon=0.20\) supports the recalibrated depth 5.
The full circuit and
finite-shot plots are reported in Supplementary
Figures~S5 and~S6;
bootstrap percentile intervals are shown in Supplementary
Figure~S8.

\subsection{Backend-derived device-like validation and matched baselines}
\label{subsec:device-like-validation}

The final main-text experiment assesses observable normal-form fitting and
sampled-depth certification under combined, hardware-like error mechanisms in a
GenericBackendV2 simulation. A calibration-derived FakeOslo study in the
Supplement provides a second backend-derived test with an archived calibration
snapshot.
Six-qubit Grover circuits are transpiled to a \texttt{GenericBackendV2} basis and
coupling graph with backend-derived gate errors, relaxation, and gate durations.
The simulated success probabilities are extracted from premeasurement density
matrices; readout-error metadata is available but readout noise is not applied
to these probabilities. At the ideal reference depth \(k_N^0=6\), the
transpiled circuit
has depth 2866 and 1915 CX gates.  Because the unscaled backend-derived model
destroys the useful first lobe, we use the controlled strength interpolation
\(\mathcal E_\alpha=(1-\alpha)\mathcal I+\alpha\mathcal E\) to resolve the
first-peak-preserving regime. The backend construction, strength scan, normal-form
fit, matched baselines, and calibration accounting are specified in Supplementary
Subsections~S4.13--S4.15.

Figure~\ref{fig:device-like-validation} summarizes the running \(N=64\) example.
At \(\alpha=0.01\), the sampled and recalibrated depths remain 6.  At
\(\alpha=0.06\), both are 5, with the neighbouring sampled values exchanging
order at \(\alpha_{\mathrm{cross}}\approx0.0235\).  The shifted normal-form fit
has \(R^2=0.998865\), and
\[
P_N(5)=0.626030,\qquad P_N(6)=0.594629,\qquad P_N(7)=0.505227.
\]
For the sampled diagnostics, let
\(\mathcal D_{13}=\{0,1,\ldots,12\}\) denote the 13-depth
evaluation set and let
\(\mathcal D_8=\{0,1,3,5,6,8,10,12\}\) denote the sparse
eight-depth calibration design. Both are distinct from the theoretical
first-peak window \(\mathcal W_N\). For a normal form fitted on D13, define
the fitted discrepancy
\[
R_{N,\mathcal D_{13}}^{\mathrm{fit}}
=
\max_{k\in\mathcal D_{13}}
\left|P_N(k)-F_N(k)\right|.
\]
This maximum bounds the fitted discrepancy at every depth in
\(\mathcal D_{13}\) and yields an \emph{a posteriori} sampled-ordering check on
that finite set. It is not by itself a bound on the theoretical spectral residual
outside the sampled set, nor a proof of the leading-pair assumptions.
The leading recalibrated gap is evaluated on this same numerical set
\(\mathcal D_{13}\), using the sampled-gap minimization of Section~4.5
with \(\mathcal D_{13}\) as its competitor domain. For the D13 fit
it is
\(\Delta_N^{\mathrm{rec}}(\mathcal D_{13})=0.024246\) and
\(2R_{N,\mathcal D_{13}}^{\mathrm{fit}}=0.021077\), so depth 5 passes
the \emph{a posteriori} ordering check on the tested sampled depths; the robust
fallback is
\(\{4,5,6,7,8\}\). Supplementary Subsection~S4.16
repeats the certificate using a normal form fitted only on D8, with the
remaining five depths held out. The recalibration gap is still evaluated
over \(\mathcal D_{13}\) using that D8-fitted normal form. For
GenericBackendV2, the D8 fit gives
\(\Delta_N^{\mathrm{rec}}(\mathcal D_{13})=0.024370\) and a
positive but narrow all-depth margin of about \(3.6\times10^{-5}\); the
calibration-derived FakeOslo case has an all-depth margin of
\(8.07\times10^{-4}\). The difference between the two displayed
\(\Delta_N^{\mathrm{rec}}(\mathcal D_{13})\) values comes from the
D13 and D8 fitting designs; both gaps use the same D13 candidate domain.

\begin{figure}[t]
    \centering
    \includegraphics[width=\textwidth]{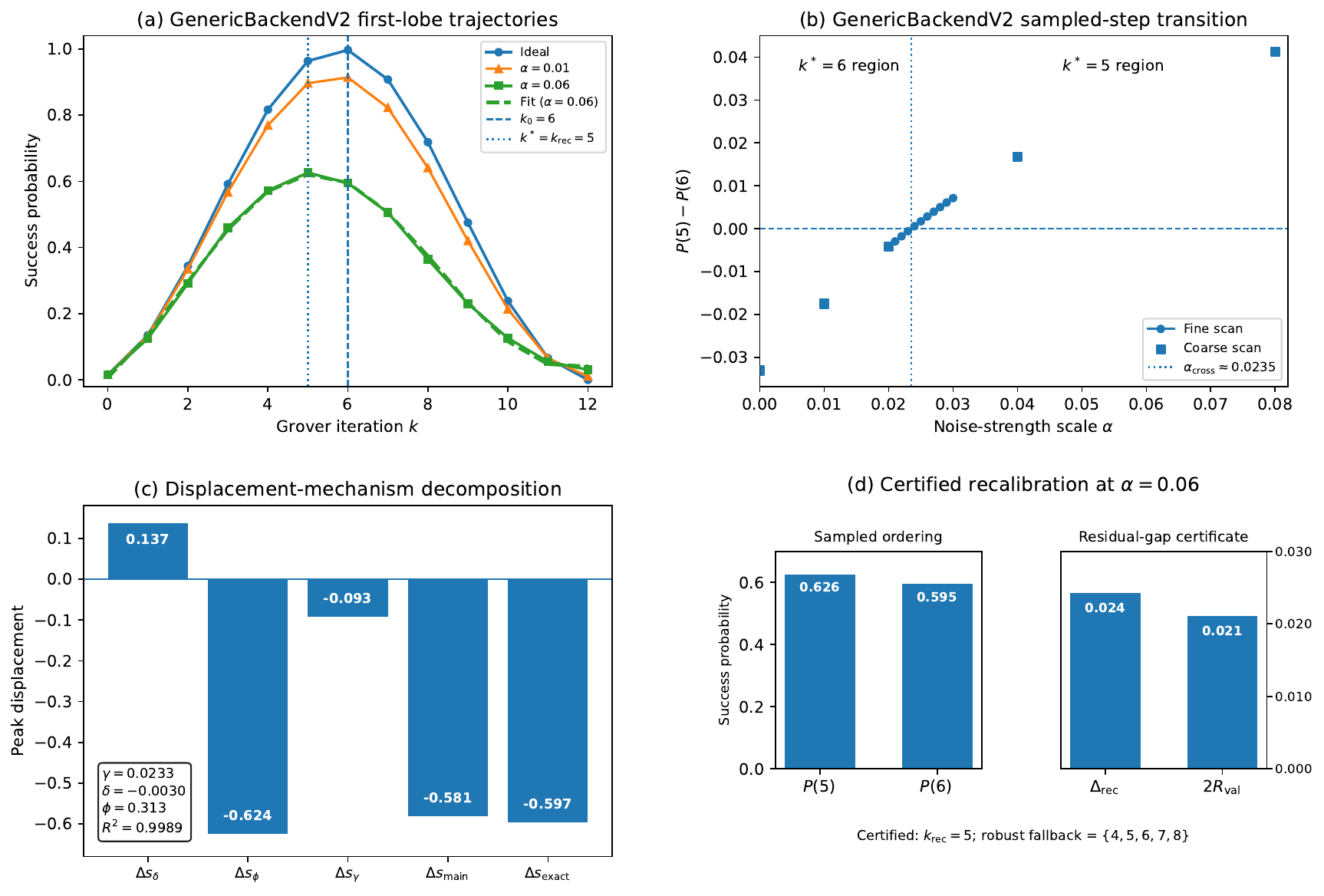}
    \caption{   
    GenericBackendV2 device-like validation for \(n=6\), \(N=64\).
    Panel (a) compares the ideal first lobe with reference and
    shifted noise strengths. Panel (b) resolves the sampled-step
    transition through the sign change of \(P_N(5)-P_N(6)\).
    Panel (c) decomposes the shifted leading-peak displacement
    into frequency, phase, and damping contributions and compares
    their sum with the fitted displacement. Panel (d) compares
    the D13 fitted-discrepancy scale
    \(2R_{N,\mathcal D_{13}}^{\mathrm{fit}}\)
    (labeled \(2R^{\mathrm{val}}\) in the panel) with
    \(\Delta_N^{\mathrm{rec}}(\mathcal D_{13})\) to perform an
    \emph{a posteriori} ordering check on the finite sampled set.
    This check neither bounds the theoretical spectral residual outside
    that set nor establishes the leading-pair assumptions.   
    }

    \label{fig:device-like-validation}
\end{figure}

\FloatBarrier

\paragraph{Matched quantitative baselines.}
Table~\ref{tab:device-like-baselines} compares the same trajectories under three
depth-selection rules.  The ``ideal reference'' is the known-fraction integer
Grover depth \(k_N^0=6\), not the randomized Boyer--Brassard--H\o yer--Tapp
(BBHT) schedule for an unknown marked fraction
\cite{boyer1998tight}.  The present framework checks the sampled first peak using the simulated
trajectory-derived discrepancy, whereas the Leng et al. rule uses a
success-probability predictor with the restart-aware
objective \(k/P(k)\); the in-hindsight minimum of that objective is shown only as
a diagnostic.

\begin{table}[t]
\centering
\caption{Matched quantitative comparison on the backend-derived device-like
trajectories. Peak retention is normalized by the actual noisy first-peak
probability in each regime; cost regret is relative to the in-hindsight minimum
of \(k/P(k)\) over the matched domain \(k=1,\ldots,6\).}
\label{tab:device-like-baselines}
\footnotesize
\setlength{\tabcolsep}{1.2pt}
\renewcommand{\arraystretch}{1.08}
\begin{tabular}{@{}>{\raggedright\arraybackslash}p{0.16\linewidth}>{\raggedright\arraybackslash}p{0.23\linewidth}*{5}{>{\centering\arraybackslash}p{0.109\linewidth}}@{}}
\toprule
Regime & Rule & \shortstack{Selected\\$k$} & \shortstack{$P_{\mathrm{noisy}}$\\$(k)$} & \shortstack{Peak\\retention} & \(k/P(k)\) & \shortstack{Cost\\regret} \\
\midrule
Reference (\(\alpha=0.01\)) & Ideal Grover step (known fraction) & 6 & 0.913853 & 1.000000 & 6.565609 & 26.25\% \\
Reference (\(\alpha=0.01\)) & Present framework & 6 & 0.913853 & 1.000000 & 6.565609 & 26.25\% \\
Reference (\(\alpha=0.01\)) & Leng et al. & 4 & 0.769176 & 0.841684 & 5.200373 & 0.00\% \\
\midrule
Shifted (\(\alpha=0.06\)) & Ideal Grover step (known fraction) & 6 & 0.594629 & 0.949841 & 10.090323 & 54.56\% \\
Shifted (\(\alpha=0.06\)) & Present framework & 5 & 0.626030 & 1.000000 & 7.986832 & 22.34\% \\
Shifted (\(\alpha=0.06\)) & Leng et al. & 4 & 0.572021 & 0.913728 & 6.992745 & 7.11\% \\
\bottomrule
\end{tabular}
\end{table}

In the reference regime, the present framework retains the first-peak depth 6,
whereas the Leng rule selects depth 4.  In the shifted regime, the present
framework restores the noisy first-peak depth 5, while Leng selects 4 and the
in-hindsight \(k/P(k)\) optimum is 3.  The first-peak and restart-aware objectives yield different choices:
the spectral stopping framework selects an \emph{a posteriori} checked first-peak
depth with a robust fallback, whereas the restart-aware rule prioritizes lower
expected
execution cost.

A calibration-derived \texttt{FakeOslo} snapshot (17 July 2022) gives the same qualitative
\(6\rightarrow5\) first-peak transition under controlled strength scaling, with
\(\alpha_{\mathrm{cross}}\approx0.011727\). At the frozen shifted point
\(\alpha=0.030\), the fit has \(R^2=0.999017\), \(k_N^*=k_N^{\mathrm{rec}}=5\),
and the \emph{a posteriori} finite-set residual-gap check remains positive when the
normal form is fitted only on the D8 depths and tested on the five excluded
depths
\(\{2,4,7,9,11\}\). The complete protocol and stress-test figure are reported
in Supplementary Subsection~S4.16 and
Figure~S10.

\section{Practical Implications and Limitations}
\label{sec:practical-algorithmic-implications}

\subsection{Operational use and amplitude-estimation connection}
\label{subsec:diagnostic-workflow}
\label{subsec:practical-decision-rule}

The observable spectral framework yields the retain--recalibrate--window
hierarchy in Figure~\ref{fig:diagnostic-hierarchy}. Under finite shots,
the application uses joint bootstrap support for fitted stopping decisions.
This empirical support is not, by itself, a rigorous confidence bound on the
unobserved structural residual \(R_N^{\mathrm{val}}\); theorems requiring a
valid residual allowance retain that condition separately.

Amplitude estimation provides a natural application because it must choose
amplification powers even when the underlying success amplitude is initially
unknown. In the tested workflow, a coarse QPE-free amplitude-estimation stage
maximizes the ideal binomial success likelihood over a small pre-specified
set of amplification depths to infer a provisional ideal reference depth.
This is a likelihood-based QPE-free estimator of the type discussed in
\cite{suzuki2020amplitudeEstimationWithoutQPE}. The estimator has no access to the true
success fraction, ideal Grover angle, ideal stopping depth, noisy optimum, or
fitted normal-form parameters. The provisional reference defines a seven-depth
guard band, within which the stopping framework fits the noisy first lobe and
returns either a unique recalibrated depth or a robust local window.
The complete application protocol is provided in Supplementary
Subsection~S4.17.

For the \(N=64\) running example, offline screening against the known
GenericBackendV2 simulated optimum selects a localization configuration, using the
fixed depth schedule \(\{0,1,2,4\}\) with 512 shots per depth. Under the shifted
GenericBackendV2 trajectory, this coarse estimator returns provisional depths 7
or 8 rather than the true noisy optimum 5, but every resulting guard band
contains both the ideal reference depth 6 and the noisy optimum. The local
stopping stage then uses 8192 shots per depth and 200 bootstrap refits per
end-to-end dataset. Across 100 independent datasets, the bootstrap modal depth
is 5 in every case: 60 datasets support a unique bootstrap-selected depth 5,
while the remaining 40 return robust windows. There are no incorrect decisive
outputs, no fit failures, and every fallback window contains the true noisy
optimum. The coarse estimator localizes the candidate region, while the fitted
stopping rule resolves the subsequent sampled-depth decision.

The same configuration is transferred without retuning to the
calibration-derived FakeOslo trajectory, providing the independent-backend
transfer test after the GenericBackendV2 design screening. It again attains
complete observed relevant-window coverage, with 78 unique correct
recalibrations and 22 safe fallbacks out of 100 datasets. Table~\ref{tab:ae-localized-application} summarizes the resulting
cost--reliability tradeoff. Under conservative accounting in which the
amplitude-estimation and local-correction measurements are independent, the
GenericBackendV2 and FakeOslo configurations cost about 72\% and 71\% less,
respectively, than the shifted D8 reference configuration. The D8 reference
uses 32768 shots per depth versus 8192 for local calibration, with higher
unique-depth return rates; the observed cost difference is primarily a
shot-budget comparison, not a saving isolated to AE localization. Reusing
overlapping observations gives only a small additional credit.

\begin{table}[h!]
\centering
\caption{AE-localized stopping application versus the shifted D8 calibration
baseline. ``Correct unique'' is the observed unconditional rate of returning
the true single depth; ``wrong decisive'' counts incorrect bootstrap-supported
single-depth outputs; fallback coverage is conditional on returning a window.
The D8 reference uses 32768 shots per depth, versus 8192 for local calibration
in the AE-localized routes; these are not equal-shot or equal-reliability
comparisons. All rates are empirical over 100 datasets per route.}
\label{tab:ae-localized-application}
\scriptsize
\setlength{\tabcolsep}{1.2pt}
\renewcommand{\arraystretch}{1.06}
\begin{tabular}{@{}>{\raggedright\arraybackslash}p{0.275\linewidth}*{6}{>{\centering\arraybackslash}p{0.108\linewidth}}@{}}
\toprule
Route & \shortstack{Correct\\unique} & \shortstack{Wrong\\decisive} & \shortstack{Fallback\\coverage} & \shortstack{Mean\\$Q$} & \shortstack{Mean\\shots} & \shortstack{$Q$ reduction\\vs D8} \\
\midrule
Shifted D8 (GenericBackendV2) & 0.99 & 0.00 & 1.00 & 1,474,560 & 262,144 & -- \\
AE-localized GenericBackendV2 & 0.60 & 0.00 & 1.00 & 407,286 & 59,392 & 72.38\% \\
AE-localized FakeOslo & 0.78 & 0.00 & 1.00 & 421,622 & 59,392 & 71.41\% \\
\bottomrule
\end{tabular}
\end{table}

\subsection{Calibration cost and when it pays}
\label{subsec:calibration-cost}

The correction rules require the observable normal form to be calibrated or
otherwise inferred.  Let \(M_j\) be the shots at sampled depth \(j\), and define
the Grover-iterate-equivalent calibration cost
\[
    Q_{\mathrm{cal}}=\sum_j M_j j.
\]
For each reported design/regime, the reliability study uses 100 independent
calibration datasets with 200 bootstrap refits per dataset. The six-depth stress design is cheaper but visibly
distorts the shifted fit: it gives \(s_N^{\mathrm{lead}}=5.091864\) and an RMSE of \(0.019880\) evaluated over the 13-depth set
\(\mathcal D_{13}\), compared with \(5.169850\) and \(0.007342\) for
the D13 reference fit. The recommended D8 design defined in Subsection~\ref{subsec:device-like-validation}
gives \(s_N^{\mathrm{lead}}=5.166592\) and RMSE \(0.007595\), nearly
matching D13 at substantially lower cost.

Across the 300 tested datasets there are no incorrect decisive
single-depth outputs, every fallback window contains the true sampled optimum,
and the fit-failure rate is zero. These are empirical frequencies, not zero-error
guarantees.

For the shifted trajectory, keeping \(k=6\) gives
\(k/P(k)=10.090323\), whereas depth \(k=5\), selected by an \emph{a posteriori} sampled-set check, gives 7.986832, a
20.85\% reduction. Against the recommended D8 cost
\(Q_{\mathrm{cal}}=1{,}474{,}560\), that saving is recovered only after about
701,007 repeated uses under approximately stationary noise. The D8 calibration cost thus requires amortized reuse of an approximately
stationary setting and is not recovered for a one-off workload. The AE-localized application in Subsection~6.1 addresses this cost bottleneck
differently: a cheap coarse localization stage first restricts the relevant depth
region, after which only that local window is calibrated. Its measured end-to-end
costs and reliability are reported in Table~\ref{tab:ae-localized-application}.
The tested lower-shot AE-localized configuration has lower calibration cost
than the higher-shot D8 reference, without isolating a localization-specific
saving. Dividing its mean query cost by the same per-use $k/P(k)$ difference
gives approximately $1.94\times10^{5}$ repeated uses for GenericBackendV2.
Since 40\% of its datasets return a window, this is only an illustrative
normalization, not an observed break-even point or an equal-reliability comparison.

\begin{table}[h!]
\centering
\caption{Finite-shot calibration reliability for the running example. Rates are
measured over 100 independent calibration datasets per case. ``Correct'' is the
unconditional observed rate of returning the true single depth; ``wrong decisive''
counts incorrect bootstrap-supported single-depth outputs; fallback coverage is
conditional on returning a robust window. Rates are empirical, not statistical
coverage guarantees.}
\label{tab:calibration-cost}
\scriptsize
\setlength{\tabcolsep}{1.2pt}
\renewcommand{\arraystretch}{1.06}
\begin{tabular}{@{}>{\raggedright\arraybackslash}p{0.23\linewidth}*{6}{>{\centering\arraybackslash}p{0.118\linewidth}}@{}}
\toprule
Regime/design & \(M\) & \(Q_{\mathrm{cal}}\) & Decisive & Correct & \shortstack{Wrong\\decisive} & \shortstack{Fallback\\coverage} \\
\midrule
Reference D8 & 2048 & 92160 & 0.98 & 0.98 & 0.00 & 1.00 \\
Shifted D8 & 32768 & 1474560 & 0.99 & 0.99 & 0.00 & 1.00 \\
Shifted D13 & 32768 & 2555904 & 0.99 & 0.99 & 0.00 & 1.00 \\
\bottomrule
\end{tabular}
\end{table}

\FloatBarrier
\subsection{Scope and limitations}
\label{subsec:scope-limitations}

The framework applies to stationary noisy standard amplitude amplification
with a known ideal success fraction and a fixed reference step, subject to an
observable-relevant semisimple leading conjugate pair, spectral separation,
and controlled non-leading residuals on the first lobe. The asymptotic scaling
laws and encoded-subspace channel specialization are developed for single-target
Grover search. Phase-modified, fixed-point, and QSP/QSVT-based amplification
schemes have different trajectory geometries; nearly defective or strongly
non-normal dynamics may also fall outside the controlled damped-cosine class.

The GenericBackendV2 and FakeOslo studies use strength-scaled simulations and
an archived calibration snapshot, not live hardware executions. Independently
transpiled circuits at each depth need not implement powers of one stationary
per-iterate map; their normal-form fits are effective trajectory diagnostics.
Fits and held-out sampled residual checks do not establish the spectral assumptions
of Section~4.1 for those device-like trajectories. The unscaled backend-derived
noise also suppresses the useful first lobe. The
calibration-cost analysis assumes a noise setting stable enough to amortize
calibration over repeated uses. The AE-localized application supplies a
provisional depth region from finite-shot data; it does not alter the
known-reference assumptions of the stopping theorems. A deployment-level
statistical certificate would additionally require a valid residual allowance,
including measurement and model-mismatch uncertainty; that was not established
in these simulations. The sampled first-peak objective is also distinct from restart-aware runtime minimization and from
the end-to-end objective of an amplitude-estimation protocol.

\section{Conclusion}
\label{sec:conclusion}

We established an observable spectral framework for stationary noisy standard
amplitude amplification with a leading observable-relevant oscillatory pair,
spectral separation, and controlled residual contributions. The resulting
normal form connects frequency, damping, and phase perturbations to first-peak
motion and integer stopping decisions across the defined dynamics class. In
single-target Grover search, frequency and damping have \(\Theta(N)\)
stopping-coordinate sensitivities, while observable phase offsets act at
\(\Theta(\sqrt N)\) scale. Sampled-gap and residual bounds distinguish
first-peak survival from same-step stability, identify residual-induced
reversals, and support retention, certified recalibration, or a robust window.

The numerical studies test these conclusions using analytic constructions,
circuit simulations, finite-shot fits, and strength-scaled backend-derived
models. In the simulated \(N=64\) example, an \emph{a posteriori}
sampled-set ordering check selects depth 5 over 6, with success probability
increasing from 0.595 to 0.626. The amplitude-estimation-localized application
illustrates a calibration-cost--decision-reliability trade-off under unequal
shot budgets: the lower measured cost relative to the higher-shot D8 reference
is driven primarily by the smaller shot budget, rather than an isolated
localization-specific saving. These device-like results are simulation based;
the spectral assumptions and empirical limitations are detailed in
Subsection~\ref{subsec:scope-limitations}.

\section*{Research statements}

\subsection*{Availability of data and materials}
The numerical protocols, parameter settings, and
supporting analyses are documented in Supplementary
Section~S4. The numerical data and reproducibility
code supporting the reported results are provided as a
separate ZIP archive to the editors and referees with
the journal submission. The archive is not included
in this arXiv preprint. No human-subject or
proprietary datasets were used.

\subsection*{Competing interests}

The authors declare that they have no competing interests.

\subsection*{Funding}

No specific funding was received for this research.

\subsection*{Authors' contributions}

Ge Zhong conceived the research, developed the
theoretical framework and mathematical derivations,
designed and conducted the numerical experiments,
performed the computational analyses, and wrote
the original manuscript.

Charith Jayasekara, Nisal De Silva, Muhammad Usman,
Muhammad Aamir Cheema, and Chetan Arora provided
substantive scientific guidance and critical review,
including evaluation of the theoretical arguments,
assessment of the numerical validation, and
recommendations for additional analyses and
verification. They also contributed to the
scientific interpretation and revision of the manuscript.

ChatGPT was used to assist with language polishing,
manuscript restructuring, LaTeX editing, consistency
checks, selected derivation cross-checks, and preliminary
drafting and debugging of numerical code. All mathematical
arguments, scripts, simulations, figures, citations,
interpretations, and conclusions were reviewed and
verified by the authors. No AI-generated numerical data
or unverified scientific claims were used. All authors
take full responsibility for the content of the work.

\bibliographystyle{unsrtnat}
\bibliography{references}

% === SUPPLEMENTARY MATERIAL ===
% Integrated appendix: the scientific text and all original S-labels are retained.
\clearpage
\phantomsection
\pdfbookmark[1]{Supplementary Material}{supplementary-material}
\begin{center}
{\Large\bfseries Supplementary Material}
\end{center}
\medskip

\noindent
This supplement is organized as a derivation and reproducibility companion to the
main text. Supplementary Table~\ref{tab:notation-summary} collects the notation used across the theory. Section~\ref{app:spectral-normal-form} derives the observable spectral
normal form, leading-peak equation, residual bound, perturbative realizability
route, and encoded-subspace specialization used in main-text
Sections~4.1--4.2.
Section~\ref{app:peak-displacement} derives the mechanism-resolved displacement
formula, its local remainder, the continuous-cell geometry, and the curvature-to-
sampled-gap bound used in main-text Sections~4.2--4.3.
Section~\ref{app:stopping-rule-correction-proofs} gives the correction-specific
containment and overhead arguments for main-text Section~4.5.
Section~\ref{app:residual-numerical-protocol} derives the residual-reversal
construction and records the numerical protocols, including the backend-derived
validation, matched baselines, calibration-cost study, the calibration-derived
FakeOslo/held-out stress test, and the AE-localized stopping application in
Subsection~\ref{app:ae-localized-application-protocol}. Finally,
Section~\ref{app:additional-figures} collects only the supporting plots that add
information beyond the main-text figures. Throughout, ``main-text'' references refer to
numbered results in Sections~1--7 of this document.

\appendix

\setcounter{section}{0}
\renewcommand{\thesection}{S\arabic{section}}
\renewcommand{\thesubsection}{\thesection.\arabic{subsection}}
\renewcommand{\thesubsubsection}{\thesubsection.\arabic{subsubsection}}
\setcounter{equation}{0}
\renewcommand{\theequation}{S\arabic{equation}}
\setcounter{figure}{0}
\renewcommand{\thefigure}{S\arabic{figure}}
\setcounter{table}{0}
\renewcommand{\thetable}{S\arabic{table}}
\renewcommand{\theHsection}{supp.\arabic{section}}
\renewcommand{\theHsubsection}{supp.\arabic{section}.\arabic{subsection}}
\renewcommand{\theHsubsubsection}{supp.\arabic{section}.\arabic{subsection}.\arabic{subsubsection}}
\renewcommand{\theHequation}{supp.\arabic{equation}}
\renewcommand{\theHfigure}{supp.\arabic{figure}}
\renewcommand{\theHtable}{supp.\arabic{table}}

\begin{table}[h!]
\centering
\caption{Core notation used throughout the paper.}
\label{tab:notation-summary}

\footnotesize
\setlength{\tabcolsep}{4pt}
\renewcommand{\arraystretch}{1.06}
\begin{tabular}{p{0.18\textwidth}p{0.76\textwidth}}
\toprule
Symbol & Meaning \\
\midrule
\(N\), \(a_N\), \(\theta_N\) &
Search size, initial success probability, and amplitude angle, with
\(\sin^2\theta_N=a_N\). In single-target Grover search, \(a_N=1/N\) and
\(\theta_N\sim N^{-1/2}\). \\

\(s_N^0\), \(k_N^0\) &
Ideal continuous first-peak location and ideal sampled stopping step:
\(s_N^0=\pi/(4\theta_N)-1/2\), \(k_N^0=\operatorname{round}(s_N^0)\). \\

\(m_N\), \(m_N^+\), \(m_N^-\) &
Rounding-cell margin and directional right/left margins from \(s_N^0\) to the
integer-cell boundaries. \\

\(\mu_N(s)\), \(m_N^{\mathrm{rem}}\) &
Local rounding margin of a continuous peak \(s\), and the conservative remaining
margin \(m_N^{\mathrm{rem}}=m_N-|\Delta s_N^{\mathrm{lead}}|\). \\

\(\mathcal W_N\), \(\mathcal K_N\) &
Continuous first-peak window and integer first-peak window,
\(\mathcal K_N=\mathcal W_N\cap\mathbb Z_{\geq0}\); the relevant
floor--ceiling candidates are required to lie in \(\mathcal K_N\). \\

\(\widetilde Q_N\), \(\lambda_{\pm,N}\) &
Noisy one-step iterate and observable-relevant leading spectral pair,
\(\lambda_{\pm,N}=\rho_Ne^{\pm i\omega_N}\). \\

\(\rho_N\), \(\beta_N\), \(g_N\) &
Leading-pair modulus, largest observable residual spectral modulus (excluding
stationary background modes; zero if none), and gap \(g_N=\rho_N-\beta_N\). \\

\(K_N\), \(R_N^{\mathrm{val}}\) &
Projector-growth constant and residual value bound
\(R_N^{\mathrm{val}}=\max_{k\in\mathcal K_N}|r_N(k)|\). \\

\(B_N\), \(C_N\), \(F_N(k)\), \(r_N(k)\) &
Background, leading oscillatory amplitude, leading damped-cosine curve, and
non-leading residual contribution in \(P_N(k)=F_N(k)+r_N(k)\). \\

\(\gamma_N\), \(\delta_N\), \(\phi_N\) &
Damping scale, effective-angle perturbation, and observable phase offset, with
\(\omega_N=4(\theta_N+\delta_N)\) and \(\chi_N=2\theta_N+\phi_N\). \\

\(s_N^{\mathrm{lead}}\), \(\Delta s_N^{\mathrm{lead}}\) &
Noisy leading continuous first peak and its displacement from \(s_N^0\). \\

\(\Delta s_N^{\mathrm{main}}\), \(\mathcal R_N^{\mathrm{disp}}\) &
First-order displacement contribution and higher-order displacement remainder. \\

\(\kappa_N\), \(c_{\mathrm{curv}}\), \(\Delta_N^F\) &
Local curvature, curvature-gap regularity constant, and leading sampled value gap
near the first peak. \\

\(k_N^*\), \(k_N^{\mathrm{rec}}\), \(W_N\), \(\mathcal I_N(W_N)\) &
Noisy sampled first-peak maximizer, recalibrated stopping step, robust-window
radius, and the corresponding robust stopping window. \\

\(D_N^r\), \(D_{N,\mathrm{crit}}^r\) &
Direct residual differential and its critical threshold for sampled reversal. \\

\(A_N^r\), \(A_{N,\mathrm{crit}}^r\), \(q_N\) &
Spectral residual coefficient, its critical value, and residual-to-leading
spectral ratio \(q_N=\beta_N/\rho_N\). \\
\bottomrule
\end{tabular}
\end{table}

\FloatBarrier

\section{Spectral Normal Form and Perturbative Realizability}
\label{app:spectral-normal-form}

This supplementary section supplies the technical basis for the spectral model in
main-text Subsections~4.1.1--4.1.2 and the observable normal form in
main-text Subsection~4.2.1. Sections~\ref{subsec:app-linear-representation}--\ref{subsec:app-background-residual-modes} derive how an
observable-relevant leading eigenvalue pair becomes a damped cosine and how the
remaining modes enter the background and residual terms; Sections~\ref{subsec:app-residual-bound-spectral-separation}--\ref{app:bauer-fike}
provide residual control and a perturbative realizability route, and Section~\ref{app:encoded-depolarizing-specialization}
gives the encoded-subspace depolarizing specialization.

\subsection{Linear representation of the noisy trajectory}
\label{subsec:app-linear-representation}

Let \(\mathsf H_N\) be the Hilbert space of the search problem and let
\(\mathsf L(\mathsf H_N)\) denote the associated operator space. After choosing any
fixed vectorization convention, the noisy superoperator
\(\widetilde{\mathcal Q}_N\) can be represented by a finite-dimensional matrix
\(\widetilde Q_N\). The initial state \(\rho_{N,0}\) is represented by a vector
\(|\rho_{N,0}\rangle\rangle\), while the success measurement operator \(M_N\)
defines a linear functional
\[
    \langle\langle M_N|
\]
such that
\[
    P_N(k)
    =
    \langle\langle M_N|
    \widetilde Q_N^k
    |\rho_{N,0}\rangle\rangle .
\]
This expression is only a matrix representation of the physical quantity
\(\operatorname{Tr}(M_N\widetilde{\mathcal Q}_N^k(\rho_{N,0}))\); it does not
depend on the particular vectorization convention.

We assume that, on the observable-relevant part of the spectrum, the leading
oscillatory pair is semisimple. The same derivation can be extended to finite
Jordan blocks, but the residual then contains additional polynomial factors in
\(k\). Since the main text treats the first-peak-preserving perturbative regime, we
use the semisimple form here.

Let the observable-relevant leading pair be
\[
    \lambda_{\pm,N}
    =
    \rho_N e^{\pm i\omega_N},
    \qquad
    \rho_N=e^{-\gamma_N}.
\]
Let \(\Pi_{+,N}\) and \(\Pi_{-,N}\) be the corresponding spectral projectors.
Because the observed success probability is real-valued, the two contributions
combine as complex conjugates.

\subsection{Leading oscillatory contribution}
\label{subsec:app-leading-oscillatory-contribution}

The leading-pair contribution to the observed success probability is
\[
    P_N^{\mathrm{osc}}(k)
    =
    \langle\langle M_N|
    \left(
        \lambda_{+,N}^k\Pi_{+,N}
        +
        \lambda_{-,N}^k\Pi_{-,N}
    \right)
    |\rho_{N,0}\rangle\rangle .
\]
Define the complex observable coefficient
\[
    A_N
    =
    \langle\langle M_N|
    \Pi_{+,N}
    |\rho_{N,0}\rangle\rangle .
\]
The conjugate mode contributes \(\overline{A_N}\) in the real observable. Therefore
\[
    P_N^{\mathrm{osc}}(k)
    =
    A_N \rho_N^k e^{ik\omega_N}
    +
    \overline{A_N}\rho_N^k e^{-ik\omega_N}
    =
    2\rho_N^k
    \operatorname{Re}\!\left(A_N e^{ik\omega_N}\right).
\]
Writing
\[
    A_N=-\frac{C_N}{2}e^{i\chi_N},
    \qquad
    C_N>0,
\]
we obtain
\[
    P_N^{\mathrm{osc}}(k)
    =
    -C_N \rho_N^k\cos(k\omega_N+\chi_N).
\]
Since \(\rho_N=e^{-\gamma_N}\), this becomes
\[
    P_N^{\mathrm{osc}}(k)
    =
    -C_Ne^{-\gamma_N k}\cos(k\omega_N+\chi_N).
\]
The sign convention is chosen so that the first success-probability peak occurs near
a cosine minimum.

\subsection{Background and residual modes}
\label{subsec:app-background-residual-modes}

The full observed trajectory also contains stationary observable contributions.
In particular, the eigenvalue-\(1\) modes of a CPTP channel contribute a
constant term \(B_N\). Other non-leading modes, including slowly varying but
nonstationary modes, cannot in general be absorbed into a constant background.

The remaining observable modes are collected into the residual
\[
    r_N(k)
    =
    \sum_{j\in\mathcal J_N}
    a_{j,N}\lambda_{j,N}^k ,
\]
where \(\mathcal J_N\) indexes observable-relevant modes other than the
leading pair and the stationary modes included in \(B_N\), and the
coefficients \(a_{j,N}\) include the spectral projector, initial-state, and
measurement overlaps.
This gives the scalar decomposition
\begin{equation}
    P_N(k)
    =
    B_N
    -
    C_N e^{-\gamma_N k}
    \cos(k\omega_N+\chi_N)
    +
    r_N(k).
    \label{eq:app-observable-normal-form}
\end{equation}
This is the observable-level cosine normal form used in the main text.

In the ideal standard amplitude-amplification case,
\[
    B_N=\frac{1}{2},
    \qquad
    C_N=\frac{1}{2},
    \qquad
    \gamma_N=0,
    \qquad
    \omega_N=4\theta_N,
    \qquad
    \chi_N=2\theta_N,
    \qquad
    r_N(k)=0.
\]
Therefore Eq.~\eqref{eq:app-observable-normal-form} reduces to
\[
    P_N^{(0)}(k)
    =
    \frac{1}{2}
    -
    \frac{1}{2}\cos((4k+2)\theta_N)
    =
    \sin^2((2k+1)\theta_N).
\]

\subsection{Residual bound from spectral separation}
\label{subsec:app-residual-bound-spectral-separation}

We now derive the residual estimate used in
main-text Subsection~4.1.2. Let
\[
    \beta_N
    =
    \max_{j\in\mathcal J_N}|\lambda_{j,N}|
\]
be the largest observable residual modulus, with \(\beta_N=0\) by
convention if \(\mathcal J_N=\varnothing\). In that case
\(r_N(k)\equiv0\), and no residual decay estimate is needed.
Otherwise the spectral separation assumption is
\[
    g_N=\rho_N-\beta_N>0.
\]
Equivalently, with
\[
    q_N=\frac{\beta_N}{\rho_N},
\]
we have \(q_N<1\).

For integer \(k\in\mathcal K_N=\mathcal W_N\cap\mathbb Z_{\geq0}\), where
\(\mathcal W_N=[c_1/\theta_N,c_2/\theta_N]\) is the continuous
first-lobe window, the residual contribution satisfies the spectral bound
\[
    |r_N(k)|
    \leq
    K_N^{(0)}\beta_N^k,
\]
where \(K_N^{(0)}\) collects the observable-overlap coefficients, spectral
projector norms, and possible eigenvector-conditioning factors. Since
\[
    q_N=\frac{\beta_N}{\rho_N},
    \qquad
    \beta_N^k=\rho_N^k q_N^k,
\]
we can also write
\[
    |r_N(k)|
    \leq
    K_N^{(0)}\rho_N^k q_N^k .
\]
In the CPTP setting considered in the main text, the leading modulus satisfies
\(\rho_N\leq 1\). Therefore
\[
    |r_N(k)|
    \leq
    K_N^{(0)}q_N^k .
\]
Renaming \(K_N^{(0)}\) as \(K_N\), we obtain
\[
    |r_N(k)|
    \leq
    K_N q_N^k .
\]
Thus no factor \(\rho_N^{c_*/\theta_N}\) is hidden inside \(K_N\); it has only been
discarded using \(\rho_N\leq 1\). Keeping this factor would give a slightly sharper
bound in damping-dominated cases.

Since \(k\geq c_*/\theta_N\) on the relevant part of the first-peak window, with
\(c_*>0\), we obtain
\[
    R_N^{\mathrm{val}}
    =
    \max_{k\in\mathcal K_N}|r_N(k)|
    \leq
    K_N q_N^{c_*/\theta_N}.
\]
Using
\[
    q_N
    =
    \frac{\beta_N}{\rho_N}
    =
    1-\frac{g_N}{\rho_N},
\]
and \(\log(1-x)\leq -x\), this gives
\begin{equation}
    R_N^{\mathrm{val}}
    \leq
    K_N
    \exp\!\left(
        -\frac{c_*g_N}{\rho_N\theta_N}
    \right).
    \label{eq:app-residual-bound}
\end{equation}

The factor \(K_N\) is necessary because the noisy iterate need not be normal. It
absorbs spectral projector norms, eigenvector conditioning, and observable-overlap
constants, but not the exponential first-window factor \(\rho_N^{c_*/\theta_N}\).
Thus spectral separation alone is not enough; the residual estimate is useful only
when this conditioning factor does not grow too quickly with \(N\).

\subsection{Connection with the main assumptions}
\label{subsec:app-main-assumptions-connection}

The derivation above explains the assumptions used in the main text. The leading
oscillatory pair gives the damped cosine component. The background modes give
\(B_N\). The remaining observable modes give \(r_N(k)\), and their contribution is
controlled by the spectral ratio \(\beta_N/\rho_N\), the first-peak window length
\(O(1/\theta_N)\), and the projector-growth factor \(K_N\).

The normal form should therefore be understood as an observable-level reduction,
not as a claim that the full noisy channel has only two relevant eigenvalues. Other
modes may be present in the full superoperator. They are harmless for fixed sampled
stopping only when their observable residual contribution is below the
curvature-scaled sampled-gap allowance used in
main-text Subsection~4.3.2.

\subsection{Conditioning regimes and observable non-degeneracy}
\label{subsec:app-conditioning-nondegeneracy}

The projector-growth factor \(K_N\) in main-text Eq.~(11) is part of
the model assumptions.  It collects the conditioning of the observable spectral
decomposition, including spectral projector norms, eigenvector-conditioning
factors, and initial-state--measurement overlaps.  A typical admissible regime is
\[
    K_N=O(N^\alpha)
\]
for a fixed \(\alpha\geq0\), or more generally
\(\log K_N=O(\log N)\).  If the observable-relevant block is normal or uniformly
diagonalizable and the relevant overlaps are bounded, then \(K_N=O(1)\).  More
generally, polynomial growth can be absorbed into the residual estimate when the
relative spectral decay across the \(O(1/\theta_N)\) first-peak window is
sufficiently strong.  Nearly defective or highly non-normal generators can instead
produce large transient growth and rapidly increasing projector norms; in that
case the Bauer--Fike or spectral-gap route may be too conservative, and the normal
form must be verified through model-specific resolvent, pseudospectral, or direct
dynamical estimates.

The background \(B_N\) in the observable normal form is constrained by the
probability range in physical CPTP realizations, but its precise value is irrelevant
to horizontal peak location: a constant background shifts the trajectory
vertically.  The leading oscillatory component must nevertheless remain
non-degenerate.  The condition used in the main text is
\[
    C_Ne^{-\gamma_Ns_N^{\mathrm{lead}}}\geq c_0
\]
for some \(c_0>0\).  In perturbative settings, \(C_N\) is controlled by the
observable overlap of the perturbed leading spectral projectors with the initial
state and success measurement.  Continuity of these projectors therefore keeps
\(C_N\) close to its ideal value whenever the relevant block remains sufficiently
well conditioned.

\subsection{Exact leading-peak calculus}
\label{app:exact-leading-peak-calculus}

For
\[
    F_N(k)
    =
    B_N-C_Ne^{-\gamma_Nk}\cos(k\omega_N+\chi_N),
\]
write \(\psi_N(k)=k\omega_N+\chi_N\).  Then
\[
    F_N'(k)
    =
    C_Ne^{-\gamma_Nk}
    \bigl(
        \gamma_N\cos\psi_N(k)
        +
        \omega_N\sin\psi_N(k)
    \bigr).
\]
The stationary branch near the ideal first peak satisfies
\[
    \psi_N(k)
    =
    \pi-\arctan\!\left(\frac{\gamma_N}{\omega_N}\right),
\]
which yields main-text Eq.~(13).  On this branch,
\[
    \cos\psi_N(s_N^{\mathrm{lead}})
    =
    -\frac{\omega_N}{\sqrt{\omega_N^2+\gamma_N^2}},
    \qquad
    \sin\psi_N(s_N^{\mathrm{lead}})
    =
    \frac{\gamma_N}{\sqrt{\omega_N^2+\gamma_N^2}}.
\]
Consequently,
\[
    F_N''(s_N^{\mathrm{lead}})
    =
    -C_Ne^{-\gamma_Ns_N^{\mathrm{lead}}}
    \omega_N\sqrt{\omega_N^2+\gamma_N^2}
    <0,
\]
and
\[
    \kappa_N
    =
    C_Ne^{-\gamma_Ns_N^{\mathrm{lead}}}
    \omega_N\sqrt{\omega_N^2+\gamma_N^2}.
\]
If \(\gamma_N=O(\omega_N)\), then
\[
    \kappa_N
    \asymp
    C_Ne^{-\gamma_Ns_N^{\mathrm{lead}}}\omega_N^2.
\]
Thus a non-collapsing first peak has
\(\kappa_N=\Theta(1/N)\) in single-target Grover scaling.

\subsection{Bauer--Fike perturbation bounds and perturbative realizability}
\label{app:bauer-fike}

This supplementary section gives a perturbative route into the spectral normal-form assumptions
used in the main text. The purpose is not to characterize all noisy CPTP channels,
but to show that a standard small-perturbation argument can produce an
observable-relevant leading oscillatory pair with controlled damping and frequency
shift.

\subsubsection{Perturbing the ideal oscillatory pair}

Let \(Q_N\) be a finite-dimensional matrix representation of the ideal
amplitude-amplification superoperator on the observable-relevant subspace, and let
\[
    \widetilde Q_N
    =
    Q_N+E_N
\]
be the corresponding noisy one-step iterate. The perturbation \(E_N\) represents
the one-step effect of the noise channel in the chosen representation.

For standard amplitude amplification, the probability-level oscillatory component
has the ideal eigenvalue pair
\[
    \mu_{\pm,N}
    =
    e^{\pm i4\theta_N}.
\]
Assume that, on the relevant block, \(Q_N\) is diagonalizable:
\[
    Q_N=V_ND_NV_N^{-1}.
\]
Bauer--Fike gives, for every eigenvalue \(\lambda\) of \(\widetilde Q_N\),
\begin{equation}
    \min_{\mu\in\sigma(Q_N)}
    |\lambda-\mu|
    \leq
    \kappa(V_N)\|E_N\|,
    \qquad
    \kappa(V_N)=\|V_N\|\|V_N^{-1}\|.
    \label{eq:app-bauer-fike}
\end{equation}

Let
\[
    \eta_N
    =
    \operatorname{dist}\!\left(
        \{\mu_{+,N},\mu_{-,N}\},
        \sigma(Q_N)\setminus\{\mu_{+,N},\mu_{-,N}\}
    \right)
\]
be the ideal spectral isolation of the oscillatory pair inside the relevant
subspace. If
\[
    \kappa(V_N)\|E_N\|
    <
    \frac{\eta_N}{2},
\]
then the noisy spectrum contains two eigenvalues
\(\lambda_{\pm,N}\) lying inside disjoint neighbourhoods of
\(\mu_{\pm,N}\). These eigenvalues are naturally written as
\[
    \lambda_{\pm,N}
    =
    \rho_N e^{\pm i\omega_N},
    \qquad
    \rho_N=e^{-\gamma_N}.
\]
This is the leading oscillatory pair used in
main-text Eq.~(8).

\subsubsection{Bounds on damping and frequency shift}

Set
\[
    \varepsilon_N
    =
    \kappa(V_N)\|E_N\|.
\]
Bauer--Fike implies
\[
    |\lambda_{+,N}-e^{i4\theta_N}|
    \leq
    \varepsilon_N,
    \qquad
    |\lambda_{-,N}-e^{-i4\theta_N}|
    \leq
    \varepsilon_N.
\]
For \(\varepsilon_N<1/2\), the modulus of the perturbed eigenvalue satisfies
\[
    1-\varepsilon_N
    \leq
    \rho_N
    \leq
    1+\varepsilon_N.
\]
In the CPTP setting the spectral radius is at most one, so typically
\(\rho_N\leq 1\). The damping scale
\[
    \gamma_N=-\log\rho_N
\]
is therefore small whenever \(\varepsilon_N\) is small. In particular, for
\(\rho_N\in[1-\varepsilon_N,1]\),
\[
    0\leq \gamma_N
    \leq
    -\log(1-\varepsilon_N)
    =
    O(\varepsilon_N).
\]

The frequency shift is also controlled. Since
\(\lambda_{+,N}\) lies within \(\varepsilon_N\) of \(e^{i4\theta_N}\), its argument
satisfies
\[
    |\omega_N-4\theta_N|
    =
    O(\varepsilon_N)
\]
as long as the perturbation neighbourhood does not cross the origin or another
branch cut. With the convention
\[
    \omega_N=4(\theta_N+\delta_N),
\]
this gives
\[
    |\delta_N|
    =
    O(\varepsilon_N).
\]
Thus a small one-step perturbation produces small normal-form damping and
frequency-shift parameters.

\subsubsection{Observable overlap and phase offset}

Eigenvalue proximity alone is not enough for the normal form. The perturbed
oscillatory pair must also be visible in the measured success probability. Let
\(\Pi_{+,N}\) be the spectral projector associated with
\(\lambda_{+,N}\), and define the observable coefficient
\[
    A_N
    =
    \langle\langle M_N|
    \Pi_{+,N}
    |\rho_{N,0}\rangle\rangle .
\]
If the ideal oscillatory coefficient is nonzero and the perturbation is small
enough for the corresponding spectral projector to vary continuously, then
\(A_N\neq 0\) for all sufficiently small perturbations. Writing
\[
    A_N=-\frac{C_N}{2}e^{i\chi_N},
    \qquad
    C_N>0,
\]
produces the observable amplitude \(C_N\) and phase \(\chi_N\) in the scalar normal
form.

The phase offset used in the main text is
\[
    \phi_N
    =
    \chi_N-2\theta_N.
\]
Under the same perturbative continuity assumptions, \(\phi_N\) remains small when
the one-step perturbation is small. The exact bound depends on the conditioning of
the spectral projector and the observable overlap, which is why the main theory
treats \(\phi_N\) as a normal-form parameter rather than deriving it from a single
universal microscopic noise model.

\subsubsection{Persistence of spectral separation}

We next record how residual separation can persist under perturbation. Let
\(\beta_N^{(0)}\) denote the largest modulus among reference residual
branches that may have nonzero observable overlap after perturbation, excluding
the leading pair and any stationary background modes that remain decoupled. Suppose that
the ideal oscillatory pair is isolated not only in angle but also in observable
dominance, so that
\[
    1-\beta_N^{(0)}
    >
    0.
\]
Assume that the isolated leading and residual spectral branches can be
tracked under the perturbation and that stationary background branches stay
separated from the residual observable block. Under the Bauer--Fike bound,
each such residual eigenvalue of the noisy iterate remains within
\(\varepsilon_N\) of a non-leading reference eigenvalue. Hence, at the
level of eigenvalue moduli,
\[
    \beta_N
    \leq
    \beta_N^{(0)}+\varepsilon_N,
    \qquad
    \rho_N
    \geq
    1-\varepsilon_N .
\]
A sufficient condition for noisy spectral separation is therefore
\[
    1-\varepsilon_N
    >
    \beta_N^{(0)}+\varepsilon_N,
\]
or
\[
    2\varepsilon_N
    <
    1-\beta_N^{(0)}.
\]
Under this condition,
\[
    g_N
    =
    \rho_N-\beta_N
    >
    0.
\]
This gives a perturbative route to the spectral separation assumption
used in main-text Eq.~(10) for the stated residual block. If a formerly
unobserved peripheral mode acquires nonzero observable overlap, its modulus
must also be checked; Bauer--Fike proximity alone does not establish
observable dominance in that situation.

In non-normal settings, the eigenvalue separation must be supplemented by
projector-growth control. This is the role of the constant \(K_N\) in
main-text Eq.~(11). Bauer--Fike controls eigenvalue locations, while
\(K_N\) controls how strongly the residual eigenmodes appear in the observed
success probability.

\subsubsection{Sufficient route into the first-peak-preserving regime}

Combining the previous observations gives the following sufficient perturbative
scenario. Suppose that the reference iterate has an observable oscillatory pair
\(\mu_{\pm,N}=e^{\pm i4\theta_N}\), that this pair is spectrally isolated on the
observable-relevant block, and that its observable coefficient is nonzero. If
\[
    \varepsilon_N=\kappa(V_N)\|E_N\|
\]
is small compared with the relevant spectral separations, then the noisy iterate
has a perturbed observable pair
\[
    \lambda_{\pm,N}
    =
    \rho_N e^{\pm i\omega_N}
\]
with
\[
    \gamma_N=O(\varepsilon_N),
    \qquad
    |\delta_N|=O(\varepsilon_N),
\]
and with a well-defined observable phase offset \(\phi_N\). If, in addition, the
projector-growth factor \(K_N\) is controlled and the residual spectral ratio
\(\beta_N/\rho_N\) is sufficiently below one on the \(O(1/\theta_N)\) first-peak
window, and any newly observable non-leading modes satisfy the same residual
separation and projector controls, then the residual bound
\[
    R_N^{\mathrm{val}}
    \leq
    K_N
    \exp\!\left(
        -\frac{c_*g_N}{\rho_N\theta_N}
    \right)
\]
places the noisy trajectory inside the spectral normal-form regime of the main
text.

Together, these conditions provide a sufficient perturbative route into
the observable normal-form regime. The stopping-stability theorems
themselves require only the leading observable pair, spectral separation,
and residual control. For strongly non-normal blocks, these assumptions
may instead be verified through model-specific projector, resolvent, or
direct dynamical estimates.

\subsection{Encoded-subspace depolarizing specialization}
\label{app:encoded-depolarizing-specialization}

We now derive the encoded-subspace depolarizing realization directly on
the single-target Grover subspace
\[
    \mathcal H_G
    =
    \operatorname{span}\{\lvert w\rangle,\lvert r\rangle\}.
\]
In the ordered basis \(\{\lvert w\rangle,\lvert r\rangle\}\), the ideal Grover
iterate is
\[
    G_N
    =
    \begin{pmatrix}
        \cos(2\theta_N) & \sin(2\theta_N)\\
        -\sin(2\theta_N) & \cos(2\theta_N)
    \end{pmatrix},
\]
and the initial state is
\[
    \lvert s\rangle
    =
    \sin\theta_N\lvert w\rangle
    +
    \cos\theta_N\lvert r\rangle .
\]
Let \(\alpha_N=1-p_N\). The encoded depolarizing channel acts as
\[
    \mathcal D_{p_N}^{(G)}(\rho)
    =
    \alpha_N\rho+(1-\alpha_N)\frac{I_G}{2}.
\]
Writing
\[
    \rho
    =
    \frac12\left(I+x\sigma_x+y\sigma_y+z\sigma_z\right),
    \qquad
    \sigma_z
    =
    \lvert w\rangle\langle w\rvert
    -
    \lvert r\rangle\langle r\rvert,
\]
one noisy iterate maps the Bloch vector according to
\[
    \begin{pmatrix}x\\y\\z\end{pmatrix}
    \longmapsto
    \alpha_N
    \begin{pmatrix}
        \cos4\theta_N & 0 & -\sin4\theta_N\\
        0 & 1 & 0\\
        \sin4\theta_N & 0 & \cos4\theta_N
    \end{pmatrix}
    \begin{pmatrix}x\\y\\z\end{pmatrix}.
\]
Thus, in the normalized Pauli basis
\(\{I,\sigma_x,\sigma_y,\sigma_z\}\), the superoperator is represented by
\[
    \widetilde Q_N
    =
    \begin{pmatrix}
        1 & 0 & 0 & 0\\
        0 & \alpha_N\cos4\theta_N & 0 & -\alpha_N\sin4\theta_N\\
        0 & 0 & \alpha_N & 0\\
        0 & \alpha_N\sin4\theta_N & 0 & \alpha_N\cos4\theta_N
    \end{pmatrix}.
\]
For \(N>2\), the oscillatory eigenvalues are distinct and
\[
    \sigma(\widetilde Q_N)
    =
    \left\{
        1,
        \alpha_N,
        \alpha_Ne^{+i4\theta_N},
        \alpha_Ne^{-i4\theta_N}
    \right\}.
\]
The observable-relevant conjugate pair therefore has
\[
    \rho_N=\alpha_N,
    \qquad
    \omega_N=4\theta_N,
\]
which yields
\[
    \gamma_N=-\log(1-p_N),
    \qquad
    \delta_N=0.
\]

The initial Bloch vector is
\[
    (x_0,y_0,z_0)
    =
    (\sin2\theta_N,0,-\cos2\theta_N),
\]
while the success measurement is
\[
    M_N
    =
    \lvert w\rangle\langle w\rvert
    =
    \frac12(I+\sigma_z).
\]
The \(\sigma_y\) mode is dynamically decoupled from the \(x\)--\(z\) block,
starts with zero amplitude, and has zero success-observable overlap. Consequently,
it makes no contribution to the scalar success trajectory. Iterating the
\(x\)--\(z\) block gives
\[
    z_k
    =
    -\alpha_N^k
    \cos\!\bigl((4k+2)\theta_N\bigr),
\]
and hence
\begin{equation}
    P_N(k)
    =
    \frac12
    -
    \frac12(1-p_N)^k
    \cos\!\bigl((4k+2)\theta_N\bigr).
    \label{eq:app-encoded-depolarizing-trajectory}
\end{equation}
Comparison with main-text Eq.~(12) gives
\[
    B_N=C_N=\frac12,
    \qquad
    \chi_N=2\theta_N,
    \qquad
    \phi_N=0,
    \qquad
    r_N(k)\equiv0.
\]
Thus \(R_N^{\mathrm{val}}=0\). The matrix \(\widetilde Q_N\) is normal, so its
spectral eigenvectors can be chosen orthonormal and the associated spectral
conditioning is unity in the Euclidean norm. More specifically, because every
non-leading observable coefficient vanishes, the observable residual set is empty
and the tight residual prefactor can be taken as
\[
    K_N=0.
\]
Here \(\mathcal J_N=\varnothing\), so the convention in S1.4 gives
\(\beta_N=0\); the stationary eigenvalue \(1\) contributes only to
\(B_N\), and the dynamically decoupled mode with eigenvalue
\(\alpha_N\) has zero observable overlap.
No residual spectral-separation estimate is required for this exact specialization;
this statement concerns the observable residual and does not assert that the full
superoperator has no additional eigenmodes.

The exact continuous leading peak follows from
main-text Eq.~(13):
\begin{equation}
    \Delta s_N^{\mathrm{lead}}
    =
    -\frac{1}{4\theta_N}
    \arctan\!\left(
        \frac{-\log(1-p_N)}{4\theta_N}
    \right).
    \label{eq:app-encoded-depolarizing-displacement}
\end{equation}
The continuous peak leaves the ideal cell on the early side when
\[
    \Delta s_N^{\mathrm{lead}}=-m_N^{-},
\]
which gives the geometric threshold
\begin{equation}
    p_{N,\mathrm{cont}}
    =
    1-
    \exp\!\left[
        -4\theta_N\tan(4\theta_Nm_N^{-})
    \right].
    \label{eq:encoded-continuous-threshold}
\end{equation}
This threshold is not, in general, the exact sampled transition because the damped
cosine is asymmetric. The exact early sampled equality is instead
\[
    F_N(k_N^0)=F_N(k_N^0-1),
\]
which, using Eq.~\eqref{eq:app-encoded-depolarizing-trajectory}, gives
\begin{equation}
    p_{N,\mathrm{samp}}
    =
    1-
    \frac{
        \cos\!\bigl((4k_N^0-2)\theta_N\bigr)
    }{
        \cos\!\bigl((4k_N^0+2)\theta_N\bigr)
    },
    \label{eq:encoded-sampled-threshold}
\end{equation}
whenever the ratio lies in \([0,1]\) on the relevant first-lobe branch. For
\(N=64\), these two values are
\[
    p_{N,\mathrm{cont}}\approx0.0652164,
    \qquad
    p_{N,\mathrm{samp}}\approx0.0665952.
\]
Thus the continuous rounding-cell crossing is a geometric diagnostic, whereas the
sampled equality is the exact finite-\(N\) discrete transition. Both have the same
leading single-target scaling. In the weak-noise limit,
\[
    p_{N,\mathrm{cont}}
    =
    16\theta_N^2m_N^{-}+O(\theta_N^4)
    =
    O\!\left(\frac{m_N^{-}}{N}\right),
\]
which recovers the damping sensitivity scale used in the main text.

\section{Displacement Formula and Sampled-Stability Derivations}
\label{app:peak-displacement}

This supplementary section supplies the details behind the main-text peak-location
and sampled-stability results. Sections~\ref{subsec:app-exact-displacement-identity}--
\ref{subsec:app-remainder-estimate} derive the first-order
peak-displacement formula used in main-text Eq.~(14),
starting from the exact leading first-peak location and controlling the higher-order
remainder. Sections~\ref{subsec:app-exact-rounding-geometry}--
\ref{subsec:app-expanded-directional-margins}
then connect the continuous displacement geometry to the sampled-gap conditions
used in main-text Section~4.3
and derive the speedup--stability scales used in main-text
Section~4.4.

\subsection{Exact displacement identity}
\label{subsec:app-exact-displacement-identity}

The leading continuous first peak is
\[
    s_N^{\mathrm{lead}}
    =
    \frac{
        \pi
        -
        \chi_N
        -
        \arctan(\gamma_N/\omega_N)
    }{
        \omega_N
    }.
\]
Using
\[
    \omega_N=4(\theta_N+\delta_N),
    \qquad
    \chi_N=2\theta_N+\phi_N,
\]
this becomes
\[
    s_N^{\mathrm{lead}}
    =
    \frac{
        \pi
        -
        2\theta_N
        -
        \phi_N
        -
        \arctan\!\left(\frac{\gamma_N}{4(\theta_N+\delta_N)}\right)
    }{
        4(\theta_N+\delta_N)
    }.
\]
Let
\[
    \ell_N=\pi-2\theta_N.
\]
Then
\[
    s_N^0=\frac{\ell_N}{4\theta_N},
\]
and the exact displacement can be written as
\[
    \Delta s_N^{\mathrm{lead}}
    =
    \frac{
        \ell_N
        -
        \phi_N
        -
        \arctan\!\left(\frac{\gamma_N}{4(\theta_N+\delta_N)}\right)
    }{
        4(\theta_N+\delta_N)
    }
    -
    \frac{\ell_N}{4\theta_N}.
\]

\subsection{First-order expansion}
\label{subsec:app-first-order-expansion}

We expand around
\[
    \delta_N=0,
    \qquad
    \phi_N=0,
    \qquad
    \gamma_N=0.
\]
For the frequency-shift contribution,
\[
    \frac{\ell_N}{4(\theta_N+\delta_N)}
    -
    \frac{\ell_N}{4\theta_N}
    =
    -\frac{\ell_N}{4\theta_N^2}\delta_N
    +
    O\!\left(\frac{\delta_N^2}{\theta_N^3}\right).
\]
Since \(s_N^0=\ell_N/(4\theta_N)\), the first-order term is
\[
    -\frac{s_N^0\delta_N}{\theta_N}.
\]

For the observable phase offset,
\[
    -\frac{\phi_N}{4(\theta_N+\delta_N)}
    =
    -\frac{\phi_N}{4\theta_N}
    +
    O\!\left(\frac{|\phi_N||\delta_N|}{\theta_N^2}\right).
\]
Thus the first-order phase contribution is
\[
    -\frac{\phi_N}{4\theta_N}.
\]

For the damping contribution, use
\[
    \arctan x = x+O(x^3)
\]
with
\[
    x=
    \frac{\gamma_N}{4(\theta_N+\delta_N)}.
\]
Then
\[
    -\frac{1}{4(\theta_N+\delta_N)}
    \arctan\!\left(\frac{\gamma_N}{4(\theta_N+\delta_N)}\right)
    =
    -\frac{\gamma_N}{16\theta_N^2}
    +
    O\!\left(
        \frac{\gamma_N|\delta_N|}{\theta_N^3}
        +
        \frac{\gamma_N^3}{\theta_N^4}
    \right).
\]
Hence the first-order damping contribution is
\[
    -\frac{\gamma_N}{16\theta_N^2}.
\]

Combining the three contributions gives
\begin{equation}
    \Delta s_N^{\mathrm{lead}}
    =
    -
    \frac{s_N^0\delta_N}{\theta_N}
    -
    \frac{\phi_N}{4\theta_N}
    -
    \frac{\gamma_N}{16\theta_N^2}
    +
    \mathcal R_N^{\mathrm{disp}} .
    \label{eq:app-main-displacement-expansion}
\end{equation}

\subsection{Remainder bound for main-text Eq.~(14)}
\label{subsec:app-remainder-estimate}

Introduce
\[
    u_N=\frac{\delta_N}{\theta_N},
    \qquad
    \varepsilon_N^{\mathrm{loc}}
    =
    |u_N|+|\phi_N|+\frac{\gamma_N}{\theta_N},
\]
and assume \(\varepsilon_N^{\mathrm{loc}}\leq\varepsilon_0\) with
\(0<\varepsilon_0<1\). Then \(1+u_N\geq1-\varepsilon_0>0\),
so the denominator is bounded away from zero uniformly in \(N\) for each
fixed \(\varepsilon_0\). Using the exact identity in
Subsection~\ref{subsec:app-exact-displacement-identity}, write
\[
\Delta s_N^{\mathrm{lead}}
=
-\frac{s_N^0u_N}{1+u_N}
-\frac{\phi_N}{4\theta_N(1+u_N)}
-\frac{1}{4\theta_N(1+u_N)}
 \arctan\!\left(
 \frac{\gamma_N}{4\theta_N(1+u_N)}
 \right).
\]
The frequency term satisfies the exact decomposition
\[
-\frac{s_N^0u_N}{1+u_N}
=
-s_N^0u_N
+
s_N^0\frac{u_N^2}{1+u_N},
\]
so its remainder is bounded by
\(C_1\delta_N^2/\theta_N^3\), because \(s_N^0=O(\theta_N^{-1})\).
Similarly,
\[
-\frac{\phi_N}{4\theta_N(1+u_N)}
=
-\frac{\phi_N}{4\theta_N}
+
\frac{\phi_Nu_N}{4\theta_N(1+u_N)},
\]
which contributes at most
\(C_2|\phi_N||\delta_N|/\theta_N^2\).

For the damping term set
\[
y_N=\frac{\gamma_N}{4\theta_N(1+u_N)}.
\]
On the local neighbourhood, \(|y_N|\leq C\varepsilon_0\), and Taylor's theorem
for \(\arctan y\) gives
\[
\arctan y_N=y_N+\eta_N,
\qquad
|\eta_N|\leq C_3|y_N|^3.
\]
Therefore
\[
-\frac{\arctan y_N}{4\theta_N(1+u_N)}
=
-\frac{\gamma_N}{16\theta_N^2}
+
O\!\left(
\frac{\gamma_N|\delta_N|}{\theta_N^3}
+
\frac{\gamma_N^3}{\theta_N^4}
\right),
\]
where the first correction comes from
\((1+u_N)^{-2}-1=O(u_N)\) and the second from \(\eta_N\).

Collecting the three exact decompositions yields
\[
    \left|\mathcal R_N^{\mathrm{disp}}\right|
    \leq
    C_{\mathrm{disp}}(\varepsilon_0)
    \left(
        \frac{\delta_N^2}{\theta_N^3}
        +
        \frac{|\phi_N||\delta_N|}{\theta_N^2}
        +
        \frac{\gamma_N|\delta_N|}{\theta_N^3}
        +
        \frac{\gamma_N^3}{\theta_N^4}
    \right).
\]
Because each dimensionless perturbation entering
\(\varepsilon_N^{\mathrm{loc}}\) is at most \(\varepsilon_0\), this also implies
the compact envelope
\[
    \left|\mathcal R_N^{\mathrm{disp}}\right|
    \leq
    C'_{\mathrm{disp}}(\varepsilon_0)
    \frac{\left(\varepsilon_N^{\mathrm{loc}}\right)^2}{\theta_N},
\]
with constants independent of \(N\) for fixed \(\varepsilon_0<1\).
They may grow as \(\varepsilon_0\) approaches 1. In the
\(N=64\) shifted device-like example, the fitted parameters give
\(\varepsilon_N^{\mathrm{loc}}\approx0.523\), which is covered by
\(\varepsilon_0=3/4\). The reported continuous peak uses the exact
stationary-point formula; the first-order decomposition is accompanied by its
computed remainder and is not alone a sampled-step certificate.

\subsection{Connection to the single-mechanism numerical limits}
\label{subsec:app-single-mechanism-limits}

The single-mechanism limits used as reference normalizations in the numerical
validation are direct specializations of the first-order displacement formula in
this supplementary section. We therefore record them only as reference limits, rather than
repeating the full derivation.

Under single-target Grover scaling,
\[
    \theta_N\sim N^{-1/2},
    \qquad
    s_N^0\sim \frac{\pi}{4}\sqrt N .
\]
Thus the three one-parameter scalings used in the large-\(N\) numerical check give
\[
\begin{array}{rcl}
    \delta_N=c/N,\quad \gamma_N=0,\quad \phi_N=0
    &\Longrightarrow&
    \Delta s_N^{(\delta)}
    \longrightarrow
    -\dfrac{\pi c}{4},
    \\[1.0em]
    \gamma_N=c/N,\quad \delta_N=0,\quad \phi_N=0
    &\Longrightarrow&
    \Delta s_N^{(\gamma)}
    \longrightarrow
    -\dfrac{c}{16},
    \\[1.0em]
    \phi_N=c/\sqrt N,\quad \delta_N=0,\quad \gamma_N=0
    &\Longrightarrow&
    \Delta s_N^{(\phi)}
    \longrightarrow
    -\dfrac{c}{4}.
\end{array}
\]
These limits are used only as reference slopes for the single-mechanism scaling
plots. The numerical section tests convergence to these values directly, while the
technical content of this supplementary section is the perturbative expansion and the uniform
remainder envelope for \(\mathcal R_N^{\mathrm{disp}}\).

\subsection{Exact continuous rounding-cell geometry}
\label{subsec:app-exact-rounding-geometry}

Writing \(s_N^0=k_N^0+d_N\), with \(|d_N|\leq1/2\), a displaced continuous peak
\(\widetilde s_N=s_N^0+\Delta s_N\) satisfies
\[
    \widetilde s_N
    =
    k_N^0+d_N+\Delta s_N.
\]
Hence \(|\Delta s_N|<1/2-|d_N|=m_N\) implies
\(|d_N+\Delta s_N|<1/2\), so \(\widetilde s_N\) remains in the same rounding
cell. The exact directional condition for this nearest-integer map is
\[
    -m_N^-<\Delta s_N<m_N^+.
\]
At equality the continuous peak lies on a rounding boundary. This statement is
exact for continuous-cell membership only. For an asymmetric leading curve it does
not, by itself, determine the discrete sampled maximizer; the latter is fixed by
sampled-value comparisons.

\subsection{Local sampled-gap regularity}
\label{app:sampled-stopping-proofs}

This subsection supplies a sufficient route from continuous peak geometry to a
positive discrete sampled gap. It imposes an additional local-skew condition beyond
the exact leading-curve sampled criterion \(\Delta_N^F>0\): continuous rounding-cell
membership alone is insufficient for an asymmetric first lobe.

For
\[
    F_N(k)
    =
    B_N-C_Ne^{-\gamma_Nk}
    \cos\!\bigl(k\omega_N+\chi_N\bigr),
\]
write \(\psi_N(k)=k\omega_N+\chi_N\). Its derivative is
\[
    F_N'(k)
    =
    C_Ne^{-\gamma_Nk}
    \left[
        \gamma_N\cos\psi_N(k)
        +
        \omega_N\sin\psi_N(k)
    \right].
\]
The stationary points satisfy
\[
    \tan\psi_N(k)=-\frac{\gamma_N}{\omega_N},
\]
and the branch
\[
    \psi_N(k)
    =
    \pi-\arctan\!\left(\frac{\gamma_N}{\omega_N}\right)
\]
is the strict local maximum \(s_N^{\mathrm{lead}}\). Choosing
\(\mathcal W_N\) inside the adjacent stationary minima therefore gives a strict
single-lobed first-peak window.

For the sampled comparison, the chosen first-lobe interval must contain the
neighbouring integer depths on either side of the leading peak; otherwise a
boundary candidate is not covered by the local sampled-gap argument.

Now suppose that the continuous peak remains in the ideal rounding cell, so
\[
    k_N^0=\operatorname{round}(s_N^{\mathrm{lead}}).
\]
Set
\[
    s=s_N^{\mathrm{lead}},
    \qquad
    d=s-k_N^0,
    \qquad
    |d|\leq\frac12,
\]
and define
\[
    \mu_N(s)
    =
    \frac12-|d|.
\]
Because the curve is single-lobed, it suffices to compare the two immediate
neighbours \(k_N^0-1\) and \(k_N^0+1\): all more distant sampled points on
either side of \(s\) have no larger leading value. Both neighbours must be in
the candidate window for this argument.

Assume the local expansion
\[
    F_N(k)
    =
    F_N(s)
    -
    \frac{\kappa_N}{2}(k-s)^2
    +
    E_N(k),
    \qquad
    \kappa_N=-F_N''(s)>0,
\]
for the two neighbouring comparisons, and suppose
\[
    |F_N^{(3)}(x)|\leq L_N
\]
throughout \(|x-s|\leq 3/2\), which contains \(k_N^0\) and both
neighbours. Taylor's theorem gives
\[
    |E_N(k)|\leq\frac{L_N}{6}|k-s|^3.
\]
For either neighbouring competitor \(k=k_N^0\pm1\), the quadratic gap is
\[
    \frac{\kappa_N}{2}\left[(k-s)^2-(k_N^0-s)^2\right]
    =\kappa_N\left(\frac12\mp d\right)
    \geq\kappa_N\mu_N(s).
\]
Since \(|k-s|\leq3/2\) and \(|k_N^0-s|\leq1/2\), the difference
of the Taylor remainders is at most
\(L_N[(3/2)^3+(1/2)^3]/6=7L_N/12\). Thus both neighbours satisfy
\[
    F_N(k_N^0)-F_N(k_N^0\pm1)
    \geq
    \kappa_N\mu_N(s)-\frac{7L_N}{12}.
\]
Single-lobedness extends this comparison to all other sampled competitors.
Therefore, if
\[
    \frac{7L_N}{12}
    \leq
    (1-c_{\mathrm{curv}})
    \kappa_N\mu_N(s)
\]
for some \(c_{\mathrm{curv}}\in(0,1]\), then
\begin{equation}
    \Delta_N^F
    \geq
    c_{\mathrm{curv}}\kappa_N\mu_N(s).
    \label{eq:app-local-sampled-gap-regularity}
\end{equation}
This is the local sampled-gap regularity assumption used to convert the geometric
rounding margin into the practical sufficient certificate of main-text Section~4.3. It is an
additional local-skew condition, not a consequence of rounding alone.

For the ideal single-target Grover curve,
\[
    P_N^{(0)}(k)
    =
    \cos^2\!\bigl(2\theta_N(k-s_N^0)\bigr)
    =
    1-4\theta_N^2(k-s_N^0)^2
    +O\!\left(\theta_N^4|k-s_N^0|^4\right),
\]
so
\[
    \kappa_N
    =
    8\theta_N^2+O(\theta_N^4)
\]
and, away from a sampled tie,
\[
    \Delta_N^F
    =
    8\theta_N^2\mu_N(s_N^0)+O(\theta_N^4).
\]
Thus \(c_{\mathrm{curv}}\) can be chosen close to one for the ideal curve at large
\(N\). For a noisy damped cosine, the estimate remains useful when the third-
derivative skew contribution is small compared with
\(\kappa_N\mu_N(s_N^{\mathrm{lead}})\). Near a sampled equality, a strongly
asymmetric lobe, or a flattened peak, \(\Delta_N^F\) should instead be evaluated
directly.

\subsection{Derivation of the speedup--stability scales in main-text Eq.~(26)}
\label{subsec:app-duality-derivation}

From the first-order displacement
\[
\Delta s_N^{\mathrm{main}}
=
-\frac{s_N^0\delta_N}{\theta_N}
-\frac{\phi_N}{4\theta_N}
-\frac{\gamma_N}{16\theta_N^2},
\]
direct differentiation gives
\[
\left|\partial_{\delta_N}\Delta s_N^{\mathrm{main}}\right|
=
\frac{s_N^0}{\theta_N},
\qquad
\left|\partial_{\gamma_N}\Delta s_N^{\mathrm{main}}\right|
=
\frac{1}{16\theta_N^2},
\qquad
\left|\partial_{\phi_N}\Delta s_N^{\mathrm{main}}\right|
=
\frac{1}{4\theta_N}.
\]
With \(S_N=\theta_N^{-1}\) and \(s_N^0=\Theta(S_N)\), these become
\(\Theta(S_N^2)\), \(\Theta(S_N^2)\), and \(\Theta(S_N)\), respectively.
For single-target Grover search \(S_N=\Theta(\sqrt N)\), giving the
\(N,N,\sqrt N\) sensitivities quoted in the main text.

The peak-survival scale follows from the leading envelope
\(C_Ne^{-\gamma_N s_N^{\mathrm{lead}}}\). If
\(s_N^{\mathrm{lead}}=\Theta(S_N)\) and the peak amplitude is required to remain
bounded below by a nonzero constant, then
\(\gamma_NS_N=O(1)\), or \(\gamma_N=O(S_N^{-1})\).
By contrast, keeping the continuous displacement within an \(O(m_N)\)
rounding margin requires the damping contribution
\(\gamma_NS_N^2=O(m_N)\), hence
\(\gamma_N=O(m_N/S_N^2)\). The intermediate interval between these two scales
is the damping realization of the speedup--stability separation.

The remaining no-correction scales follow from the same displacement formula.
Since \(s_N^0=\Theta(S_N)\), requiring each first-order mechanism to consume at
most an \(O(m_N)\) continuous rounding margin gives
\[
    |\delta_N|S_N^2=O(m_N),
    \qquad
    \gamma_NS_N^2=O(m_N),
    \qquad
    |\phi_N|S_N=O(m_N).
\]
Equivalently,
\[
    |\delta_N|=O\!\left(\frac{m_N}{S_N^2}\right),
    \qquad
    \gamma_N=O\!\left(\frac{m_N}{S_N^2}\right),
    \qquad
    |\phi_N|=O\!\left(\frac{m_N}{S_N}\right).
\]
For single-target Grover search, \(S_N=\Theta(\sqrt N)\), so these become
\[
    |\delta_N|=O\!\left(\frac{m_N}{N}\right),
    \qquad
    \gamma_N=O\!\left(\frac{m_N}{N}\right),
    \qquad
    |\phi_N|=O\!\left(\frac{m_N}{\sqrt N}\right),
\]
as quoted in the main text.

\subsection{Expanded directional margins}
\label{subsec:app-expanded-directional-margins}

Main-text Subsection~4.3.2 gives the direction-aware certificate and
defines the residual margin loss
\(E_N^{\mathrm{res}}=2R_N^{\mathrm{val}}/(c_{\mathrm{curv}}\kappa_N)\).
Here we record only the useful mechanism-specific specializations. For a
damping-only perturbation,
\[
    \Delta s_N^{(\gamma)}
    =
    -\frac{\gamma_N}{16\theta_N^2}\leq0,
\]
so the relevant geometric boundary is the left margin and a sufficient condition
is
\[
    \frac{\gamma_N}{16\theta_N^2}
    +|\mathcal R_N^{\mathrm{disp}}|
    +E_N^{\mathrm{res}}
    <m_N^-.
\]
Effective-angle and observable-phase perturbations instead have signed
contributions
\[
    \Delta s_N^{(\delta)}
    =-\frac{s_N^0\delta_N}{\theta_N},
    \qquad
    \Delta s_N^{(\phi)}
    =-\frac{\phi_N}{4\theta_N}.
\]
Hence a positive \(\delta_N\) shifts the peak earlier and consumes the left margin,
whereas a negative \(\delta_N\) shifts it later and consumes the right margin. A
sufficient direction-aware condition is therefore
\[
    \frac{s_N^0|\delta_N|}{\theta_N}
    +|\mathcal R_N^{\mathrm{disp}}|
    +E_N^{\mathrm{res}}
    <
    \begin{cases}
        m_N^-, & \delta_N>0,\\
        m_N^+, & \delta_N<0.
    \end{cases}
\]
Likewise, a positive observable phase offset shifts the peak earlier and a negative
offset shifts it later, so a sufficient condition is
\[
    \frac{|\phi_N|}{4\theta_N}
    +|\mathcal R_N^{\mathrm{disp}}|
    +E_N^{\mathrm{res}}
    <
    \begin{cases}
        m_N^-, & \phi_N>0,\\
        m_N^+, & \phi_N<0.
    \end{cases}
\]
These are the effective-angle and phase specializations of the directional
certificate in main-text Eq.~(25). The sign dependence is the basis of the
early/late comparisons in main-text Subsection~5.1 and Supplementary
Section~\ref{app:additional-figures}.

\subsection{Assumption-use map for the stopping formulas}
\label{subsec:app-assumption-use-map}

The derivations above use the main assumptions at distinct points. The local
perturbation condition
\(
|\delta_N|/\theta_N+|\phi_N|+\gamma_N/\theta_N\leq\varepsilon_0
\)
is used only to make the Taylor expansion leading to
main-text Eq.~(14) uniform and to control
\(\mathcal R_N^{\mathrm{disp}}\). The non-degenerate visible-lobe condition
\(C_Ne^{-\gamma_N s_N^{\mathrm{lead}}}\geq c_0\) is what prevents the local
curvature from collapsing. The single-lobed first-peak window is used when the
continuous peak is reduced to neighbouring integer candidates. The additional
sampled-gap regularity condition is required only for the curvature-based
sufficient certificate; it does not follow from continuous rounding-cell
membership. Finally, the value residual \(R_N^{\mathrm{val}}\) enters only when
leading-curve ordering is transferred to the full sampled trajectory. Keeping
these roles separate is essential: the displacement assumptions locate the leading
lobe, while the residual and sampled-gap assumptions certify the integer decision.

\section{Recalibrated and Robust-Window Stopping}
\label{app:stopping-rule-correction-proofs}

The main text uses one abstract sampled-gap preservation lemma for fixed-step,
recalibrated, and robust-window decisions. Its gap-versus-remainder ordering
argument is inherited from Ref.~\cite{zhong2026responseThreshold} and is not
re-derived in this supplement. The correction-specific material begins after that
ordering step: a displaced continuous peak must be converted into a finite integer
candidate set, and uncertainty in that displacement must be converted into a
window with controlled overhead. This section gives those new
correction-specific containment and overhead arguments in full.

\subsection{Symmetric robust-window containment}
\label{subsec:app-symmetric-window-containment}

Suppose
\[
    s_N^{\mathrm{lead}}
    =
    s_N^0+\Delta s_N^{\mathrm{lead}},
    \qquad
    |\Delta s_N^{\mathrm{lead}}|
    \leq
    \widehat\Delta_N .
\]
For any admissible continuous peak location, the sampled maximizer of a strict
single-lobed leading curve lies among
\[
    \left\{
        \left\lfloor s_N^{\mathrm{lead}}\right\rfloor,
        \left\lceil s_N^{\mathrm{lead}}\right\rceil
    \right\}.
\]
Define
\[
    \ell_N
    =
    \left\lfloor s_N^0-\widehat\Delta_N\right\rfloor,
    \qquad
    u_N
    =
    \left\lceil s_N^0+\widehat\Delta_N\right\rceil .
\]
Then every admissible floor--ceiling candidate pair is contained in the integer
interval \([\ell_N,u_N]\). Therefore the symmetric window
\[
    \mathcal I_N(W_N)
    =
    \{k\in\mathbb Z:\ |k-k_N^0|\leq W_N\}
\]
contains \(k_N^{\mathrm{rec}}\) whenever
\[
    W_N
    \geq
    \max\{
        k_N^0-\ell_N,
        u_N-k_N^0
    \}.
\]
This is the containment condition stated in
main-text Eq.~(29). It does not assume that the corrected sampled
step is obtained by rounding the continuous maximum.

Once containment holds, the robust-window guarantee follows from the same sampled
value-gap comparison used in the main text. If \(k_N^{\mathrm{rec}}\) is the unique
leading sampled maximizer inside \(\mathcal I_N(W_N)\cap\mathcal K_N\) and
\[
    2R_N^{\mathrm{val}}
    <
    \min_{\substack{
        k\in\mathcal I_N(W_N)\cap\mathcal K_N\\
        k\neq k_N^{\mathrm{rec}}
    }}
    \bigl[
        F_N(k_N^{\mathrm{rec}})-F_N(k)
    \bigr],
\]
then the full noisy windowed maximizer is \(k_N^{\mathrm{rec}}\).

\subsection{Direction-aware containment}
\label{subsec:app-direction-aware-containment}

Suppose the continuous displacement is bounded asymmetrically:
\[
    -\widehat\Delta_N^-
    \leq
    \Delta s_N^{\mathrm{lead}}
    \leq
    \widehat\Delta_N^+ .
\]
Define the extremal sampled candidates
\[
    \ell_N^-
    =
    \left\lfloor s_N^0-\widehat\Delta_N^-\right\rfloor,
    \qquad
    u_N^+
    =
    \left\lceil s_N^0+\widehat\Delta_N^+\right\rceil .
\]
The direction-aware window
\[
    \mathcal I_N(W_N^-,W_N^+)
    =
    \{k\in\mathbb Z:
        k_N^0-W_N^-
        \leq k
        \leq k_N^0+W_N^+\}
\]
contains the floor--ceiling pair for every admissible continuous peak whenever
\[
    W_N^-
    \geq
    k_N^0-\ell_N^-,
    \qquad
    W_N^+
    \geq
    u_N^+-k_N^0.
\]
Hence it contains \(k_N^{\mathrm{rec}}\). This construction can be substantially
asymmetric, but a strictly one-sided window is justified only if the extremal
candidate interval itself does not extend to the opposite side of \(k_N^0\).
Knowledge of the sign of the continuous displacement alone is not sufficient to
remove the opposite neighbouring integer for a general asymmetric peak.

For damping-dominated perturbations,
\[
    \Delta s_N^{(\gamma)}
    =
    -\frac{\gamma_N}{16\theta_N^2}.
\]
If the combined higher-order and non-damping displacement uncertainty is bounded by
\(\eta_N\), one may take
\[
    \widehat\Delta_N^-
    =
    \frac{\gamma_N}{16\theta_N^2}+\eta_N,
    \qquad
    \widehat\Delta_N^+
    =
    \eta_N,
\]
and compute the two radii from the extremal floor and ceiling above. Signed
frequency and phase perturbations are handled in the same way.

The residual-gap argument is unchanged. If
\[
    k_N^{\mathrm{dir}}
    \in
    \arg\max_{k\in\mathcal I_N(W_N^-,W_N^+)\cap\mathcal K_N}
    P_N(k),
\]
and
\[
    2R_N^{\mathrm{val}}
    <
    \min_{\substack{
        k\in\mathcal I_N(W_N^-,W_N^+)\cap\mathcal K_N\\
        k\neq k_N^{\mathrm{rec}}
    }}
    \bigl[
        F_N(k_N^{\mathrm{rec}})-F_N(k)
    \bigr],
\]
then
\[
    k_N^{\mathrm{dir}}=k_N^{\mathrm{rec}}.
\]

\subsection{Window-overhead scale}
\label{subsec:app-window-overhead-scale}

The robust-window correction preserves the amplitude-amplification query scale
whenever the window is small compared with the Grover-scale stopping time. In the
general amplitude-amplification notation, this means
\[
    W_N=o(1/\theta_N)
\]
for a symmetric window, or
\[
    W_N^-+W_N^+=o(1/\theta_N)
\]
for a direction-aware window. Under these conditions, the number of additional
candidate stopping steps is asymptotically smaller than the
\(O(1/\theta_N)\) amplitude-amplification runtime.

In the constant-displacement regime,
\[
    |\Delta s_N^{\mathrm{lead}}|=O(1),
\]
one may choose
\[
    W_N=O(1),
\]
so the correction overhead is a constant-size local search around the ideal
stopping step. For single-target Grover search, \(1/\theta_N=\Theta(\sqrt N)\), so
any \(O(1)\) robust or direction-aware window has negligible asymptotic overhead
relative to the Grover runtime.

\section{Residual-Reversal Theory and Numerical Protocols}
\label{app:residual-numerical-protocol}

This supplementary section supports the residual-reversal theory in main-text
Section~4.6 and the numerical study in main-text
Section~5.
Sections~\ref{subsec:app-residual-quantities-protocol}--\ref{subsec:app-largeN-residual-validation}
specify the residual constructions and threshold tests.
Sections~\ref{subsec:app-single-mechanism-scaling-protocol}--\ref{app:finite-shot-protocol}
record the displacement, margin, damping, circuit-level, and finite-shot
protocols, while
Subsections~\ref{app:device-like-protocol}--\ref{app:ae-localized-application-protocol}
record the GenericBackendV2, matched-baseline, calibration-cost,
calibration-derived FakeOslo/held-out, and AE-localized application protocols.

\subsection{Residual quantities used in the numerical protocols}
\label{subsec:app-residual-quantities-protocol}

The value-to-step conversion, sampled-reversal condition, and spectral
interpretation are established in
main-text Subsections~4.6.1--4.6.3.
For the numerical protocols below, \(D_N^r\) denotes the direct residual
differential between competing sampled steps, whereas \(A_N^r\) denotes
the coefficient of a parametrized residual spectral component. These are
distinct quantities: the spectral-shape factor introduced below maps
\(A_N^r\) to the corresponding value-level differential \(D_N^r\).
The following subsections specify the local constructions and parameter
choices used in the residual-validation figures.

\subsection{Residual-induced reversal construction}
\label{subsec:app-residual-reversal-construction}

The residual reversal experiment isolates the value-level failure mode described in
main-text Subsection~4.6.2. The leading curve is chosen so
that its continuous first peak remains inside the ideal rounding cell, and hence
there is no leading displacement-induced stopping failure. The sampled reversal is
then produced only through a residual differential between the ideal sampled step
and a competing sampled step.

Let \(k_N^0\) be the ideal sampled stopping step and let
\(k_{\mathrm{comp}}\) be the nearest competing sampled step in the direction of the
smallest leading value gap. Define
\[
    \Delta_N^F
    =
    F_N(k_N^0)-F_N(k_{\mathrm{comp}}).
\]
The residual is constructed so that its relevant local differential is controlled
directly by a differential parameter \(D_N^r\):
\[
    r_N(k_{\mathrm{comp}})-r_N(k_N^0)=D_N^r.
\]
Then the noisy sampled ordering between these two steps is
\[
    P_N(k_N^0)-P_N(k_{\mathrm{comp}})
    =
    \Delta_N^F-D_N^r.
\]
Thus the critical local residual differential is
\begin{equation}
    D_{N,\mathrm{crit}}^r
    =
    \Delta_N^F,
    \label{eq:app-residual-critical-differential}
\end{equation}
with a tie at \(D_N^r=D_{N,\mathrm{crit}}^r\) and a sampled reversal when
\[
    D_N^r>D_{N,\mathrm{crit}}^r.
\]
This \(D_N^r\) is a direct value-level residual differential. It is not the same
object as the spectral residual coefficient used in the parametrized construction
of main-text Subsection~4.6.3. The construction is
deliberately local: it demonstrates that a residual differential comparable to the
local sampled value gap can change the sampled stopping step, even when the leading
continuous peak remains geometrically stable.

\subsection{Spectral residual parametrization}
\label{subsec:app-spectral-residual-parametrization}

At the observable level, the non-leading contribution has the schematic spectral
expansion
\[
    r_N(k)=\sum_{j\in\mathcal J_N} a_{j,N}\lambda_{j,N}^k,
\]
where the coefficients absorb the relevant initial-state and measurement overlaps.
With
\[
    q_N=\frac{\beta_N}{\rho_N}=1-\frac{g_N}{\rho_N},
\]
the residual bound from Supplementary Section~\ref{subsec:app-residual-bound-spectral-separation}
reads
\[
    R_N^{\mathrm{val}}
    \leq
    K_Nq_N^{c_*/\theta_N}
    \leq
    K_N\exp\!\left(-\frac{c_*g_N}{\rho_N\theta_N}\right).
\]
Thus a positive spectral gap is not sufficient by itself; the relative decay must
suppress the residual below the local sampled gap over the first-peak window.

For a dominant residual component, write locally
\[
    r_N^{\mathrm{dom}}(k)
    =
    A_N^r q_N^{k-k_{\mathrm{ref}}}
    \cos\!\left((k-k_{\mathrm{ref}})\nu_N+\zeta_N\right)
    =
    A_N^r q_N^{k-k_{\mathrm{ref}}}u_N(k).
\]
For a leading sampled maximizer \(k_{\mathrm{lead}}\) and competitor
\(k_{\mathrm{comp}}\), a destabilizing phase alignment gives the critical
spectral coefficient
\begin{equation}
    A_{N,\mathrm{crit}}^r
    =
    \frac{\Delta_N^F(k_{\mathrm{comp}})}{
        \left|
        q_N^{k_{\mathrm{comp}}-k_{\mathrm{ref}}}u_N(k_{\mathrm{comp}})
        -
        q_N^{k_{\mathrm{lead}}-k_{\mathrm{ref}}}u_N(k_{\mathrm{lead}})
        \right|}.
    \label{eq:app-critical-residual-amplitude}
\end{equation}
If the denominator vanishes or has the stabilizing sign, that component does not
produce reversal.

In single-target Grover scaling,
\[
    \Delta_N^F(k_{\mathrm{comp}})
    =\Theta(\kappa_Nm_N)
    =\Theta\!\left(\frac{m_N}{N}\right).
\]
A sufficient spectral condition for same-step stability is therefore
\[
    K_N\exp\!\left(-\frac{c_*g_N}{\rho_N\theta_N}\right)
    =o(\kappa_Nm_N),
\]
or equivalently
\[
    \frac{c_*g_N}{\rho_N\theta_N}
    -\log K_N
    -\log\!\left(\frac{1}{\kappa_Nm_N}\right)
    \longrightarrow +\infty.
\]
This makes the spectral interpretation of residual-induced reversal explicit: a
leading first peak may remain visible while a sufficiently persistent non-leading
mode changes the sampled ordering. Equation~\eqref{eq:app-critical-residual-amplitude}
is used only as a parametrized realization of the value-level reversal criterion;
the direct residual differential remains the primary threshold quantity.

\subsection{Large-\texorpdfstring{$N$}{N} residual validation}
\label{subsec:app-largeN-residual-validation}

For the large-\(N\) residual validation, the search size is taken to be large while
the leading first peak is kept fixed in the correct sampled cell. The residual
ratio
\[
    q_N=\beta_N/\rho_N
\]
is varied, and for each value of \(q_N\) the residual differential is compared with
the local sampled gap:
\[
    \frac{
        r_N(k_{\mathrm{comp}})-r_N(k_N^0)
    }{
        \Delta_N^F
    }.
\]
A sampled reversal occurs when this ratio exceeds one. The validation therefore
directly tests the threshold
\[
    r_N(k_{\mathrm{comp}})-r_N(k_N^0)
    =
    \Delta_N^F.
\]
For the ideal non-degenerate single-target Grover curve, the nearest-competitor
gap has the leading scale
\[
    \Delta_N^F
    \approx
    \frac{8m_N}{N}.
\]
If the residual is parameterized symmetrically by lowering \(k_N^0\) by an amount
\(A\) and raising \(k_{\mathrm{comp}}\) by the same amount, then the direct
differential is \(2A\), and the one-sided symmetric amplitude threshold is
\[
    A_{\mathrm{crit}}^{\mathrm{sym}}
    \approx
    \frac{4m_N}{N}.
\]
This large-\(N\) test confirms that residual reversal is not a small-\(N\)
artifact: even at large search sizes, a residual mode that survives to the
first-peak window can reverse the sampled ordering if its local differential is
comparable to the curvature-scaled sampled gap.

\subsection{Single-mechanism scaling protocol}
\label{subsec:app-single-mechanism-scaling-protocol}

For the single-mechanism scaling experiment in
main-text Figure~2, the search sizes range over powers
of two and the three perturbations are applied one at a time as
\(\delta_N=c/N\), \(\gamma_N=c/N\), and \(\phi_N=c/\sqrt N\). For each case we
compute the exact continuous first peak of the corresponding one-mechanism normal
form and normalize its displacement by the analytic limits derived in
Supplementary Subsection~\ref{subsec:app-single-mechanism-limits}. Convergence of
the normalized curves to one is the numerical test.

\subsection{Combined displacement protocol}
\label{subsec:app-combined-displacement-protocol}

For the combined displacement validation in
Supplementary Figure~\ref{fig:combined-displacement}, the search size was fixed at \(N=256\). This value was selected because the ideal first peak is visibly
off-centre in its rounding cell, so that
\(m_N^+\neq m_N^-\), while the problem size still permits a dense
two-parameter scan.
The effective-angle and observable phase perturbations were scanned jointly using
the normalized displacement coordinates
\[
    D_\delta
    =
    -
    \frac{s_N^0\delta_N}{\theta_N},
    \qquad
    D_\phi
    =
    -
    \frac{\phi_N}{4\theta_N}.
\]
The linear prediction is
\[
    \Delta s_N^{\mathrm{linear}}
    =
    D_\delta+D_\phi.
\]
For each parameter pair, the exact continuous first-peak location was computed and
compared with this linear prediction. Because this \((\delta_N,\phi_N)\) scan has
\(\gamma_N=0\), the leading curve is an undamped cosine and local symmetry makes
nearest-integer rounding equivalent to direct comparison of the adjacent sampled
values. The resulting sampled outcome produces the observed diagonal stability
band.

An additional \((\delta_N,\gamma_N)\) slice fixes \(\phi_N=0\) and uses
\[
    D_\delta
    =
    -
    \frac{s_N^0\delta_N}{\theta_N},
    \qquad
    D_\gamma
    =
    -
    \frac{\gamma_N}{16\theta_N^2}.
\]
The corresponding first-order continuous rounding-cell slab is
\[
    -m_N^-
    <
    D_\delta+D_\gamma
    <
    m_N^+ .
\]
For this damped slice, the sampled outcome is evaluated directly from the integer
values of \(F_N(k)\), rather than inferred by rounding the continuous peak. The
comparison therefore tests the predicted signed-displacement scale and cancellation
structure while allowing the exact finite-\(N\) sampled boundary to differ slightly
from the continuous slab. This protocol generates the supplementary slice in
Figure~\ref{fig:app-delta-gamma-slice}.

\subsection{Margin-sensitive \texorpdfstring{$N$}{N}-scan protocol}
\label{subsec:app-margin-sensitive-scan-protocol}

For the margin-sensitive scan in Supplementary Figure~\ref{fig:margin-scan}, the same normalized
effective-angle perturbation scale was applied across a range of search sizes:
\[
    \delta_N=\pm\frac{0.60}{N}.
\]
For each \(N\), the exact continuous displacement was computed and compared with
the appropriate directional margin. The plotted surplus is of the form
\[
    m_N^{\mathrm{dir}}-|\Delta s_N|,
\]
where \(m_N^{\mathrm{dir}}\) is \(m_N^{-}\) for early shifts and \(m_N^{+}\) for
late shifts. This scan uses an undamped cosine leading form, for which local
symmetry makes the continuous rounding-cell crossing coincide with the adjacent
sampled-value transition. Positive surplus therefore corresponds to unchanged
sampled stopping in this particular scan, while negative surplus corresponds to a
sampled step change. The same-step fractions are dense-scan fractions over \(N\),
not probabilities of theorem success.

\subsection{Peak-height versus same-step protocol}
\label{subsec:app-peak-height-vs-same-step-protocol}

For the peak-height survival experiment in main-text Figure~3, the
damping-only normal form was used. Two damping scales were compared.

The displayed peak-height survival scale uses a \(10\%\) loss criterion
for the leading envelope:
\[
    e^{-\gamma_N s_N^{\mathrm{lead}}}
    =
    0.9.
\]
This gives the \(N^{-1/2}\)-type scaling in single-target Grover search.

The first-order directional scale is
\[
    \frac{\gamma_N}{16\theta_N^2}=m_N^-.
\]
The displayed directional continuous rounding-cell threshold is obtained by solving
\[
    s_N^{\mathrm{lead}}
    =
    k_N^0-\frac{1}{2}
\]
using the exact normal-form peak location in
main-text Eq.~(13). This produces an
\(N^{-1}\)-type continuous-cell no-correction scale modulated by the arithmetic margin.
For the asymmetric damped curve, the exact finite-\(N\) sampled transition is
instead determined by equality of neighbouring sampled values, as illustrated by
the encoded-subspace depolarizing specialization.

\subsection{Encoded-subspace depolarizing validation protocol}
\label{app:encoded-depolarizing-validation-protocol}

To validate the closed-form specialization independently of circuit decomposition,
we simulate the two-dimensional marked--unmarked density matrix directly. At each
Grover iterate we apply
\[
    \rho
    \longmapsto
    (1-p_N)G_N\rho G_N^\dagger
    +
    p_N\frac{I_G}{2}.
\]
For \(n=6\), \(N=64\), trajectories are generated over the first-lobe window for
\[
\begin{aligned}
 p_N\in\{&0,0.001,0.005,0.01,0.02,0.04,0.06,0.064,\\
          &0.065,0.066,0.0665,0.0666,0.067,0.07,0.08,0.10\}.
\end{aligned}
\]
The numerical density-matrix probabilities are compared directly with
Supplementary Eq.~\eqref{eq:app-encoded-depolarizing-trajectory} and are also fitted to the full
observable normal form to recover \(\gamma_N\), \(\delta_N\), and \(\phi_N\).
The first-lobe sampled maximizer is evaluated directly rather than inferred by
rounding the fitted continuous peak.

For \(N=64\), the analytic thresholds are those derived in Supplementary
Subsection~\ref{app:encoded-depolarizing-specialization}. The transition scan
includes \(p_N=0.0665\) and \(p_N=0.0666\) to resolve the exact sampled change.
Direct density-matrix evolution agrees with the closed-form trajectory to machine
precision and recovers the same continuous and sampled thresholds; no separate
figure is needed for this machine-precision check.

\subsection{Coherent oracle-error fitting protocol}
\label{app:coherent_oracle_protocol}

For the coherent-error validation in
Supplementary Figure~\ref{fig:coherent_oracle_validation}, we use a single-target
Grover search with $n=6$ and $N=64$.  The ideal marked-state phase
reflection is replaced at every iterate by
\[
O_{\varepsilon}
=
I+
\left(
e^{i(\pi+\varepsilon)}-1
\right)
\lvert w\rangle\langle w\rvert ,
\]
while the diffuser remains ideal.  The success probabilities are
computed using exact Qiskit Aer statevector simulation, without shot
noise.

For each over-rotation strength, the sampled trajectory is fitted to
\[
P_N(k)
=
B_N-C_N\cos(\omega_N k+\chi_N),
\]
and the normal-form parameters are extracted as
\[
\delta_N=\frac{\omega_N-4\theta_N}{4},
\qquad
\phi_N=\chi_N-2\theta_N.
\]
The observed sampled maximizer is restricted to the first-lobe interval
$0\leq k\leq2k_N^0$.  The coarse scan uses
\[
\varepsilon\in
\{0,0.05,0.10,0.15,0.20,0.25\},
\]
while the rounding transition is resolved on
\[
0.140\leq\varepsilon\leq0.180
\]
with spacing $0.002$.  Nonlinear least squares is used for the
zero-damping normal-form fit. For each \(\varepsilon\), we record the fit residual, the fitted
parameters \(\delta_N\) and \(\phi_N\), the exact fitted peak displacement,
the first-lobe sampled maximizer, and the sampled first-peak height.

\subsection{Statevector scaling protocol}
\label{app:coherent_scaling_protocol}

For the scaling experiment in
Figure~\ref{fig:coherent_statevector_scaling}, we use
\[
n\in\{6,8,10,12\},
\qquad
N=2^n,
\]
with the coherent oracle over-rotation scaled as
\[
\varepsilon_N=0.8\theta_N.
\]
For each search size, a single $n$-qubit Qiskit Aer statevector circuit
stores the trajectory after every Grover iterate up to
\[
k_{\max}=2k_N^0.
\]
The resulting success probabilities are fitted to the same zero-damping
normal form used in
main-text Subsection~5.4.

The relative fitted parameters, displacement components, and mechanism
ratio are computed as
\[
\frac{\delta_N}{\theta_N},
\qquad
\frac{\phi_N}{2\theta_N},
\qquad
D_{\delta,N}
=
-\frac{s_N^0\delta_N}{\theta_N},
\qquad
D_{\phi,N}
=
-\frac{\phi_N}{4\theta_N},
\]
and
\[
\left|
\frac{D_{\delta,N}}{D_{\phi,N}}
\right|.
\]
The reported finite-range exponents are obtained by ordinary least
squares fits of $\log |Y_N|$ against $\log N$ over the four displayed
search sizes. The sampled first-peak height and the first-lobe sampled maximizer are
also recorded for each \(N\).

\subsection{Finite-shot normal-form inference protocol}
\label{app:finite-shot-protocol}

The finite-shot experiment in
main-text Subsection~5.4 uses the exact coherent
statevector trajectories from
main-text Subsection~5.4 as probability baselines. For
\(n=6\), \(N=64\), the ideal geometry is
\[
    \theta_N=0.1253278312,
    \qquad
    s_N^0=5.7667498199,
    \qquad
    k_N^0=6,
\]
with directional margins
\[
    m_N^-=0.2667498199,
    \qquad
    m_N^+=0.7332501801.
\]
We use coherent oracle over-rotations
\[
    \varepsilon\in\{0.10,0.15,0.20\}
\]
and shot budgets
\[
    M\in\{512,2048,8192,32768\}
\]
per sampled depth.

For each exact statevector probability \(P_N^{\mathrm{ex}}(k)\), a finite-shot
observation is generated as
\[
    X_k\sim\operatorname{Binomial}
    \!\left(M,P_N^{\mathrm{ex}}(k)\right),
    \qquad
    \widehat P_N(k)=\frac{X_k}{M}.
\]
The observed frequencies are fitted to the full normal form
\[
    B_N-C_Ne^{-\gamma_N k}\cos(k\omega_N+\chi_N),
    \qquad
    \gamma_N\geq0,
\]
without fixing the damping parameter to its zero-damping statevector value. A
parametric bootstrap with \(B_{\mathrm{boot}}=1000\) generates binomial resamples
from the fitted trajectory and refits all five normal-form parameters. Percentile
intervals are then propagated to
\(\delta_N=(\omega_N-4\theta_N)/4\),
\(\phi_N=\chi_N-2\theta_N\),
\(s_N^{\mathrm{lead}}\), and the sampled stopping quantities.

The deterministic sampled-fit discrepancy is computed between the exact
statevector probabilities and the fitted normal form on the simulation grid;
it is not statistical shot noise. It is an offline, \emph{a posteriori} finite-set
diagnostic, distinct from the theoretical spectral residual bound
\(R_N^{\mathrm{val}}\). Parameter bootstrap refits capture model-based
sampling variability but do not establish coverage of unmodelled residuals.
The exact baselines for
\(\varepsilon=0.10,0.15,0.20\) have leading displacements approximately
\[
    -0.118604,
    \qquad
    -0.257801,
    \qquad
    -0.437693,
\]
with first-lobe sampled maximizers \(6,6,5\), respectively, and statevector fit
residuals at or below \(10^{-9}\).

For the finite-shot decision experiment, each bootstrap refit is mapped to the
left-hand side of main-text Eq.~(24), using the simulation-derived discrepancy
allowance. We require at least \(95\%\) of
bootstrap replicates to satisfy the certificate and, conservatively, require the
empirical \(97.5\)th percentile of the certificate left-hand side to lie below the
relevant margin. For recalibrated stopping, each refit constructs
\(\mathcal C_N^{\mathrm{lead}}\), evaluates the fitted sampled values on that
candidate set, and obtains \(k_N^{\mathrm{rec}}\) from
main-text Eq.~(27). A single recalibrated step is reported
only if it has at least \(95\%\) bootstrap support and the corresponding sampled-gap
condition is satisfied with at least \(95\%\) support.

When neither single-step decision reaches the bootstrap support threshold, a
robust window is built from the empirical \(95\%\) bootstrap displacement bound
\[
    |\Delta s_N^{\mathrm{lead}}|
    \leq
    \widehat\Delta_N
\]
and main-text Eq.~(29). For the intermediate
\(\varepsilon=0.15\) case this gives the three-step set
\(\{5,6,7\}\) across the displayed shot budgets. Because the true statevector
baseline has \(\gamma_N=0\) while the fit imposes \(\gamma_N\geq0\), the bootstrap
damping distribution has the expected boundary behavior; it is therefore
interpreted as contracting toward the zero-damping baseline rather than as a
symmetric interior confidence interval.

\subsection{Backend-derived device-like validation protocol}
\label{app:device-like-protocol}

The final main-text validation uses the same single-target \(n=6\), \(N=64\)
Grover problem as the running example. Each depth-\(k\) circuit is independently
transpiled to the basis and coupling graph of a \texttt{GenericBackendV2}
instance with hardware-like gate errors, relaxation parameters, gate durations,
and readout metadata. At the ideal reference depth \(k_N^0=6\), the circuit
has depth 2866 and 1915 CX gates. Success probabilities are extracted from
simulated premeasurement density matrices; readout noise is not applied to them.
Let \(\mathcal E\) denote a backend-derived gate-level error channel.

The unscaled channel is too strong for the first-lobe theory because it destroys
the useful first peak. To expose the first-peak-preserving regime without changing
the relative composition of the backend-derived errors, we use the interpolation
\[
    \mathcal E_\alpha
    =
    (1-\alpha)\mathcal I+\alpha\mathcal E,
\]
where this interpolation is applied to the noise channels associated with
circuit instructions, not once to the entire circuit. Here \(\alpha=1\) is
the unscaled device-like model and \(\alpha\rightarrow0\) approaches the
noiseless circuit. Independently transpiled depths need not implement a single
stationary noisy Grover iterate. A coarse scan in
\(\alpha\) is first used to locate the sign change of
\(P_N(5)-P_N(6)\); a finer scan around that crossing resolves
\(\alpha_{\mathrm{cross}}\approx0.0235\). The reference and shifted operating
points used in the main text are \(\alpha=0.01\) and \(\alpha=0.06\),
respectively.

For each selected trajectory, the first lobe is fitted to the observable normal
form
\[
    B_N-C_Ne^{-\gamma_N k}\cos(k\omega_N+\chi_N),
\]
and converted to
\(\delta_N=(\omega_N-4\theta_N)/4\) and
\(\phi_N=\chi_N-2\theta_N\). The fitted leading peak, sampled candidate pair,
first-peak sampled maximizer, residual bound, and recalibrated sampled gap are
then evaluated on the 13-depth numerical sampling set
\(\mathcal D_{13}=\{0,1,\ldots,12\}\), which is distinct from the
theoretical set \(\mathcal K_N=\mathcal W_N\cap\mathbb Z_{\geq0}\).
For the 13-depth sampled fit, we evaluate
\[
R_{N,\mathcal D_{13}}^{\mathrm{fit}}
=
\max_{k\in\mathcal D_{13}}
\left|P_N(k)-F_N(k)\right|,
\]
where \(P_N(k)\) is the exact simulated noisy probability and \(F_N(k)\) is the
fitted leading normal form. At \(\alpha=0.06\) the fit has
\(R^2=0.998865\), the sampled maximizer is \(5\), and the recalibrated
sampled-set certificate uses
\(\Delta_N^{\mathrm{rec}}(\mathcal D_{13})=0.024246\), where the leading
gap is minimized over the same 13-depth set, and
\(2R_{N,\mathcal D_{13}}^{\mathrm{fit}}=0.021077\). This is an \emph{a posteriori} diagnostic on \(\mathcal D_{13}\), not a
verification of the spectral assumptions or a bound outside the sampled set;
the independent held-out test is given in
Supplementary Subsection~\ref{app:fakeoslo-heldout-protocol}. The strength
interpolation is a controlled device-like validation, not a claim to reproduce
one named hardware device directly.

\subsection{Matched quantitative-baseline protocol}
\label{app:matched-baseline-protocol}

The comparison in main-text Table~2 uses the same
backend-derived trajectories and the common matched depth domain
\(k=1,\ldots,6\). Three rules are evaluated without changing the underlying noisy
probabilities. The ideal reference retains the known-fraction integer Grover depth
\(k_N^0=6\). The present framework applies the first-peak certificate and, when
needed, the recalibrated sampled maximizer. The Leng et al. baseline is evaluated
using the published success-probability predictor and restart-aware objective
\(r(k)=k/s(k)\) from Ref.~\cite{leng2025noiseTolerant} on the same candidate
domain. The predictor uses the mean native two-qubit error
\(E_{\mathrm{two,native}}=0.00177897\), scaled consistently with each
regime as \(E_{\mathrm{two}}(\alpha)=\alpha E_{\mathrm{two,native}}\).
The fixed two-qubit-count increment is \(c_{\mathrm{two}}=324\), the median
CX-count increment across the transpiled circuit family. No noisy-trajectory
samples from the present
framework are fed into the Leng predictor. The in-hindsight minimum of
\(k/P(k)\) is reported only as an oracle diagnostic and is not treated as an
executable baseline.

For a selected depth \(k\), the reported peak-retention metric is
\[
    \mathrm{Ret}(k)
    =
    \frac{P_{\mathrm{noisy}}(k)}{\max_{1\leq j\leq6}P_{\mathrm{noisy}}(j)},
\]
and the restart-aware diagnostic is \(k/P_{\mathrm{noisy}}(k)\). Cost regret is
computed relative to the minimum of that diagnostic over the same matched domain.
This protocol keeps the optimization objectives explicit: first-peak certification
and restart-aware running-time minimization need not select the same depth.

\subsection{Calibration-cost and amortization protocol}
\label{app:calibration-cost-protocol}

The calibration study quantifies the cost of inferring a normal form accurately
enough to make a discrete stopping decision. For a depth design \(D\) and a common
shot count \(M\) per sampled depth, define
\[
    S_{\mathrm{cal}}=M|D|,
    \qquad
    Q_{\mathrm{cal}}=M\sum_{j\in D}j.
\]
The six-depth stress design
\(D6=\{1,3,5,6,8,10\}\) gives
\(s_N^{\mathrm{lead}}=5.091864\) and full-window RMSE \(0.019880\) in the
shifted regime. The recommended sparse design is
\[
    D8=\{0,1,3,5,6,8,10,12\},
\]
which gives \(s_N^{\mathrm{lead}}=5.166592\) and RMSE \(0.007595\), while the
full comparison design \(D13=\{0,1,\ldots,12\}\) gives \(5.169850\) and
\(0.007342\), respectively. Each reliability
point uses 100 independent synthetic calibration datasets generated from the
corresponding device-like trajectory. Each dataset is fitted to the same observable
normal form and processed with 200 parametric-bootstrap refits. The stopping
decision uses the same fixed-step, recalibrated, and robust-window logic as the
finite-shot protocol in Supplementary Subsection~\ref{app:finite-shot-protocol}.
Simulation-derived residual allowances and bootstrap support are used for these
empirical decisions; they are not a data-only certificate of the spectral assumptions.
A run is counted as decisive when it returns a bootstrap-supported single depth
rather than the fallback window. Because the synthetic data-generating
trajectory is known,
we additionally record whether each decisive depth equals the true sampled
maximizer and whether each fallback window contains it.

For the reference D8 case with \(M=2048\), the formulas above give
\[
    S_{\mathrm{cal}}=16384,
    \qquad
    Q_{\mathrm{cal}}=92160,
\]
and the decisive rate is 0.98. For the shifted D8 case with \(M=32768\),
\[
    S_{\mathrm{cal}}=262144,
    \qquad
    Q_{\mathrm{cal}}=1474560,
\]
with decisive rate 0.99. The full D13 fit reaches the same 0.99 decisive rate at
\(Q_{\mathrm{cal}}=2555904\), which is why D8 is used for the main cost
accounting. Across the reference D8, shifted D8, and shifted D13 cases there are
zero wrong decisive outputs in 300 datasets, all fallback windows contain the
true sampled optimum, and the fit-failure rate is zero. These observed frequencies
do not establish a zero-error probability. For a single 100-dataset regime,
zero events yield only an approximate one-sided 95\% upper bound of 3\% on
the unconditional error-event rate under independent sampling.

The Grover repeat-until-success amortization uses the per-success proxy
\(k/P(k)\). In the shifted regime, retaining depth 6 costs 10.090323 in this
metric, while depth 5, selected by an \emph{a posteriori} sampled-set check, costs
7.986832. The saving per repeated use is
therefore
\[
    \Delta C
    =
    10.090323-7.986832
    =
    2.103491.
\]
Equating cumulative savings with the D8 calibration cost gives
\[
    L_{\mathrm{Grover}}
    =
    \frac{Q_{\mathrm{cal}}}{\Delta C}
    \approx
    701007
\]
repeated uses.
This is a workload-specific amortization count rather than a universal complexity
constant; it assumes that the calibrated noise setting remains approximately
stationary across repeated uses.

\subsection{Calibration-derived FakeOslo and held-out-certificate protocol}
\label{app:fakeoslo-heldout-protocol}

To test whether the stopping transition is specific to the synthetic
\texttt{GenericBackendV2} construction, we repeat the analysis with the archived
\texttt{FakeOslo} calibration snapshot dated 17 July 2022
(22:01:55+02:00) from the Qiskit Runtime fake-provider library. We fix the
physical layout \(\{0,1,2,3,4,5\}\), transpiler
\texttt{optimization\_level=1}, transpiler seed 20261010, density-matrix
simulation, and simulator seed 20261011. At \(k_N^0=6\) the transpiled circuit
has depth 3104 and 2244 two-qubit gates; the selected layout has mean and maximum
two-qubit errors \(0.007978\) and \(0.009009\), respectively.

The raw calibration snapshot is too noisy for the first-peak theory:
\(k_N^*=1\) and \(P_{\mathrm{peak}}=0.017448\), close to the uniform background
\(1/N=0.015625\). We therefore use the same controlled interpolation
\(\mathcal E_\alpha=(1-\alpha)\mathcal I+\alpha\mathcal E\).
A fine scan gives
\(\alpha_{\mathrm{cross}}=0.011727\) for the exchange of \(P_N(5)\) and
\(P_N(6)\). The frozen reference and shifted points are
\(\alpha=0.005\) and \(0.030\). At the shifted point,
\[
k_N^*=k_N^{\mathrm{rec}}=5,\qquad
s_N^{\mathrm{lead}}=5.147415,\qquad
R^2=0.999017.
\]

For the out-of-sample certificate test, the fit uses only
\[
D8=\{0,1,3,5,6,8,10,12\},
\]
while
\[
H=\{2,4,7,9,11\}
\]
is held out completely from fitting, model selection, and tuning. Define
\[
r_{D8}(k)=P_N(k)-F_{D8}(k),
\]
and
\[
R_{\mathrm{train}}=\max_{k\in D8}|r_{D8}(k)|,\qquad
R_{\mathrm{hold}}=\max_{k\in H}|r_{D8}(k)|,\qquad
R_{\mathrm{all}}=\max_{0\leq k\leq12}|r_{D8}(k)|.
\]
The same stress test is applied to the original GenericBackendV2 shifted
trajectory. The results are
\begin{center}
\begin{tabular}{lcccc}
\toprule
Case & \(\Delta_N^{\mathrm{rec}}\) & \(R_{\mathrm{train}}\) &
\(R_{\mathrm{hold}}\) & \(R_{\mathrm{all}}\)\\
\midrule
GenericBackendV2 (\(\alpha=0.06\))
&0.024370&0.012167&0.011022&0.012167\\
FakeOslo (\(\alpha=0.030\))
&0.022905&0.011049&0.009607&0.011049\\
\bottomrule
\end{tabular}
\end{center}
so \(2R_{\mathrm{train}}\), \(2R_{\mathrm{hold}}\), and
\(2R_{\mathrm{all}}\) are all below \(\Delta_N^{\mathrm{rec}}\) in both
cases. The all-depth margins are approximately
\(3.6\times10^{-5}\) for GenericBackendV2 and \(8.07\times10^{-4}\) for
FakeOslo. The GenericBackendV2 all-depth margin is therefore narrow, while the
calibration-derived FakeOslo case retains a larger positive margin.

\subsection{AE-localized stopping application protocol}
\label{app:ae-localized-application-protocol}

This protocol supplies the end-to-end application reported in
Subsection~6.1. The purpose is to start without the
true success fraction and use a deliberately coarse QPE-free amplitude-estimation
stage only to localize the relevant depth region before applying the existing
stopping framework. The true values \(a=1/64\), \(\theta_N\), \(k_N^0=6\), and
the noisy optimum \(k_N^*=5\) are withheld from the estimator and used only for
offline screening and scoring.

For an AE depth \(k\), the ideal likelihood model is
\[
    p_k(\theta)=\sin^2((2k+1)\theta).
\]
Given \(y_k\) successes from \(M_A\) shots, the estimator maximizes the joint
binomial log-likelihood over \(0<\theta<\pi/2\), converts the result to a
provisional ideal peak location, and rounds that location to a provisional depth
\(\widehat{k}_0\). We fix in advance the schedules
\[
    A4=\{0,1,2,4\},
    \qquad
    A5=\{0,1,2,4,8\},
\]
with \(M_A\in\{512,2048,8192\}\) shots per AE depth. Among the screened AE
schedules and shot budgets, the selected configuration is the cheapest one
achieving complete coverage of both the ideal reference and the true noisy optimum
after the fixed guard-band rule
\[
    L7(\widehat{k}_0)
    =
    \{\max(0,\widehat{k}_0-3),\ldots,\widehat{k}_0+3\}.
\]
This criterion selects A4 with \(M_A=512\). This use of the known simulated
truth is confined to offline design screening; the selected AE configuration is
then frozen. After the local-budget screen below, the complete pipeline is also
transferred to FakeOslo without retuning.

On the shifted GenericBackendV2 trajectory, the selected AE stage returns
\(\widehat{k}_0=7\) in 96 datasets and 8 in 4 datasets out of 100. Thus AE alone
never selects the true noisy optimum 5, but every guard band contains both depths
5 and 6. Before finite-shot testing, both observed local windows pass an
exact-probability feasibility check: the constrained normal-form fit returns
candidate pair \(\{5,6\}\), recalibrated depth 5, and a positive certificate
margin.

Local finite-shot calibration is then screened at
\(M_L\in\{2048,8192,32768\}\) shots per local depth. The frozen choice
\(M_L=8192\) gives the best cost--reliability compromise before the final
bootstrap study. The final experiment uses 100 independent end-to-end datasets,
independent AE and local measurements, 200 bootstrap refits per dataset, and a
pre-specified joint-support threshold of 0.95 for a unique depth. On
GenericBackendV2, 60 datasets return a bootstrap-supported unique depth of 5,
while 40 return a robust window;
there are zero wrong decisive outputs, zero point-fit failures, and every
fallback window contains depth 5. All 100 datasets have bootstrap modal depth 5.
The reported unique-depth rate of 0.60 is measured in this final
bootstrap study, after the budget-screening stage. Each finite-shot fit uses
its observed-frequency discrepancy and a model-based bootstrap support rule;
these are empirical decision diagnostics, not a high-confidence bound on the
exact unobserved trajectory residual.

For conservative query accounting we define
\[
    Q=\sum_j M_j j,
    \qquad
    S=\sum_j M_j,
\]
and add the independently collected AE and local-correction costs. The selected
GenericBackendV2 route has mean
\[
    Q_{\mathrm{total}}=407285.76,
    \qquad
    S_{\mathrm{total}}=59392,
\]
which is 72.38\% lower query-equivalent calibration cost and 77.34\% fewer
shots than the shifted D8 reference (32768 shots per depth). This cost difference
is driven mainly by the lower local shot budget of 8192, not by AE localization
alone, and comes with a lower unique-depth rate (0.60 versus 0.99). For
comparison, D8 at 8192 shots per depth has the arithmetic cost
\(Q=368640\), below the AE-localized total; its decision reliability at that
budget was not tested. Reusing observations changes the reported cost saving only
from 72.38\% to 72.51\%.

Finally, the full pipeline is transferred without retuning to the frozen
FakeOslo shifted trajectory at \(\alpha=0.030\). The provisional-depth mixture
changes to 71 occurrences of 7 and 29 of 8, but relevant-window coverage remains
one. The final decision gives 78 unique correct recalibrations and 22 safe
fallbacks, with zero wrong decisive outputs and complete fallback coverage. Its
conservative mean query cost is 421621.76, 71.41\% below the higher-shot D8
reference. These are empirical finite-shot application results at different
budgets and return rates, not a controlled AE-only cost reduction or a guarantee
for arbitrary noise models or hardware.

\section{Supporting Numerical Results}
\label{app:additional-figures}

This section collects supporting plots for directional geometry, large-\(N\)
residual persistence, signed multi-mechanism displacement, arithmetic margin
effects, coherent circuit transitions, finite-shot decisions and parameter
uncertainty, coherent-error scaling, the joint damping--angle slice, and the
calibration-derived held-out test.

\subsection{Directional rounding-cell diagnostics}
\label{subsec:app-directional-rounding-diagnostics}

\begin{figure}[H]
    \centering
    \includegraphics[width=\linewidth]{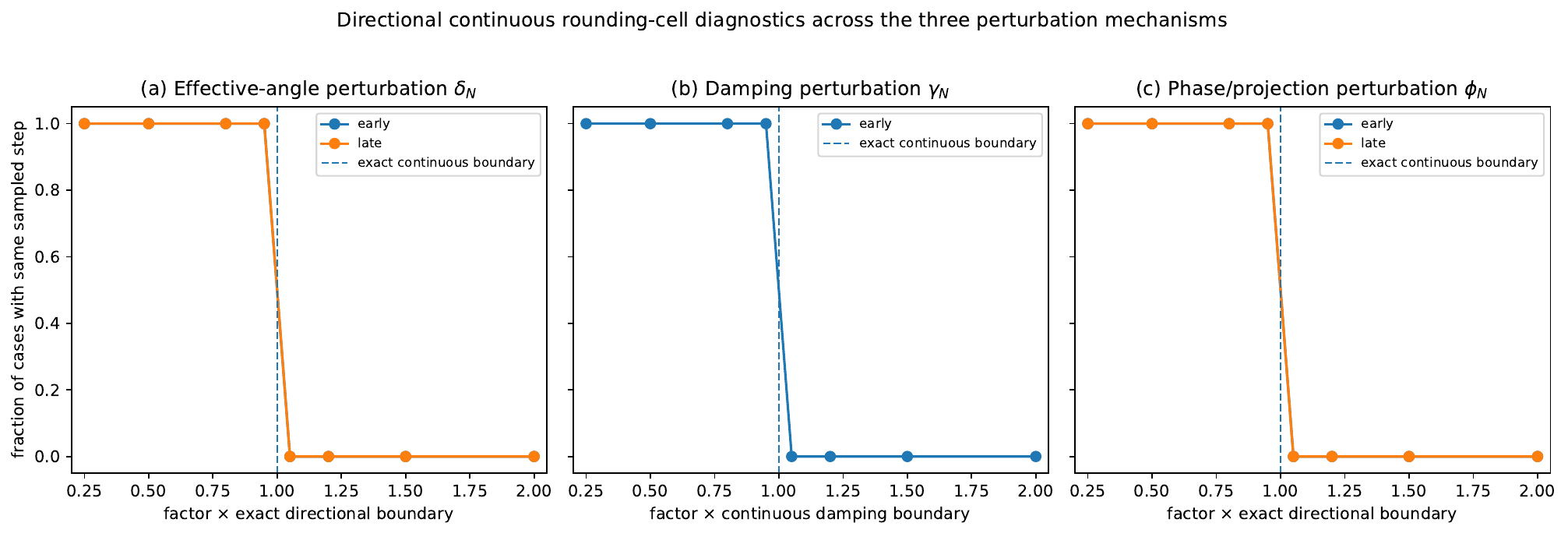}
    \caption{
    Directional continuous rounding-cell diagnostics for the effective-angle
    perturbation \(\delta_N\), damping perturbation \(\gamma_N\), and observable
    phase offset \(\phi_N\). Perturbation strengths are normalized by the
    directional geometric boundary. Undamped cosine slices have an exact
    nearest-sample interpretation; the damped slice illustrates the same
    characteristic scale but need not share the identical finite-\(N\) sampled
    transition.
    }
    
% Figure description (not typeset):
% Directional-boundary scans for effective-angle, damping, and observable-phase
% perturbations, with each perturbation normalized by its corresponding geometric
% rounding-cell boundary.

\label{fig:app-directional-tightness}
\end{figure}

Figure~\ref{fig:app-directional-tightness} examines the directional continuous
rounding-cell scale for the three perturbation mechanisms. For each mechanism, the
perturbation strength is normalized by the corresponding directional geometric
boundary. For the undamped frequency- and phase-shift slices, local cosine symmetry
makes this boundary coincide with the adjacent sampled transition. For damping, the
plot should be read as a geometric scale diagnostic; the exact asymmetric sampled
transition is instead obtained from neighbouring sampled-value equality, as
described in Supplementary Subsection~\ref{app:encoded-depolarizing-validation-protocol}.

\subsection{Large-\texorpdfstring{$N$}{N} spectral residual validation}
\label{subsec:app-largeN-spectral-residual-validation}

\begin{figure}[t]
    \centering
    \includegraphics[width=\linewidth]{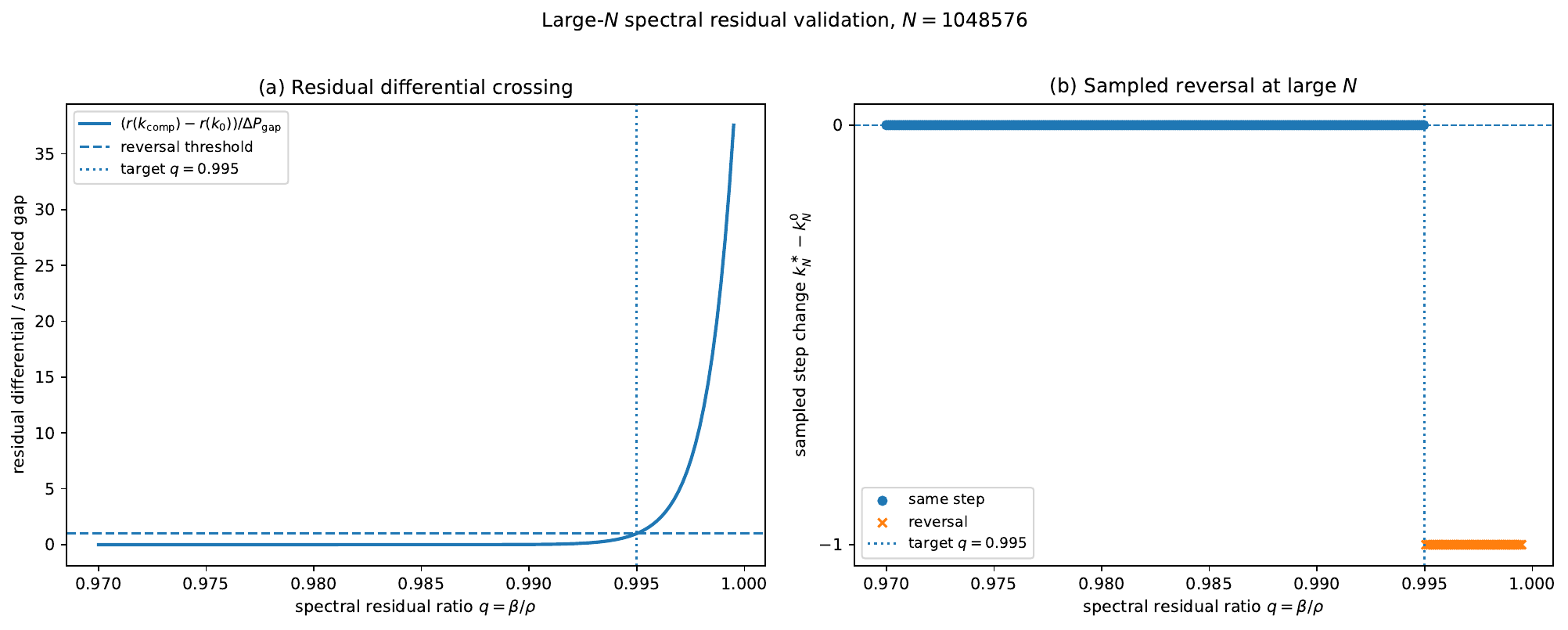}
    \caption{
    Large-\(N\) spectral residual validation. As the residual spectral ratio
    \(q=\beta/\rho\) approaches one, the residual mode survives longer across the
    first-peak window. A sampled reversal occurs when the residual differential
    exceeds the local sampled gap.
    }
    
% Figure description (not typeset):
% Large-\(N\) residual scan showing that residual persistence increases as
% \(q=\beta/\rho\) approaches one and that sampled reversal occurs when the
% residual differential exceeds the local sampled gap.

\label{fig:app-largeN-residual}
\end{figure}

Figure~\ref{fig:app-largeN-residual} extends the residual-induced reversal test to
a large search size. The residual differential is compared with the local sampled
gap as the residual spectral ratio \(q=\beta/\rho\) is varied. The sampled reversal
appears when the residual differential crosses the local gap, matching the
condition in main-text Eq.~(31). This confirms that the
residual mechanism is not a small-\(N\) artifact.

\subsection{Signed-displacement and arithmetic-margin scans}
\label{subsec:app-signed-displacement-margin-scans}

\begin{figure}[t]
    \centering
    \includegraphics[width=\linewidth]{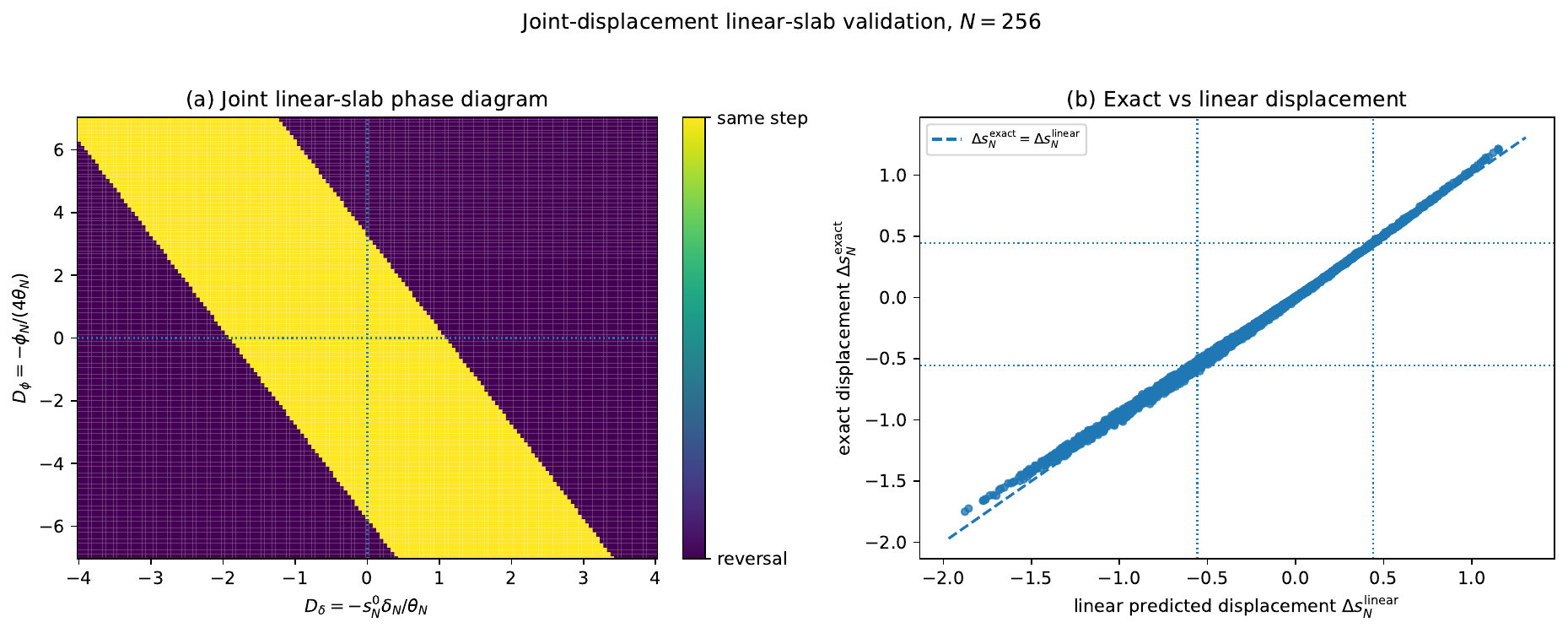}
    \caption{
    Combined displacement at \(N=256\).  The stable region follows the diagonal
    coordinate \(D_\delta+D_\phi\), and the exact continuous displacement agrees
    with the first-order prediction in the perturbative regime.
    }
    
% Figure description (not typeset):
% Two-parameter scan at \(N=256\) showing a diagonal stability band governed by
% the signed coordinate \(D_\delta+D_\phi\), together with close agreement between
% the exact and first-order continuous displacements in the perturbative region.

\label{fig:combined-displacement}
\end{figure}

\begin{figure}[t]
    \centering
    \includegraphics[width=\linewidth]{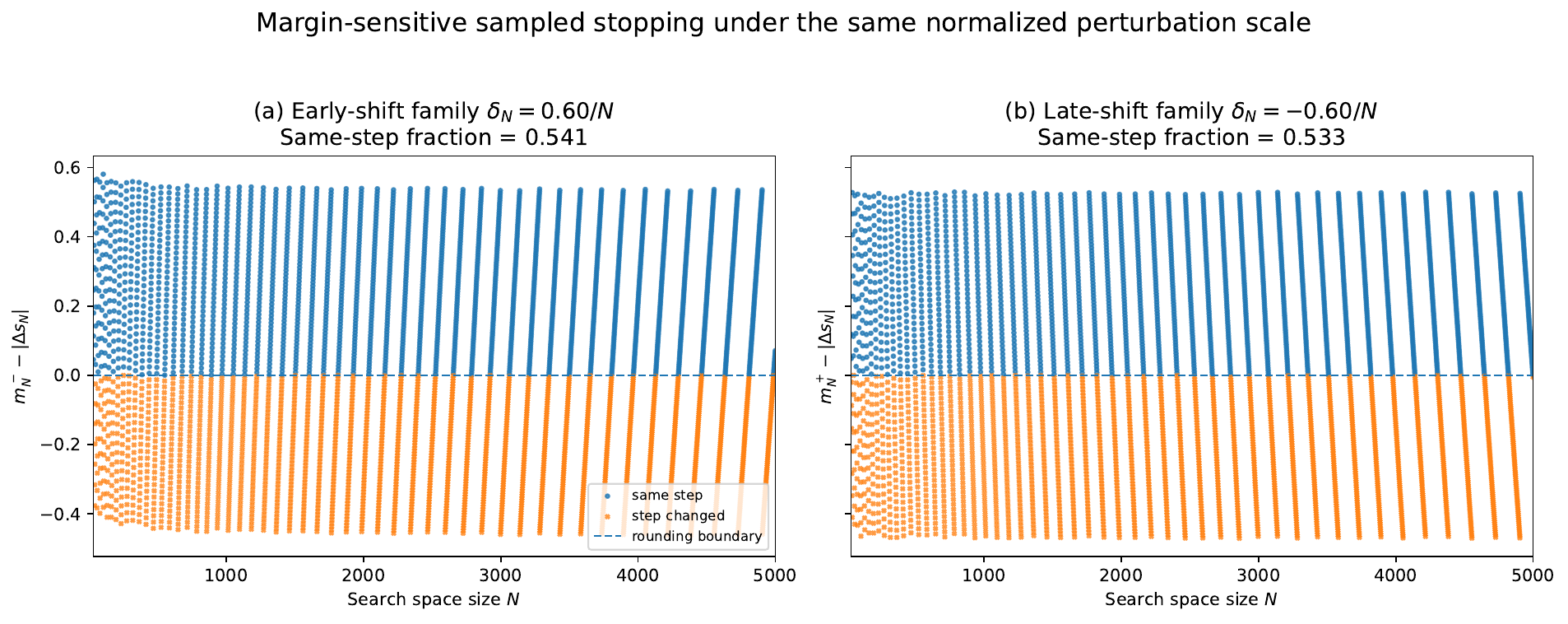}
    \caption{
    Margin-sensitive stopping for \(\delta_N=\pm0.60/N\).  The same normalized
    perturbation produces stable and unstable instances according to the
    directional arithmetic margin.  The title fractions are dense-scan fractions
    over \(N\), not success probabilities.
    }
    
% Figure description (not typeset):
% Dense scan over search size \(N\) for positive and negative
% \(\delta_N=0.60/N\), showing that the same perturbation scale yields both stable
% and unstable sampled steps according to the directional rounding margin.

\label{fig:margin-scan}
\end{figure}

The two plots below support the compact mechanism-and-margin discussion in
main-text Subsection~5.1.
Figure~\ref{fig:combined-displacement} shows that the joint
\((\delta_N,\phi_N)\) scan at \(N=256\) collapses onto the signed coordinate
\(D_\delta+D_\phi\), including cancellation between mechanisms.
Figure~\ref{fig:margin-scan} then shows that an \(O(1)\) displacement generated
by the same \(1/N\) perturbation scale can either preserve or change the sampled
step according to the finite-\(N\) directional margin.

\subsection{Coherent-error circuit transition}
\label{subsec:app-coherent-circuit-transition}

\begin{figure}[t]
    \centering
    \includegraphics[width=\textwidth]{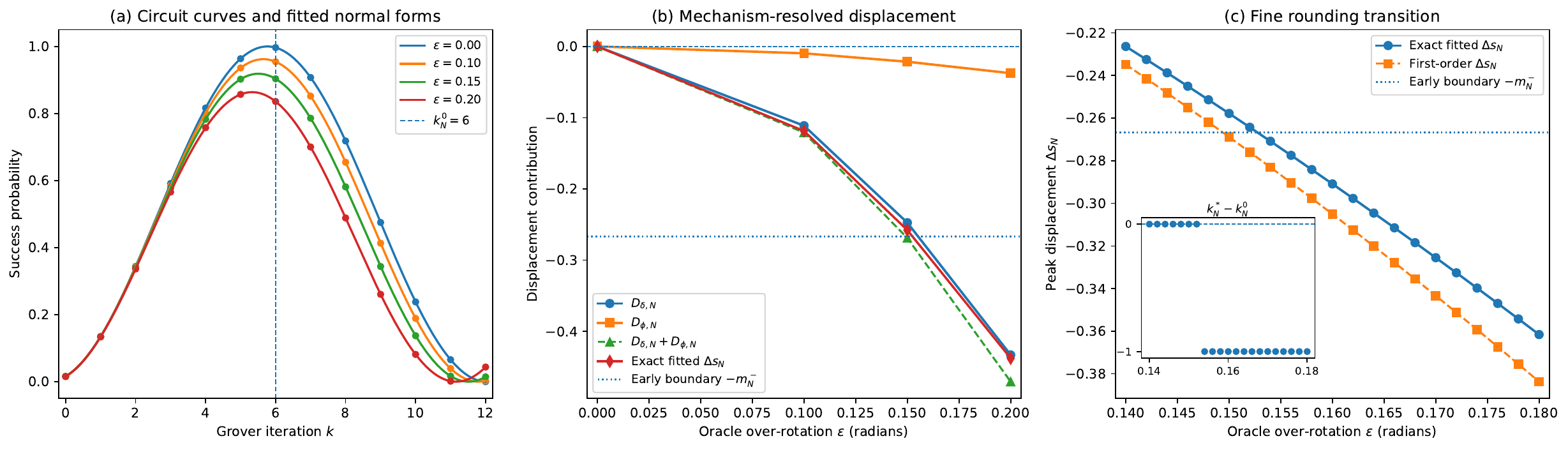}
    \caption{
    Coherent-error circuit validation for \(n=6\), \(N=64\).
    (a) Statevector probabilities and zero-damping normal-form fits.
    (b) Frequency and phase displacement components and the early rounding
    boundary.  (c) Fine scan of the transition from \(k_N^*=k_N^0\) to
    \(k_N^*=k_N^0-1\).
    }
    
% Figure description (not typeset):
% Three panels showing coherent-error Grover trajectories, their fitted
%         displacement components, and the one-step sampled transition.

    \label{fig:coherent_oracle_validation}
\end{figure}

Figure~\ref{fig:coherent_oracle_validation} gives the full circuit-level result
summarized in main-text Subsection~5.4.
The fitted trajectories have negligible residual, the correlated frequency and
phase contributions move the peak toward the early boundary, and the observed
sampled transition occurs near \(\varepsilon\simeq0.154\).

\subsection{Finite-shot stopping decisions}
\label{subsec:app-finite-shot-stopping-decisions}

\begin{figure}[t]
    \centering
    \includegraphics[width=\textwidth]{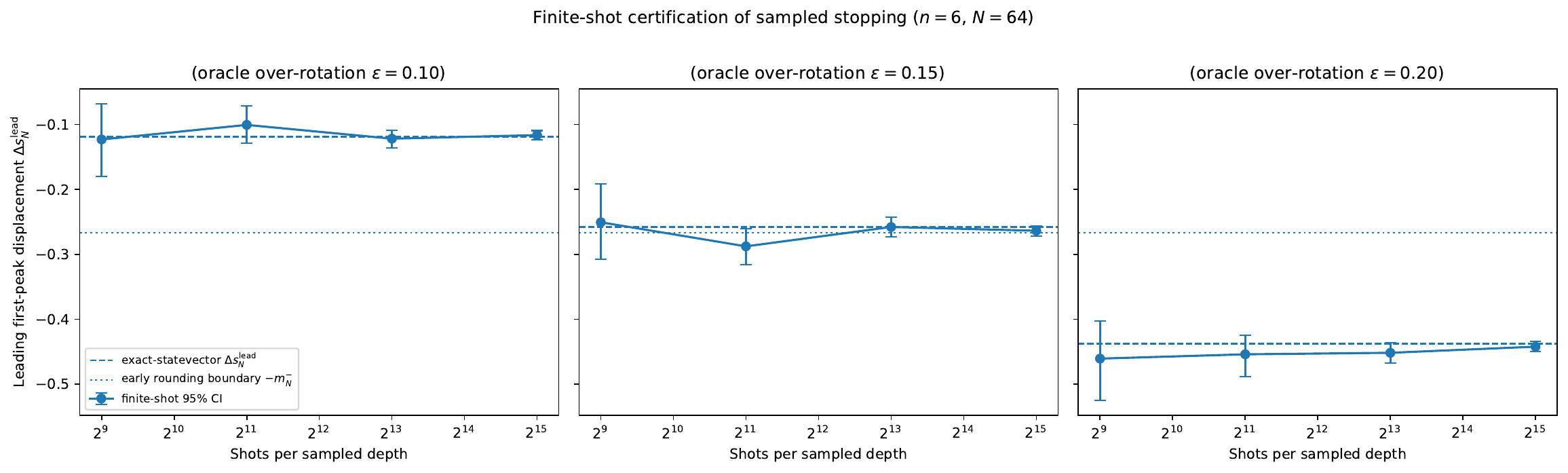}
    \caption{
    Finite-shot stopping decisions for \(n=6\), \(N=64\), using
    \(B_{\mathrm{boot}}=1000\) parametric-bootstrap refits at each shot budget.
    The three coherent-error cases illustrate the stopping hierarchy:
    \(\varepsilon=0.10\) retains the bootstrap-supported ideal step \(k_N^0=6\);
    \(\varepsilon=0.15\) returns the robust window \(\{5,6,7\}\); and
    \(\varepsilon=0.20\) supports the recalibrated step
    \(k_N^{\mathrm{rec}}=5\). Sampling uncertainty is propagated through the
    fitted normal-form parameters rather than absorbed into the deterministic
    sampled-fit discrepancy.
    }
    
% Figure description (not typeset):
% Finite-shot bootstrap stopping decisions for three coherent-error strengths
%         and four shot budgets, illustrating fixed-step certification, robust-window
%         coverage, and recalibrated stopping.

    \label{fig:finite-shot-stopping}
\end{figure}

Figure~\ref{fig:finite-shot-stopping} gives the full bootstrap decision plot for
the finite-shot experiment summarized in main-text
Subsection~5.4.  It separates the
three operational outcomes---retain the ideal step, return a robust local window,
and support a recalibrated depth---across the displayed shot budgets.

\subsection{Coherent-error statevector scaling across qubit counts}
\label{subsec:coherent_statevector_scaling}

\begin{figure}[t]
    \centering
    \includegraphics[width=\textwidth]{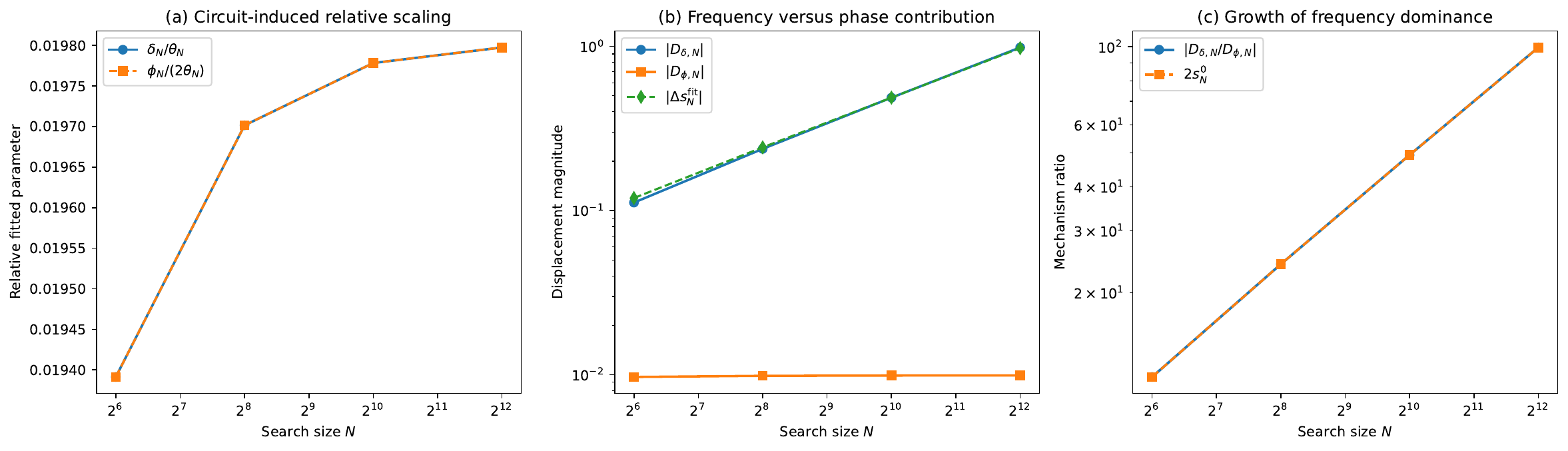}
    \caption{
    Coherent-error statevector scaling for \(n\in\{6,8,10,12\}\) with
    \(\varepsilon_N=0.8\theta_N\).
    (a) The relative fitted parameters approach a common constant.
    (b) The frequency contribution is consistent with \(\Theta(\sqrt N)\) growth,
    while the phase contribution remains consistent with \(\Theta(1)\).
    (c) Their ratio agrees with \(2s_N^0\).
    }
    
% Figure description (not typeset):
% Three scaling panels showing the fitted relative parameters, their
%         stopping-coordinate contributions, and the mechanism ratio.

    \label{fig:coherent_statevector_scaling}
\end{figure}

Using the protocol in Supplementary
Subsection~\ref{app:coherent_scaling_protocol}, all four statevector trajectories
fit the zero-damping normal form with residual below \(6\times10^{-14}\).
Figure~\ref{fig:coherent_statevector_scaling}(a) shows
\[
    \frac{\delta_N}{\theta_N}
    \approx
    \frac{\phi_N}{2\theta_N}
    \approx \text{constant},
\]
with finite-range exponents \(-0.496\) for both fitted parameters. Combining this
with \(s_N^0=\Theta(\sqrt N)\) predicts
\[
    D_{\delta,N}=\Theta(\sqrt N),
    \qquad
    D_{\phi,N}=\Theta(1),
\]
and panel~(b) gives exponents \(0.523\) and \(0.005\), respectively. Since
\(\phi_N=2\delta_N\), the exact mechanism ratio is
\[
    \left|\frac{D_{\delta,N}}{D_{\phi,N}}\right|=2s_N^0,
\]
which panel~(c) recovers with fitted exponent \(0.518\).

The sampled first-peak height remains between \(0.954\) and \(0.962\), while the
two larger instances select
\[
    k_N^*=24\neq k_N^0=25\quad(n=10),
    \qquad
    k_N^*=49\neq k_N^0=50\quad(n=12).
\]
Thus the explicit coherent circuit provides a supporting realization of the
speedup--stability separation: a high Grover-scale first peak can survive after
fixed same-step stability has failed.

\subsection{Finite-shot normal-form parameter uncertainty}
\label{subsec:app-finite-shot-parameters}

\begin{figure}[t]
    \centering
    \includegraphics[width=\linewidth]{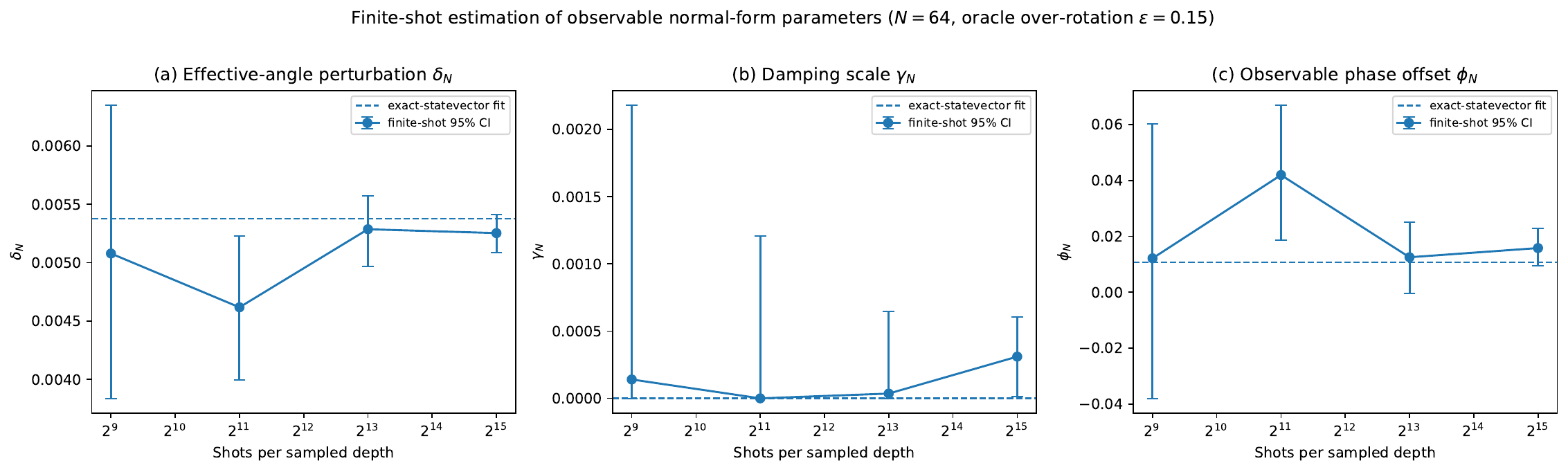}
    \caption{
    Finite-shot normal-form parameter uncertainty for the representative
    intermediate case \(n=6\), \(N=64\), and \(\varepsilon=0.15\).
    Percentile intervals from \(B_{\mathrm{boot}}=1000\) parametric-bootstrap
    refits are shown across the four shot budgets. The \(\delta_N\) and
    \(\phi_N\) intervals contract toward the exact statevector-fit baselines,
    while the constrained damping estimate contracts toward \(\gamma_N=0\).
    }
    
% Figure description (not typeset):
% Bootstrap percentile intervals for the fitted normal-form parameters across four
% shot budgets in the intermediate coherent-error case. The angle and phase
% intervals contract around their statevector baselines, while the constrained
% damping estimate contracts toward zero.

\label{fig:app-finite-shot-parameters}
\end{figure}

Figure~\ref{fig:app-finite-shot-parameters} reports the bootstrap uncertainty in
the fitted normal-form parameters for the representative intermediate case
\(\varepsilon=0.15\) used in the main finite-shot experiment. As the shot budget
grows, the intervals for \(\delta_N\) and \(\phi_N\) contract around their exact
statevector-fit values. The damping parameter contracts toward the zero-damping
baseline, with the expected one-sided behavior caused by the constraint
\(\gamma_N\geq0\). This representative case illustrates the parameter uncertainty
propagated through the bootstrap stopping-decision procedure summarized in
main-text Subsection~5.4 and shown in
Supplementary Figure~\ref{fig:finite-shot-stopping}.

\subsection{Joint damping--angle displacement slice}
\label{subsec:app-delta-gamma-slice}

\begin{figure}[t]
    \centering
    \includegraphics[width=\textwidth]{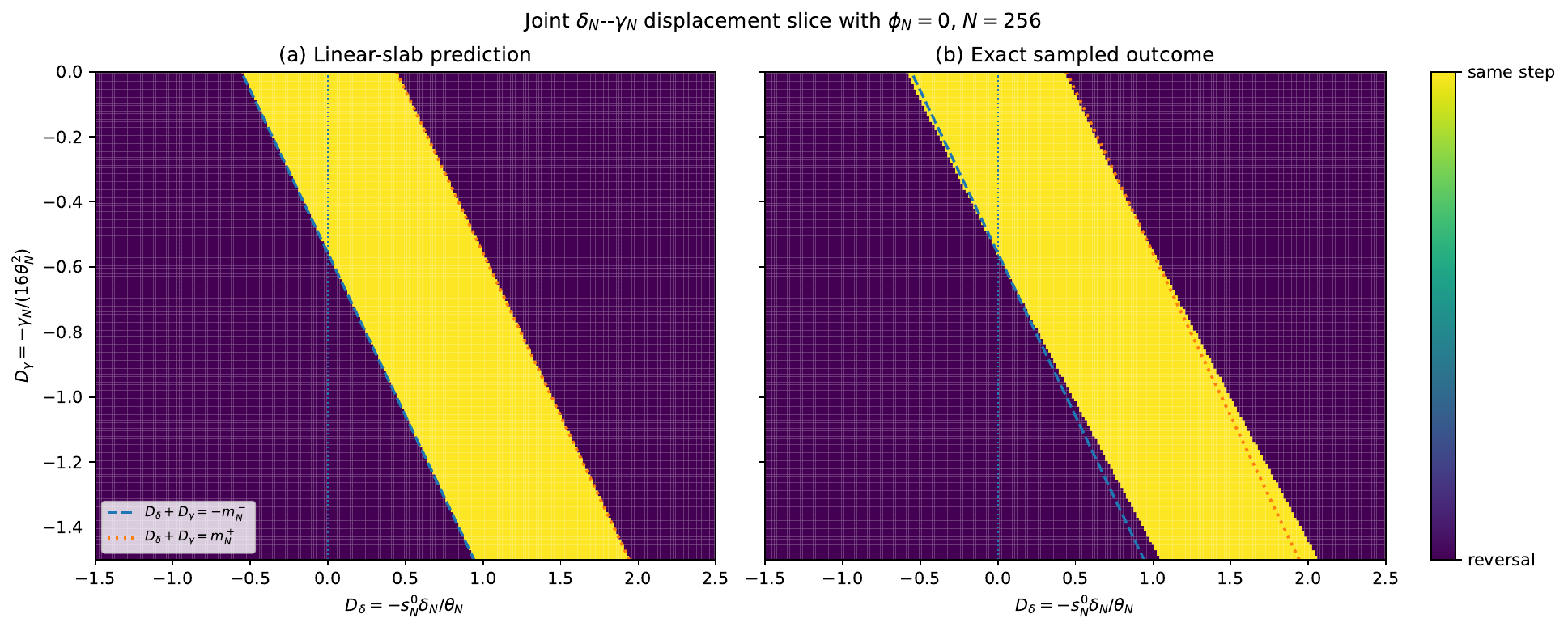}
    \caption{Joint effective-angle and damping displacement slice with
    \(\phi_N=0\) and \(N=256\). Panel (a) shows the linear-slab prediction in the
    normalized displacement coordinates
    \(D_\delta=-s_N^0\delta_N/\theta_N\) and
    \(D_\gamma=-\gamma_N/(16\theta_N^2)\). Panel (b) shows the directly
    evaluated sampled outcome. The comparison confirms the predicted signed
    displacement and cancellation structure. Because damping makes the lobe
    asymmetric, the continuous slab is a geometric no-correction diagnostic rather
    than an exact finite-\(N\) sampled boundary.}
    
% Figure description (not typeset):
% Two-panel joint effective-angle and damping scan at \(N=256\). The first panel
% shows the predicted affine continuous-cell slab in \(D_\delta+D_\gamma\); the
% second shows the directly evaluated sampled outcome and the effect of damping-induced
% asymmetry near the boundary.

\label{fig:app-delta-gamma-slice}
\end{figure}

Figure~\ref{fig:app-delta-gamma-slice} provides an additional two-parameter
validation of the combined signed-displacement condition. Supplementary
Figure~\ref{fig:combined-displacement} displays the
\((\delta_N,\phi_N)\) slice with \(\gamma_N=0\). Here we instead fix
\(\phi_N=0\) and vary the effective-angle and damping contributions jointly. In
the normalized displacement coordinates
\[
    D_\delta
    =
    -\frac{s_N^0\delta_N}{\theta_N},
    \qquad
    D_\gamma
    =
    -\frac{\gamma_N}{16\theta_N^2},
\]
the leading continuous-cell no-correction condition becomes the affine slab
\[
    -m_N^-
    <
    D_\delta+D_\gamma
    <
    m_N^+ .
\]
Thus damping participates in the same signed displacement mechanism as the
coherent frequency and phase parameters: it can add to or cancel an
effective-angle-induced shift depending on the sign of \(D_\delta\).

\subsection{Calibration-derived FakeOslo and held-out certificate}
\label{subsec:app-fakeoslo-heldout-figure}

\begin{figure}[t]
    \centering
    \includegraphics[width=\textwidth]{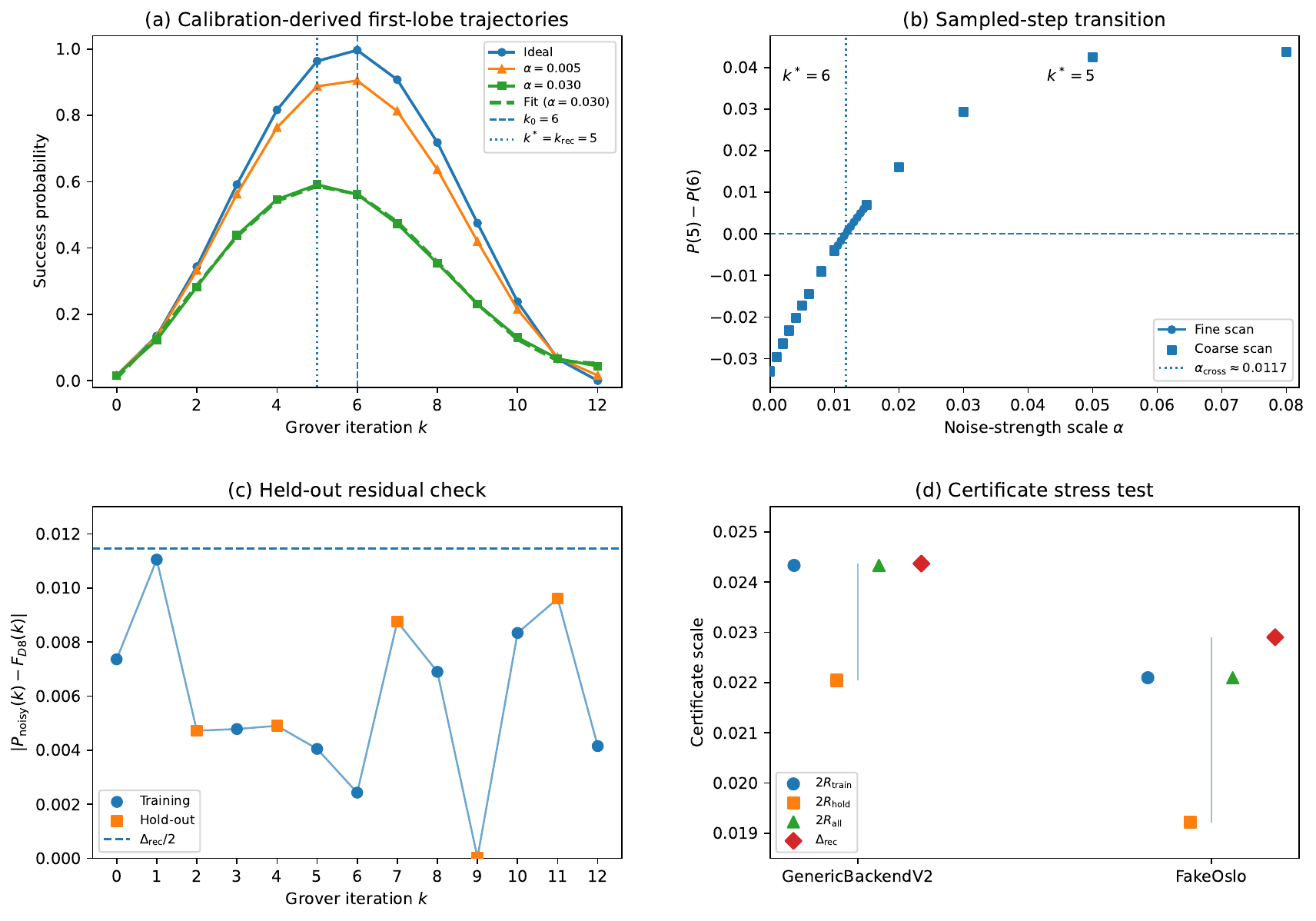}
    \caption{
    Calibration-derived FakeOslo validation and held-out certificate stress
    test for \(n=6\), \(N=64\).
    (a) Ideal, reference \((\alpha=0.005)\), and shifted
    \((\alpha=0.030)\) first-lobe trajectories, with the shifted D8 fit.
    (b) The sampled ordering changes from depth 6 to 5 at
    \(\alpha_{\mathrm{cross}}\approx0.0117\).
    (c) Absolute D8 fit residuals by depth; the orange points
    \(H=\{2,4,7,9,11\}\) are excluded from fitting, and the dashed line is
    \(\Delta_N^{\mathrm{rec}}/2\).
    (d) Train, held-out, and all-depth residual scales for the
    GenericBackendV2 and FakeOslo shifted cases. Both satisfy the residual-gap
    certificate; the GenericBackendV2 case has only a narrow positive all-depth margin.
    }
    
% Figure description (not typeset):
% Four-panel supplementary validation showing calibration-derived FakeOslo
%     first-lobe trajectories, the 6-to-5 sampled-step transition, residuals on
%     D8 training and held-out depths, and a certificate stress-test comparison
%     between GenericBackendV2 and FakeOslo.

    \label{fig:fakeoslo-heldout-validation}
\end{figure}

Figure~\ref{fig:fakeoslo-heldout-validation} summarizes the calibration-derived
validation from
Supplementary Subsection~\ref{app:fakeoslo-heldout-protocol}. It shows the
first-lobe transition under the FakeOslo calibration snapshot and separates the
D8 training residuals from the five depths excluded from fitting. The final
panel compares the train, hold-out, and all-depth residual scales against the
recalibrated gap for both backend constructions.

\FloatBarrier % place outstanding supplementary figures before References

\end{document}